\documentclass[a4paper,11pt]{article}
\pdfoutput=1 % if your are submitting a pdflatex (i.e. if you have
\usepackage{jheppub} % for details on the use of the package, please
\usepackage{amsmath}
\usepackage{slashed}
\usepackage{graphicx}
\usepackage{caption}
\usepackage{subcaption}
\usepackage{amssymb}
\usepackage{mathrsfs}
\usepackage{bm}
\usepackage{xspace}
\usepackage{booktabs}
\usepackage{xcolor}
\usepackage{listings}
\usepackage{verbatim}
\usepackage{mathtools}
\usepackage{hyperref}
\usepackage{breqn}
\usepackage{tikz}
\usetikzlibrary{positioning}
\usetikzlibrary{shapes.geometric, arrows}
\usetikzlibrary{arrows.meta}
\usepackage{multirow}

\newcommand\Eqn[1]     {Eq.\,(\ref{#1})}

\newcommand\eqn[1]     {eq.\,(\ref{#1})}

\newcommand{\fig}[1]{Fig.~\ref{fig:#1}}

\def\beq{\begin{equation}}
\def\eeq{\end{equation}}
\def\beqa{\begin{eqnarray}}
\def\eeqa{\end{eqnarray}}

\newcommand{\be}{\begin{equation}}
\newcommand{\ee}{\end{equation}}
\newcommand{\bea}{\begin{eqnarray}}
\newcommand{\eea}{\end{eqnarray}}

\newcommand{\ord}[1]{{\mathcal O}(#1)}

\newcommand{\as}{\alpha_s}

\newcommand{\nn}{\nonumber}

\newcommand{\ga}{\gamma}
\newcommand{\Ga}{\Gamma}
\newcommand{\de}{\delta}

\newcommand{\eps}{\epsilon}

\newcommand{\la}{\lambda}

\newcommand{\cF}{\mathcal{F}}

\newcommand{\cI}{\mathcal{I}}

\newcommand{\cO}{\mathcal{O}}

\allowdisplaybreaks[2]

\title{Region analysis of QED massive fermion form factor}
\author[2,4] {Jaco ter Hoeve,}
\author[1,2,3]{Eric Laenen,}
\author[1,2]{Coenraad Marinissen,}
\author[5]{Leonardo Vernazza,}
\author[1,6]{Guoxing Wang}

\affiliation[1]{Institute of Physics/Institute for Theoretical Physics Amsterdam, University of Amsterdam, Science Park 904, 1098 XH Amsterdam, The Netherlands}
\affiliation[2]{Nikhef, Theory Group, Science Park 105, 1098 XG, Amsterdam, The Netherlands}
\affiliation[3]{Intitute for Theoretical Physics, Utrecht University, Leuvenlaan 4, 3584 CE Utrecht, The Netherlands}
\affiliation[4]{Department of Physics and Astronomy, VU Amsterdam, 1081HV Amsterdam, The Netherlands}
\affiliation[5]{INFN Sezione di Torino, Via Pietro Giuria, 1, 10125 Torino TO, Italy}
\affiliation[6]{Laboratoire de Physique Th\'eorique et Hautes Energies (LPTHE), UMR 7589, Sorbonne Universit\'e et CNRS, 4 place Jussieu, 75252 Paris Cedex 05, France}

\abstract{
\noindent
We perform an analysis of the one- and two-loop massive quark form factor in QED in a region expansion, up to next-to-leading power in the quark mass. This yields an extensive set of regional integrals, categorized into three topologies, against which factorization theorems at next-to-leading power could be tested. Our analysis reveals a number of subtle aspects involving rapidity regulators, as well as additional regions that manifest themselves only beyond one loop, at the level of single diagrams, but which cancel in the form factor.}
\begin{document}

\maketitle

%%%%%%%%%%%%%%%%%%%%%%%%%%%%%%%%%%%%%%%%%%%%%%%%%%%%%%%%%%%%%%
\section{Introduction}
\label{sec:intro}
%%%%%%%%%%%%%%%%%%%%%%%%%%%%%%%%%%%%%%%%%%%%%%%%%%%%%%%%%%%%%%

The study of power corrections in scattering processes 
at hadron colliders has received increasing attention
in the past few years due to its importance for precision 
physics. Power corrections become relevant every time a 
scattering process involves two, or more, widely separated 
scales. This is a very common situation at hadron colliders: 
different scales arise not only due to the presence
of particles with very different masses. Often, 
different scales have a dynamical origin, related 
to the physical cuts necessary to select a given 
observable, or follow from the value of kinematic 
observables. Consider for instance a $n$-particle
scattering process in QED, with the emission of 
an additional soft photon with momentum $k$, 
and denote the corresponding amplitude 
${\cal M}_{n+1}$. This kinematic configuration 
is typical of processes occurring near threshold, 
where almost all the energy of the initial state particles 
goes into the required final state, such that extra 
radiation is constrained to be soft. The amplitude
is conveniently described as a power expansion in 
the ratio $\xi \sim E/Q \ll 1$, where $E$ 
represents the energy of the soft photon, and 
$Q$ is the invariant mass of the final state
\begin{equation}\label{powerexpansion}
{\cal M}_{n+1} = {\cal M}^{\rm LP}_{n+1} 
+ {\cal M}^{\rm NLP}_{n+1} + {\cal O}(\xi)\,,
\end{equation}
where the leading power (LP) term scales as
${\cal M}^{\rm LP}_{n+1} \sim 1/\xi$, and the 
next-to-leading power (NLP) contribution is
of order ${\cal M}^{\rm NLP}_{n+1} \sim \xi^0$.
In perturbation theory each coefficient of the 
power expansion contains large logarithms of the 
type $\as^n \log^m \xi \sim 1$, with $m$ up to 
$2n-1$, which need to be resummed to obtain 
precise predictions. 

Key to the resummation program
is the determination of how the scattering 
amplitude factorizes into simpler (single 
scale) objects, involving the corresponding 
non-radiative amplitude ${\cal M}_n$ as well 
as collinear and soft matrix elements describing 
soft and collinear radiation. Factorization 
analyses can be developed within the 
original theory (i.e., QED or QCD), 
see e.g.~\cite{Sterman:1987aj,Catani:1989ne,Catani:1990rp,Korchemsky:1993xv,Korchemsky:1993uz} for seminal papers. 
Alternatively this can be done by means of 
an effective theory, such as the soft-collinear 
effective field theory 
(SCET) \cite{Bauer:2000yr,Bauer:2001yt,Beneke:2002ph},
constructed to correctly reproduce
soft and collinear modes in the scattering 
process. The factorization structure of 
the LP amplitude in \Eqn{powerexpansion}
has been known for a long time, while
the study of ${\cal M}^{\rm NLP}_{n+1}$
is more recent. Within QCD, factorization 
theorems have been developed for specific 
cases, such as the case $n = 2$ corresponding
to Drell-Yan like processes~\cite{DelDuca:1990gz,Laenen:2010uz,Bonocore:2015esa,Bonocore:2016awd}
(see also \cite{Qiu:1990xxa,Qiu:1990xy} for 
factorization studies at NLP in 
$\Lambda^2_{\rm QCD}/Q^2$);
the study of ${\cal M}^{\rm NLP}_{n+1}$ 
for general $n$ has been initiated in 
\cite{Laenen:2020nrt} (see also 
\cite{Gervais:2017yxv}). Within SCET 
we refer to \cite{Beneke:2019oqx} for 
studies of the factorization structure 
of Drell-Yan at general subleading power, 
and to~\cite{Larkoski:2014bxa,Beneke:2017ztn,Beneke:2018rbh,Beneke:2019kgv} 
for various aspects of the factorization 
properties of the scattering amplitude 
${\cal M}^{\rm NLP}_{n+1}$.

As discussed in \cite{Laenen:2020nrt}, 
when studying the factorization of 
${\cal M}^{\rm NLP}_{n+1}$, it is useful 
to distinguish two contributions: one in 
which the radiation is emitted from the 
external legs, and another in which the 
radiation is emitted internally, from a particle within 
the hard scattering kernel. Schematically, 
this corresponds to separating the amplitude 
into two parts
\be\label{LBKeq}
{\cal M}_{n+1} = {\cal M}_{n+1}^{\rm ext} 
+ {\cal M}_{n+1}^{\rm int}\,.
\ee
This decomposition is useful, because the amplitude 
${\cal M}_{n+1}^{\rm int}$ can actually 
be obtained by means of the Ward identity 
from ${\cal M}_{n+1}^{\rm ext}$. In turn, 
it was shown that understanding the 
factorization properties of 
${\cal M}_{n+1}^{\rm ext}$ requires one
to understand the factorization 
properties of the corresponding 
non-radiative amplitude, ${\cal M}_{n}$. 
This is best seen by considering for 
instance a soft photon emission from 
an outgoing fermion $i$. In this case 
${\cal M}_{n+1}^{\rm ext}$ takes the 
form 
\be\label{LBKwrong}
{\cal M}_{n+1}^{\rm ext} = \bar u(p_i) (i e q_i \gamma^{\mu})
\, \frac{i(\slashed{p}_i+\slashed{k}+m)}{(p_i+k)^2 - m^2} 
\, {\cal M}_n(p_1,\ldots, p_i+k, \ldots p_n)\,,
\ee
where ${\cal M}_n$ represents the elastic 
amplitude (with the spinor $\bar{u}(p_i)$ 
stripped off).
Understanding the factorization of 
${\cal M}_{n+1}^{\rm ext}$ requires
the expansion of the non-radiative 
amplitude ${\cal M}_{n}$ for small 
$k$, but such expansion become 
non-trivial in presence of 
massless external particles, 
or external particles whose 
mass $m$ is much smaller compared 
to the other momentum invariants 
in the scattering, $m^2 \ll s_{ij}$
with $s_{ij} = (p_i + p_j)^2$. In 
this case, one needs to take into 
account that the non-radiative 
amplitude factorizes into 
non-trivial functions involving
configurations of virtual soft 
and collinear momenta. According 
to the power counting analysis 
developed in \cite{Laenen:2020nrt},
focusing only on the configurations 
which involve non-trivial collinear 
matrix elements, up to NLP 
the elastic amplitude factorizes
according to 
\begin{align} \label{NLPfactorization}
\mathcal{M}_{n}\big|^{\rm LP+NLP}_{\rm coll}
&= \bigg(\prod_{i=1}^n J_{(f)}^i\bigg) 
\otimes H \, S
+\sum_{i=1}^n \bigg(\prod_{j\neq i}J_{(f)}^j\bigg)
\Big[J^i_{(f\gamma)} \otimes H^i_{(f\gamma)}
+J^i_{(f\partial \gamma)} \otimes H^i_{(f\partial \gamma)}\Big]\,S 
\nonumber \\
&\quad + \sum_{i=1}^n \bigg(\prod_{j\neq i}J_{(f)}^j\bigg)
J^i_{(f\gamma\gamma)} \otimes H^i_{(f\gamma\gamma)}\, S
+ \sum_{i=1}^n \bigg(\prod_{j\neq i}J_{(f)}^j\bigg)
J^i_{(f\!f\!f)} \otimes H^i_{(f\!f\!f)}\, S 
\nonumber \\
&\quad+ \sum_{1\leq i \leq j \leq n} 
\bigg(\prod_{k\neq i,j}J_{(f)}^k \bigg)
J^i_{(f\gamma)}J^j_{(f\gamma)} \otimes H^{ij}_{(f\gamma)(f\gamma)}\, S\,,
\end{align}
where the functions $J_I$ and $S$ describe
long-distance collinear and soft virtual 
radiation in ${\cal M}_n$, and $H_{I}$ 
are hard functions, representing the 
contribution due to hard momenta 
configurations. In \Eqn{NLPfactorization}
the first term represents the LP 
contribution, while the second term in 
the first line starts at ``$\sqrt{\rm NLP}$", 
and the contributions in the second and 
third line start at NLP. \Eqn{NLPfactorization} 
is expected to be valid to all orders in
perturbation theory. In \cite{Laenen:2020nrt} 
some explicit checks have been provided at one 
loop, however, a more thorough test of the factorization 
formula requires at least a two-loop computation, 
since the functions in the second and third
lines of \Eqn{NLPfactorization} appear 
for the first time at NNLO.

The purpose of this paper is to provide 
data that can be used to validate 
\Eqn{NLPfactorization}. To this end
we consider the simplest QED process
that gives non-trivial contributions
to all the jet functions appearing in 
\Eqn{NLPfactorization}, namely, the 
annihilation of a massive fermion
anti-fermion pair of mass $m$ into 
an off-shell photon of invariant mass
$Q$ (or the time-reversed process of photon decay). More specifically, we consider the 
matrix element of the QED massive vector current 
$\bar \psi \gamma^{\mu} \psi$, 
which in turn is expressed in terms 
of two form factors $F_1(Q^2, m^2)$ and 
$F_2(Q^2, m^2)$. The two-loop result is 
known 
\cite{Bernreuther:2004ih,Gluza:2009yy}, 
(See also \cite{Blumlein:2020jrf}
for an earlier calculation of the corresponding 
contribution to the $e^+e^- \to \gamma^*/Z^{0*}$ 
cross section.)
and in recent years a lot of effort has 
been devoted to the calculation of the 
three loop correction \cite{Henn:2016kjz,Henn:2016tyf,Ablinger:2017hst,Lee:2018nxa,Lee:2018rgs,Ablinger:2018yae,Blumlein:2018tmz,Blumlein:2019oas,Fael:2022rgm,Fael:2022miw,Fael:2023zqr,Blumlein:2023uuq},
although no complete analytic result 
as yet exists.

For our purposes we need the two-loop 
small mass expansion of the form factors,
i.e. $m \ll Q$, which is also given in 
\cite{Bernreuther:2004ih,Gluza:2009yy}.
However, the small mass expansion alone 
does not provide enough information 
to compare with the corresponding 
factorized expression that one 
would obtain evaluating the form factor 
according to \Eqn{NLPfactorization}. 
Indeed, in the small mass limit 
it is possible to calculate the 
form factors with the method of 
expansion by momentum regions, 
\cite{Beneke:1997zp,Smirnov:2002pj}.
Within this approach, one assigns to
the loop momentum a specific scaling, 
which can be hard, collinear, soft, etc, 
with respect to the scaling of the external 
particle momenta. Each term defines a 
momentum region, and it is then possible 
to expand the form factors directly at 
the level of the integrand in the small 
parameters appearing in each region. The 
full result is recovered by summing over 
all regions. This approach is particularly
useful because we expect the jet functions 
in \Eqn{NLPfactorization} to be 
directly related with the collinear
and anti-collinear region 
contributions. (See
\cite{Bonocore:2014wua,Bahjat-Abbas:2018hpv}
for previous applications of the method of 
regions to study the correspondence between 
collinear regions and jet functions.)

To this end,  we 
present in this paper the calculation of the 
two-loop massive form factors 
evaluated within the method of 
regions. To the best of our 
knowledge this is the first 
time that the two-loop result 
is evaluated entirely within
this method. Thus far in literature
only single integrals involved in 
the two-loop massive form factors 
have been evaluated within the 
method of regions, see e.g. 
\cite{Smirnov:1998vk,Smirnov:1999bza,Smirnov:2002pj}.\footnote{Let us notice that a similar regional analysis involving the heavy-to-light form factor has been considered at two loops in \cite{Engel:2018fsb}.} 
Besides providing more data
for the comparison with 
\Eqn{NLPfactorization} 
(which will be considered 
in a forthcoming work),  this calculation 
actually has intrinsic value of its own. 

For instance, one feature of 
the region expansion, which was found 
in \cite{Smirnov:1998vk,Smirnov:1999bza,Smirnov:2002pj},
is that, at the level of single integrals, 
more regions appear at two loop, 
that were not present in the calculation 
at one loop. This is problematic from the 
point of view of an effective field theory
description, and in general for the 
derivation of factorization theorems 
valid at all orders in perturbation 
theory. It is clear that if new 
momentum modes appear at each order 
in perturbation theory, no factorization 
theorem can be expected to be valid to 
all orders in perturbation theory. 
However, it was already observed 
in \cite{Becher:2007cu} that, although 
new regions appear in single integrals 
at two loops, their contribution cancels 
when summing all diagrams, i.e. at the 
level of the form factors. The analysis 
in \cite{Becher:2007cu} considered only 
the LP terms in the small mass expansion;
our calculation shows that the ultra-collinear 
regions cancel also at NLP, at the level of 
the form factors. This result restores 
confidence in the all-order validity of 
factorization formulae such as 
\Eqn{NLPfactorization}, whose derivation 
is based on power counting arguments 
\cite{Laenen:2020nrt} using the momentum 
modes appearing at one loop. 

Another well-known feature of the method 
of regions is that expansion of the integrand 
in certain regions may render the integral 
divergent, even in dimensional regularization, 
such that additional analytic regulators are 
necessary in order to make the integral 
calculable. We check that this is indeed 
the case for the massive form factor, starting
at two loops. Analytic regulators 
were already applied in the past to 
the calculation of single integrals. 
In this work we 
need to apply analytic regulators to 
the calculation of different diagrams, 
which gives us the opportunity 
to discuss a few different regulators in detail
and verify their consistency by checking 
that the dependence on the analytic 
regulator cancel at the level of single
integrals, given that the full result 
does not require analytic regulators. 

The paper is structured as follows.
In Sect.~\ref{sec:FF}, we set up 
our notation, introduce the momentum 
region expansion and present the 
result for the regions contributing 
at one loop. We move then to 
considering the calculation at two 
loops. Sect.~\ref{sec:flow} describes the general approach we adopt for its expansion by regions, while Sect.~\ref{sec:details} presents the corresponding technical details. As will be discussed there, we compute diagrams categorized into three different topologies depending on the flow of the internal momenta, which we denote by $A$, $B$ and $X$. The main results are provided in Sect.~\ref{sec:results}, where we list explicit expressions of the form factor specified per region, up to NLP. We conclude and discuss our results in Sect.~\ref{sec:discussion}, pointing to several interesting subtleties we encountered and which can be relevant for future developments. 

In App.~\ref{sec:app-rapidity-reg} and \ref{sec:regions_X}, we present further technical details related to the use of rapidity regulators and the regional analysis performed in topology $X$, which is the most challenging of the three topologies.

%%%%%%%%%%%%%%%%%%%%%%%%%%%%%%%%%%%%%%%%%%%%%%%%%%%%%%%%%%%%%%
\section{Massive form factors}
\label{sec:FF}
%%%%%%%%%%%%%%%%%%%%%%%%%%%%%%%%%%%%%%%%%%%%%%%%%%%%%%%%%%%%%%

\begin{figure}[t]
\centering
\subfloat[]{
    \includegraphics[width=.3\textwidth]{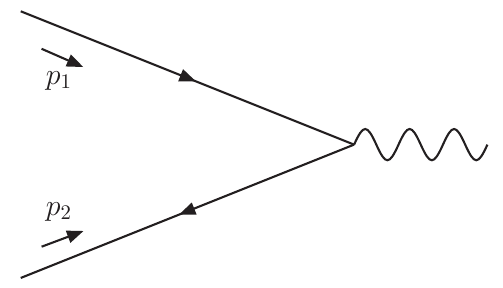}
}
\hspace{2cm}
\subfloat[]{
    \includegraphics[width=.3\textwidth]{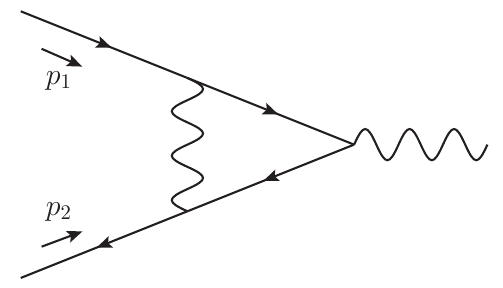}
}
\caption{Tree and one loop diagrams 
  contributing to the massive quark form factor in QED.}
\label{fig:FF-tree+one-loop}
\end{figure}
In this paper we consider the quark-antiquark
annihilation process
\be
q(p_1) + \bar q(p_2) \to \gamma^*(q),
\ee
whose tree and one loop contribution 
in QED is given respectively in Fig.~\ref{fig:FF-tree+one-loop} (a) and (b). 
Following \cite{Bernreuther:2004ih}, 
the corresponding vertex function $V^{\mu}(p_1,p_2)$ 
is expressed in terms of two form factors, 
$F_1$ and $F_2$, as follows:
\bea\label{VertexFF}
V^{\mu}(p_1,p_2) &=& 
\bar v(p_2) \, \Gamma^{\mu}(p_1,p_2)
\, u(p_1), \\[0.1cm]
\label{FFsDef}
\Gamma^{\mu}(p_1,p_2) &=& -i \, e \, e_q \,
\bigg[F_{1}\big(s,m^2\big)\gamma^{\mu}
+ \frac{1}{2m} F_{2}\big(s,m^2\big) \,
i \, \sigma^{\mu\nu} q_{\nu} \bigg],
\eea
where $\sigma^{\mu\nu} = \frac{i}{2}[\gamma^{\mu},\gamma^{\nu}]$;
furthermore, $s = (p_1 + p_2)^2$ represents the center of mass 
energy, and $m$ is the quark mass. The form factors $F_{i},\, i=1,2$ 
can be extracted by applying projection 
operators:
\be
F_{i}\big(s,m^2\big) 
= {\rm Tr}\big[P_{i}^{\mu}(m,p_1,p_2) 
\, \Gamma_{\mu}(p_1,p_2) \big],
\ee
where 
\be
P_{i}^{\mu}(m,p_1,p_2)
= \frac{\slashed{p}_1 + m}{m}
\bigg[i\, g_1^{(i)} \gamma^{\mu} 
+ \frac{i}{2m} g_{2}^{(i)} 
\big( p_2^{\mu} - p_1^{\mu}\big) \bigg]
\frac{\slashed{p}_2 - m}{m},
\ee
and\footnote{The definition of 
$g_2^{(2)}$ in \cite{Bernreuther:2004ih}
has a typo; here we follow the definition 
given in Eq.~(2.7) of \cite{Ablinger:2017hst}.}
\bea\label{projectors} \nn
g_1^{(1)} &=& -\frac{1}{e N_c} 
\frac{1}{4(1-\eps)} \frac{1}{(s/m^2-4)}, \\ \nn
g_2^{(1)} &=& \frac{1}{e N_c} 
\frac{3-2\eps}{(1-\eps)} \frac{1}{(s/m^2-4)^2}, \\ \nn
g_1^{(2)} &=& \frac{1}{e N_c} 
\frac{1}{(1-\eps)} \frac{1}{s/m^2(s/m^2-4)}, \\ 
g_2^{(2)} &=& -\frac{1}{e N_c} 
\frac{1}{(1-\eps)} \frac{1}{(s/m^2-4)^2}
\bigg[\frac{4 m^2}{s} + 2 - 2\eps \bigg].
\eea
In this paper we work in $d=4-2\eps$ dimensions, compute the unsubtracted form factor and omit counterterm insertions, which makes the mass $m$ the unrenormalized mass.
The form factors have the perturbative expansion
\begin{align}
F_{i}\left(s,m^2,\mu^2\right)&= 
F^{(0)}_{i}\left(s,m^2\right) 
+ \frac{e_q^2\,\alpha_{\rm EM}}{4\pi} \, 
F^{(1)}_{i}\left(s,m^2,\mu^2\right) \nn\\
&\hspace{3cm}+ \bigg(\frac{e_q^2\,\alpha_{\rm EM}}{4\pi}\bigg)^2 
F^{(2)}_{i}\left(s,m^2,\mu^2\right) 
+ {\cal O}\left(\alpha_{\rm EM}^3\right)\, ,\label{asexpansion}
\end{align}
where
\be
F^{(0)}_{1}\left(s,m^2\right) = 1, \qquad \qquad \qquad 
F^{(0)}_{2}\left(s,m^2\right) = 0.
\ee
We are interested in computing the higher order 
corrections in the small mass (or high-energy) 
limit $m^2/s \sim \lambda^2 \ll 1$. 
We assume the center of mass frame, with the 
incoming quark moving along the positive $z$-axis. The 
momenta of the quark and anti-quark can then 
be decomposed along two light-like directions, 
$n_\pm=(1,0,0,\mp 1)$
as follows:
\bea \nn
p^{\mu}_1 &=& \left(\sqrt{p^2+m^2},0,0,p\right)
= p_1^+ \frac{n_-^{\mu}}{2} + p_1^-\frac{n_+^{\mu}}{2}, \\ 
p^{\mu}_2 &=& \left(\sqrt{p^2+m^2},0,0,-p\right)
= p_2^+ \frac{n_-^{\mu}}{2} + p_2^-\frac{n_+^{\mu}}{2}\,.
\eea
In the small mass limit the $p_{i}^{\pm}$
components have the scaling properties
\begin{align} \nn
p_1^+ &=  n_+ \cdot p_1 = p_2^- = n_- \cdot p_2 = \sqrt{p^2+m^2}+p\sim \sqrt{s}, \\
p_1^- &=  n_- \cdot p_1 = p_2^+ = n_+ \cdot p_2 = \sqrt{p^2+m^2}-p\sim \la^2\sqrt{s}. 
\label{eq:scaling_rel}
\end{align}
In what follows it will prove useful to 
define a variable 
\be 
\hat{s} \equiv p_1^+p_2^- 
= \left(\sqrt{m^2+p^2}+p\right)^2\,,
\ee
such that 
\begin{align}
s %=(p_1+p_2)^2  
= 2m^2 + p_{1}^+ p_{2}^- 
+ p_{1}^- p_{2}^+  = 2m^2 + \hat{s} 
+ \frac{m^4}{\hat{s}}.
\end{align}
We calculate the higher order corrections 
to the form factor in dimensional regularization, 
and use the method of expansion by regions 
\cite{Beneke:1997zp,Smirnov:2002pj,Jantzen:2011nz}
to evaluate the loop integrals in the limit 
$m^2 \ll s$. In general one expects several regions
to contribute. It is possible to use geometric 
methods (see e.g. \cite{Pak:2010pt,Jantzen:2012mw,Gardi:2022khw})
to reveal all regions contributing to an integral. 
In case of the problem at hand we find it is still 
possible to find all regions contributing up 
to two loops by straightforward inspection of the 
propagators in the loops. One advantage of
this method is that the regions are directly
associated to the scaling of the loop momenta,
rather than to the scaling of Feynman parameters,
as when geometric methods are used. 
This will allow us to relate more easily our results
to the construction of an effective field theory
description of the quark form factor, as effective
field theories are typically constructed to reproduce
the momentum regions of the present problem. 

In what follows we decompose a generic momentum 
$k$ along the light-like directions $n_{\pm}$:
\be\label{eq:SudakovDecomposition}
k^{\mu} = k^+ \frac{n_-^{\mu}}{2} + k^-\frac{n_+^{\mu}}{2} + k_{\perp}^{\mu}, 
\qquad \qquad \quad 
k^{\mu} = (k^+, k^-, k_{\perp}),
\ee
where $k_{\pm} = n_{\pm} \cdot k$. The second 
identity in the equation above provides a compact notation to indicate scaling relations. Up to one 
loop the following loop momentum modes contribute:
\begin{align}
\text{Hard ($h$):} \qquad 
&k\sim\sqrt{\hat{s}}\,
(\lambda^0,\lambda^0,\lambda^0)\,, \nn \\
\text{Collinear ($c$):} \qquad  
&k\sim\sqrt{\hat{s}}\,
(\lambda^0,\lambda^2,\lambda^1)\,, \label{eq:momentumModes1L} \\ 
\text{anti-Collinear ($\bar{c}$):} \qquad 
&k\sim\sqrt{\hat{s}}\,
(\lambda^2,\lambda^0,\lambda^1)\,. \nn 
\end{align}
As we will see in what follows, 
beyond one loop we find that 
two additional momentum modes
are necessary, which scale as
\begin{align}
\text{Ultra-Collinear ($uc$):} \qquad 
&k\sim\sqrt{\hat{s}}\,
(\lambda^2,\lambda^4,\lambda^3)\,, \nn \\ 
\text{Ultra-anti-Collinear ($\overline{uc}$):} \qquad 
&k\sim\sqrt{\hat{s}}\,
(\lambda^4,\lambda^2,\lambda^3)\,. \label{eq:MomentumModesUc}
\end{align}
One might also expect the following modes
to contribute:\footnote{In literature 
the modes of \eqn{eq:MomentumModesS} are 
sometimes referred to as \emph{soft}
and \emph{ultra-soft} respectively
\cite{Beneke:2002ph}.}
\begin{align} \label{eq:MomentumModesS}
\text{Semi-Hard ($sh$):} \qquad 
&k\sim\sqrt{\hat{s}}\,
(\lambda^1,\lambda^1,\lambda^1)\,, \nn \\ 
\text{Soft ($s$):} \qquad 
&k\sim\sqrt{\hat{s}}\,
(\lambda^2,\lambda^2,\lambda^2)\,,
\end{align}
however, these turn out to give rise to 
scaleless loop integrals and therefore 
do not contribute up to two-loop level.\footnote{In the presence of rapidity divergences this depends on the type of rapidity regulator, which we discuss in detail in Sec.~\ref{sec:details} and App.~\ref{sec:app-rapidity-reg}.}

Throughout the paper we define the loop integration measure  in dimensional
regularization as follows:
\be
\int [d k] \equiv 
\bigg(\frac{\mu^2 e^{\gamma_E}}{4\pi}\bigg)^{\eps}
\int \frac{d^{4-2\eps}k}{(2\pi)^{4-2\eps}}. 
\ee
In the limit $m^2\ll s$ we express the form 
factors as a power expansion in $m^2/\hat s$,
and calculate the first two terms in the 
expansion. In general, the form factor is 
given as a sum over the contributing regions.
At one loop we have for instance
\be
F_i^{(1l)}\bigg(\frac{\mu^2}{\hat s}
,\frac{\mu^2}{m^2},\eps\bigg) = 
F_i^{(1l)}\Big|_h\bigg(\frac{\mu^2}{\hat{s}},\eps\bigg) 
+ F_i^{(1l)}\Big|_c\bigg(
\frac{\mu^2}{m^2},\eps\bigg)
+F_i^{(1l)}\Big|_{\bar c}\bigg(
\frac{\mu^2}{m^2},\eps\bigg).
\ee
Each region is expected 
to depend non-analytically (logarithmically) on a single 
scale, which is dictated by the kinematics
of the process. We find that the 
non-analytic dependence of the hard 
region is conveniently given in terms
of the factor $\hat s$, while the 
non-analytic structure of the collinear
and anti-collinear regions is given
in terms of the mass $m$. At one loop
only the single diagram of Fig.~\ref{fig:FF-tree+one-loop}(b) contributes 
to the form factors and the region expansion 
is easy. The expansion of the corresponding 
scalar integral has been discussed at length 
in appendix A of \cite{Laenen:2020nrt}, to 
which we refer for further details. Here
we simply report the result, and postpone 
a more technical discussion concerning the 
region expansion for the two-loop calculation 
to Sec.~\ref{sec:flow}. For the form factor 
$F_1$ we have 
\bea\label{eq:FF1-1l-h} \nn
F_1^{(1l)}\Big|_h  &=& 
\bigg(\frac{\mu^2}{-\hat s -i0^+}\bigg)^{\eps}
\bigg\{ - \frac{2}{\eps^2} - \frac{3}{\eps} 
- 8 + \zeta_2 + \eps \bigg(-16 + \frac{3\zeta_2}{2} 
+ \frac{14\zeta_3}{3}\bigg) \\
&&\hspace{2.0cm}
+\, \frac{m^2}{\hat s} \bigg[- \frac{2}{\eps} -6 
+ \eps \big(-16 + \zeta_2\big) \bigg] 
+{\cal O}(\eps^2)+ {\cal O}(\lambda^4) \bigg\},
\eea
for the hard region, and 
\bea\label{eq:FF1-1l-c} \nn
F_1^{(1l)}\Big|_c  &=& 
\bigg(\frac{\mu^2}{m^2-i0^+}\bigg)^{\eps}
\bigg\{\frac{1}{\eps^2} + \frac{2}{\eps} + 4  
+ \frac{\zeta_2}{2} + \eps \bigg(8 + \zeta_2 
- \frac{\zeta_3}{3}\bigg) \\ 
&&\hspace{2.0cm} 
+\, \frac{m^2}{\hat s} \bigg[\frac{1}{\eps} 
+ 5 + \eps \bigg(13 + \frac{\zeta_2}{2}\bigg) \bigg] 
+{\cal O}(\eps^2)+ {\cal O}(\lambda^4) \bigg\},
\eea
for the collinear region, and
\be \label{eq:FF1-1l-cb}
F_1^{(1l)}\Big|_{\bar c} = F_1^{(1l)}\Big|_c.
\ee
In the prefactors in Eqs.~\eqref{eq:FF1-1l-h} and~\eqref{eq:FF1-1l-c}, we have explicitly written the Feynman prescription $i0^+$, which upon expanding in $\eps$ give logarithms $\log(\mu^2/(-\hat{s}-i0^+))$ and $\log(\mu^2/(m^2-i0^+))$. For $\mu^2>0, m^2>0$ and $\hat{s}> 0$ these can be rewritten using $\log(\mu^2/(-\hat{s}-i0^+))\rightarrow \log(\mu^2/\hat{s}) + i\pi$ and $\log(\mu^2/(m^2-i0^+))\rightarrow \log(\mu^2/m^2)$ to obtain the imaginary parts. For notational convenience, we will drop the Feynman prescription in what follows and note that these can always be reinstated by $\hat{s}\rightarrow \hat{s} + i0^+$ and $m^2\rightarrow m^2 -i0^+$ after which the imaginary parts can be retrieved adopting the rule described above.

The form factor $F_2$ starts 
at NLP. We have 
\bea\label{eq:FF2-1l-h} 
F_2^{(1l)}\Big|_h &=&
\bigg(\frac{\mu^2}{-\hat s}\bigg)^{\eps}
\bigg\{\frac{m^2}{\hat s} \bigg[\frac{4}{\eps} 
+ 16 + \eps \big(32 - 2\zeta_2\big) \bigg]
+{\cal O}(\eps^2)+ {\cal O}(\lambda^4) \bigg\}, 
\eea
and
\bea\label{eq:FF2-1l-c} 
F_2^{(1l)}\Big|_c &=&
\bigg(\frac{\mu^2}{m^2}\bigg)^{\eps}
\bigg\{\frac{m^2}{\hat s} \bigg[- \frac{2}{\eps} 
- 8 + \eps \big(-16 - \zeta_2 \big) \bigg] 
+{\cal O}(\eps^2)+ {\cal O}(\lambda^4) \bigg\}, 
\eea
and
\be\label{eq:FF2-1l-cb}
F_2^{(1l)}\Big|_{\bar c} = F_2^{(1l)}\Big|_c.
\ee
These results can be compared directly with 
\cite{Bernreuther:2004ih} by extracting the 
coefficients ${\cal F}^{(1)}_{i}(s,\mu)$ 
as defined in Eq. (18) there, as follows
\be
{\cal F}^{(1)}_{i}(s,\mu) = 
\frac{e^{-\eps \gamma_E}}{\Gamma(1 + \eps)} 
\bigg(\frac{\mu^2}{m^2}\bigg)^{-\eps} F^{(1)}_{i}(s,\mu).
\ee
Summing over the regions and expanding  also the scale factors in 
powers of $\eps$ we find 
\bea \nn
{\cal F}^{(1)}_{1}(s,\mu) &=& 
\bigg\{ \frac{1}{\eps} 
\bigg[1-2\ln\bigg(-\frac{m^2}{\hat s}\bigg) \bigg]
- 3 \ln\bigg(-\frac{m^2}{\hat s}\bigg)
- \ln^2\bigg(-\frac{m^2}{\hat s}\bigg) + 2 \zeta_2 \\ 
&&\hspace{1.0cm}
+\, \frac{m^2}{\hat s} \bigg[4 
- 2 \ln\bigg(-\frac{m^2}{\hat s}\bigg) \bigg]
+{\cal O}(\eps)+ {\cal O}(\lambda^4) \bigg\},
\eea
and 
\bea \nn
{\cal F}^{(1)}_{2}(s,\mu) &=& 
4 \frac{m^2}{\hat s} 
\ln\bigg(-\frac{m^2}{\hat s}\bigg)
+{\cal O}(\eps)+ {\cal O}(\lambda^4),
\eea
in agreement with the high-energy expansion
$s \gg m^2$ of eqs. (19) and (20) of 
\cite{Bernreuther:2004ih}.

The massive form factor at two-loop was first computed in QED in \cite{Bonciani:2003ai}, followed by \cite{Bernreuther:2004ih}, which considered its generalization to QCD. The former provides the full result at the level of the individual diagrams, while the latter presents results with all diagrams combined, see Eqs.~(22) and (23) in \cite{Bernreuther:2004ih} for $F_1$ and $F_2$ respectively for more details.

%%%%%%%%%%%%%%%%%%%%%%%%%%%%%%%%%%%%%%%%%%%%%%%%%%%%%%%%%%%%%%
\section{Calculational steps}
\label{sec:flow}
%%%%%%%%%%%%%%%%%%%%%%%%%%%%%%%%%%%%%%%%%%%%%%%%%%%%%%%%%%%%%%
Here we describe the general approach that we adopt throughout the rest of this work, deferring a discussion of technical details  to Sec.~\ref{sec:details}. In Sec.~\ref{subsec:diagrams}, we present the diagrams that contribute to the two-loop massive form factor, followed by a discussion of their associated integrals and their classification into three different topologies\footnote{We adopt the definition of topology as given in \cite{Studerus:2009ye} in the context of IBP reductions for Feynman integrals.} denoted by $A$, $B$ and $X$ in  Sec.~\ref{subsec:topologies}. We conclude the section with a brief summary in Sec.~\ref{subsec:summary_details} where we preview various subtleties that we shall encounter in Sec.~\ref{sec:details}.

%%%%%%%%%%%%%%%%%%%%%%%%%%%%%%%%%%%%%%%%%%%%%%%%%%%%%%%%%%%%%%
\subsection{Diagrams contributing to the two-loop form factor}
\label{subsec:diagrams}
%%%%%%%%%%%%%%%%%%%%%%%%%%%%%%%%%%%%%%%%%%%%%%%%%%%%%%%%%%%%%%
\begin{figure}
\centering
\subfloat[]{
    \includegraphics[width=.22\textwidth]{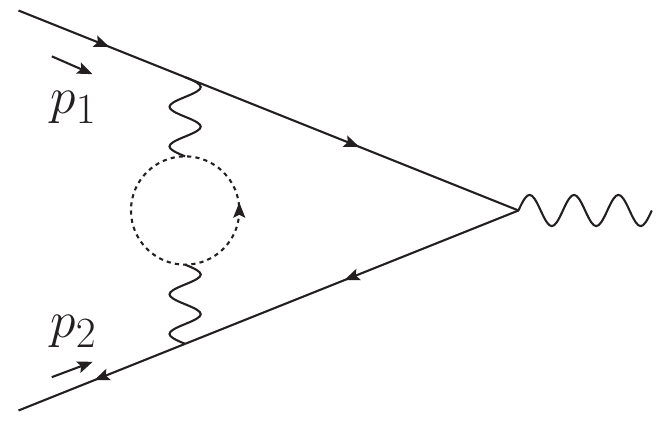}
}
\label{fig:QED_8}
\subfloat[]{
    \includegraphics[width=.22\textwidth]{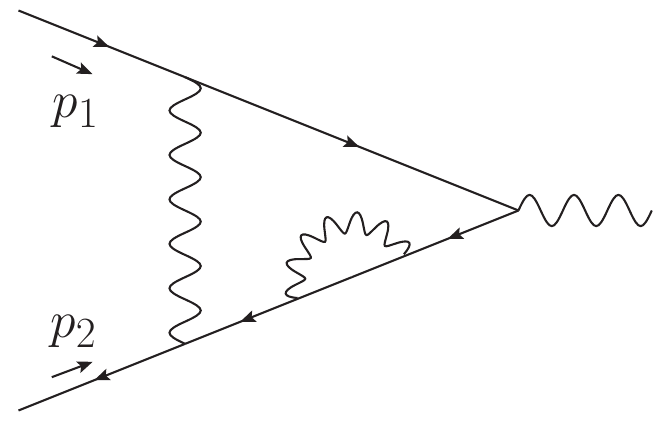}
}
\label{fig:QED_5}
\subfloat[]{
    \includegraphics[width=.22\textwidth]{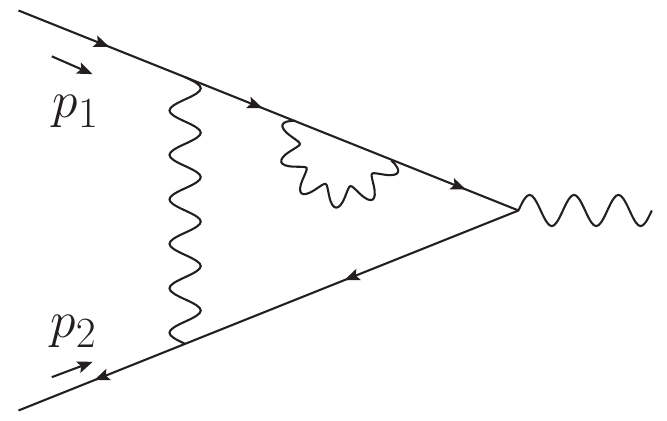}
}
\label{fig:QED_6}
\subfloat[]{
    \includegraphics[width=.22\textwidth]{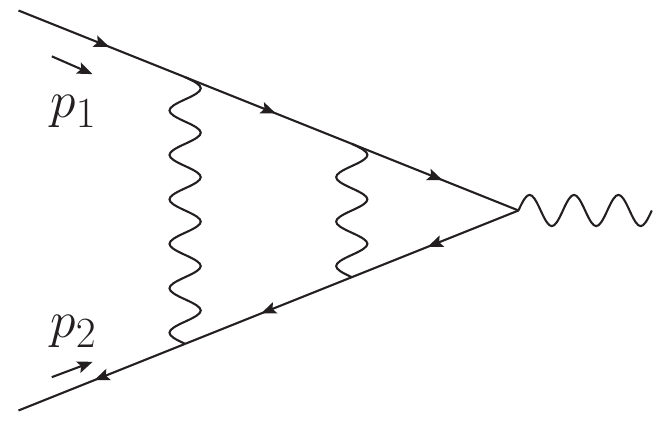}
}
\label{fig:QED_2}
\subfloat[]{
    \includegraphics[width=.22\textwidth]{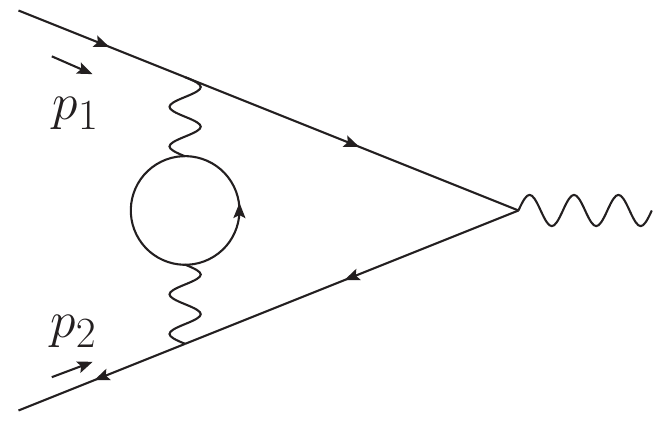}
}
\label{fig:QED_7}
\subfloat[]{
    \includegraphics[width=.22\textwidth]{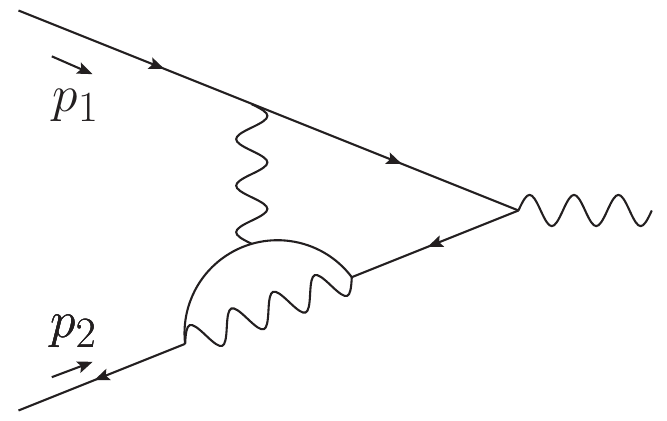}
}
\label{fig:QED_3}
\subfloat[]{
    \includegraphics[width=.22\textwidth]{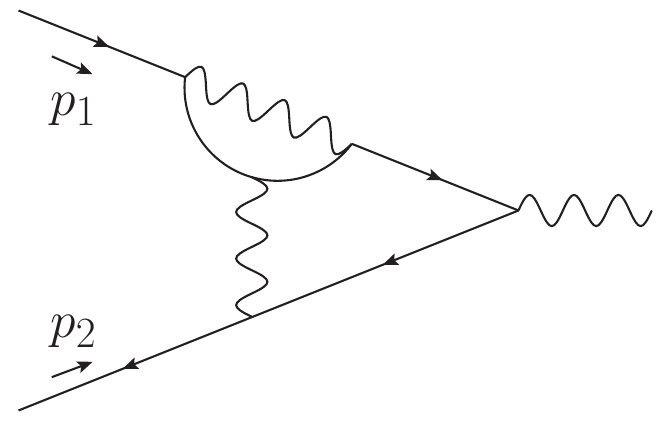}
}
\label{fig:QED_4}
\subfloat[]{
    \includegraphics[width=.22\textwidth]{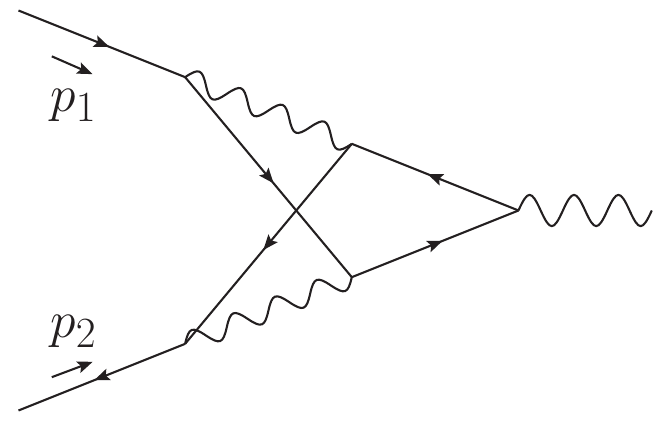}
}
\label{fig:QED_1}
\caption{Diagrams that contribute to the massive form factor at two-loop in QED. Dashed lines represent massless fermions.}
\label{fig:QED_diagrams}
\end{figure}

\begin{figure}
     \centering
     \begin{subfigure}[b]{0.45\textwidth}
         \centering
         \includegraphics[width=.60\textwidth]{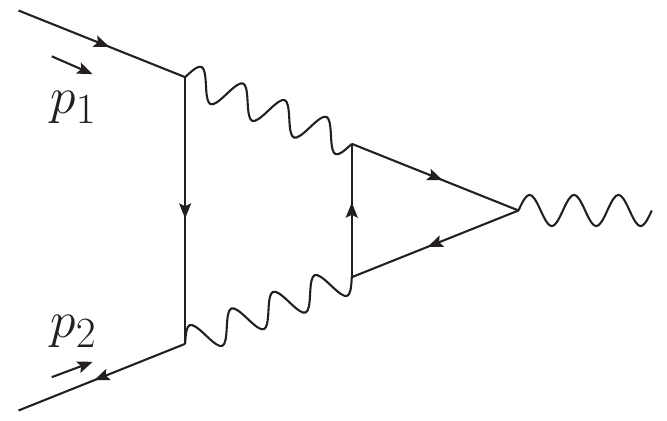}
        \label{fig:QED_9_Furry}
     \end{subfigure}
     \begin{subfigure}[b]{0.45\textwidth}
         \centering
         \includegraphics[width=.60\textwidth]{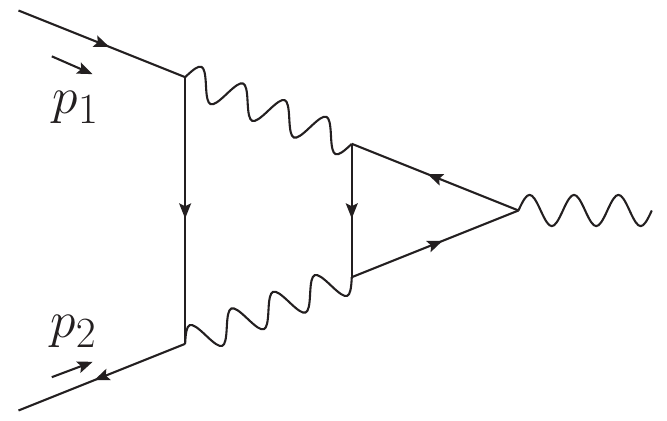}
     \end{subfigure}
\caption{Diagrams with a closed fermion loop that cancel in QED by Furry's theorem \cite{PhysRev.51.125}.}
      \label{fig:QED_Furry}
\end{figure}

The diagrams that contribute to the massive form factor at two-loop are displayed in \fig{QED_diagrams}. There are eight diagrams in total, labeled (a) - (h), with $p_1$ and $p_2$ denoting the external momenta of the two incoming fermions. Solid and dashed lines are used to represent massive and massless fermions respectively. Fig.~\ref{fig:QED_Furry} suggests there are two additional diagrams to account for, but these diagrams cancel by Furry's theorem \cite{PhysRev.51.125}.
Concerning diagram (e), note that the fermion running inside the loop does not need to correspond to those on the external lines, but here we ignore this possibility for simplicity as it would introduce an additional hierarchy of scales that makes the power counting much more involved. Diagrams (b) and (c), as well as (f) and (g), are related by exchanging $p_1$ and $p_2$ and therefore we expect this symmetry to be present also during an expansion by regions. 

In anticipation of our regional analysis, it is convenient to classify diagrams (a) to (h) into three different topologies labeled $A$, $B$ and $X$ which are distinguished by the flow of their internal momenta. 

%%%%%%%%%%%%%%%%%%%%%%%%%%%%%%%%%%%%%%%%%%%%%%%%%%%%%%%%%%%%%%
\subsection{Topology classification}
\label{subsec:topologies}
%%%%%%%%%%%%%%%%%%%%%%%%%%%%%%%%%%%%%%%%%%%%%%%%%%%%%%%%%%%%%%
Starting with diagrams (a)-(d), we note that the Feynman integrals contributing to these diagrams share the following parameterization 
\begin{align}
\nn I_{A;\{n_i\}}
&\equiv \int[dk_1][dk_2]
\frac{1}{k_1^{2n_1}}
\frac{1}{k_2^{2n_2}}
\frac{1}{[(k_1-k_2)^2]^{n_3}}
\frac{1}{[(k_1+p_1)^2-m^2]^{n_4}}
\frac{1}{[(k_2+p_1)^2-m^2]^{n_5}}\\
&\hspace{5cm}\times
\frac{1}{[(k_1-p_2)^2-m^2]^{n_6}}
\frac{1}{[(k_2-p_2)^2-m^2]^{n_7}}\,,
\label{eq:topoA_full}
\end{align}
which we define as topology $A$. In Eq.~\eqref{eq:topoA_full}, $k_1$ and $k_2$ denote the internal loop momenta and the integer $n_i$ represents the generic power associated to the $i$th propagator. 

In a similar way, the integrals associated to diagrams (e), (f) and (g) in Fig.~\ref{fig:QED_diagrams} can be parameterized by
\begin{align}
I_{B;\{n_i\}}^{b_3,b_4,b_5,b_6} 
&\equiv \int[dk_1][dk_2]
\frac{1}{[k_1^2-m^2]^{n_1}}
\frac{1}{[k_2^2-m^2]^{n_2}}
\frac{\mu_3^{2b_3}}{[(k_1-p_1)^2]^{n_3+b_3}}\label{eq:topoB_full}\\
&\hspace{-1.5cm}\times
\frac{\mu_4^{2b_4}}{[(k_1+k_2-p_1)^2-m^2]^{n_4+b_4}}
\frac{\mu_5^{2b_5}}{[(k_1+p_2)^2]^{n_5+b_5}}
\frac{\mu_6^{2b_6}}{[(k_1+k_2+p_2)^2-m^2]^{n_6+b_6}}
\frac{1}{[(k_1+k_2)^2]^{n_7}}\,,
\nn
\end{align}
and this defines topology $B$. An important distinction compared to topology $A$ is the appearance of the complex numbers $b_i$ associated to propagators $3,4,5$ and 6, where artificial scales $\mu_i$ with unit mass dimensions have been introduced on dimensional grounds. The need for the powers $b_i$ can be seen as follows. When expanding in momentum regions, one finds eikonal propagators that contain only the $k_i^+$ or $k_i^-$ momentum components. As a result, additional divergences may arise from the $k_i^+$ and $k_i^-$ integrals because the dimensional parameter $\eps$ regulates only the transverse momentum components $k_{i,\perp}$. 
Various regulators have been introduced in the literature to tame these rapidity divergences, e.g.~space-like Wilson-lines \cite{collins_2023}, $\delta$ regulators \cite{Echevarria:2011epo,Echevarria:2012js,Echevarria:2015usa,Echevarria:2015byo,Echevarria:2016scs}, $\eta$ regulators \cite{Chiu:2011qc,Chiu:2012ir}, exponential regulators \cite{Li:2016axz}, analytic regulators \cite{Beneke:2003pa,chiu:2007yn,Becher:2010tm,Becher:2011dz} and pure rapidity regulators \cite{Ebert:2018gsn}. In this work, we adopt the analytic regulator \cite{Beneke:2003pa}, meaning that we raise the relevant propagators to complex powers $b_i$. The rapidity divergences then manifest themselves as poles in $b_i$, similar to poles in $\eps$ that one encounters in dimensional regularization. As will be described in greater detail in Sec.~\ref{sec:topoB}, one does not need to add all four regulators $b_i$ simultaneously to regulate the rapidity divergences present in diagrams (e), (f) and (g). However,  we do need to make different choices per diagram, and therefore the parameterization in Eq.~\eqref{eq:topoB_full} captures all three diagrams at once. Furthermore, we point out that the $b_i$'s do not need to be different; in fact, we will see that only a single regulator is sufficient in topology $B$. We refer to App.~\ref{sec:app-rapidity-reg} for more details, as well as for a study on the use of other rapidity regulators. Note that the rapidity divergences show up only as a result of the expansion by regions, since the corresponding full Feynman integral gets fully regularized by the dimensional regulator $\eps$ alone. This observation provides us with a valuable cross-check: all dependence on $b_i$ must cancel once all regions are combined. 

Finally, we come to topology $X$, which corresponds to diagram (h) and is characterized by the parameterization  
\begin{align}
\nn I_{X;\{n_i\}}^{b_3,b_4,b_5,b_6}
&= \int[dk_1][dk_2]\frac{1}{k_1^{2n_1}}\frac{1}{k_2^{2n_2}}\frac{\mu_3^{2b_3}}{[(k_2-p_1)^2-m^2]^{n_3+b_3}}\frac{\mu_4^{2b_4}}{[(k_1+k_2-p_1)^2-m^2]^{n_4+b_4}}\\
&\hspace{1cm}
\times\frac{\mu_5^{2b_5}}{[(k_1+p_2)^2-m^2]^{n_5+b_5}}\frac{\mu_6^{2b_6}}{[(k_1+k_2+p_2)^2-m^2]^{n_6+b_6}}\frac{1}{[(k_1+k_2)^2]^{n_7}}\, ,
\label{eq:topoX_full}
\end{align}
where again we need $b_i\ne 0$ in order to regulate rapidity divergences that show up once we expand by region.

%%%%%%%%%%%%%%%%%%%%%%%%%%%%%%%%%%%%%%%%%%%%%%%%%%%%%%%%%%%%%%
\subsection{Summary of technical details}
\label{subsec:summary_details}
%%%%%%%%%%%%%%%%%%%%%%%%%%%%%%%%%%%%%%%%%%%%%%%%%%%%%%%%%%%%%%
A common approach in the computation of higher order loop diagrams is to reduce the many Feynman integrals to master integrals using integration by parts (IBP) identities. When it comes to calculating these integrals following the method of regions one has two alternative options. Either one first reduces the large number of integrals in each topology, Eqs.~\eqref{eq:topoA_full}-\eqref{eq:topoX_full}, to master integrals before expanding by regions. However, as we are interested in an expansion up to NLP, the expansion of a master integral into its momentum regions might lead to many additional integrals, so that again a new reduction to master integrals is recommended per momentum region. Therefore, one might as well first expand by regions and only perform an IBP reduction at the very end of the calculational steps. Ultimately, these two ways are equivalent and cannot lead to different final results. We will discuss these alternative approaches further in Sec.~\ref{sec:topoA}, where we also point out the subtleties that enter while expanding the topologies in different momentum regions.

Another difficulty we encountered concerns the analytic regulators that we added to topologies $B$ and $X$, Eqs.~\eqref{eq:topoB_full} and~\eqref{eq:topoX_full}. Although the analytic regulator is a convenient regulator when computing Feynman integrals, it has the downside that the usual IBP reduction programs cannot handle non-integer powers of the propagators. To the best of our knowledge, only \texttt{Kira} \cite{Klappert:2020nbg} is suitable for this, which we therefore adopt as our standard IBP reduction program. In topology $A$, which does not require rapidity regulators, we also use LiteRed \cite{Lee:2013mka} as an independent crosscheck of our results. 

As discussed in Sec.~\ref{sec:FF}, the one-loop form factor contains just three momentum regions: hard, collinear and anti-collinear. However, the number of momentum regions is much larger at the two-loop level. First, the two loop momenta $k_1$ and $k_2$ can have different scalings, which gives already nine possibilities that combine the hard, collinear and anti-collinear momentum regions. Second, we find the appearance of two new momenta scalings: ultra-collinear and ultra-anti-collinear; the contribution from such regions was already observed in \cite{Smirnov:1999bza}. Finally for topology $B$ and X, new regions might appear when one shifts the loop momenta \emph{before} expanding in regions. Although such a shift leaves the full integral invariant, it can lead to additional regions when expanding. 

A summary of our work flow is given in Fig.~\ref{fig:pipeline_flowchart}, which shows the steps we have discussed so far. Not shown is the final step, which consists of verifying whether the small mass limit as given in \cite{Bonciani:2003ai,Bernreuther:2004ih} is reproduced after collecting all regions.

%------------------------------------------------------------
\tikzstyle{io} = [trapezium, trapezium left angle=70, trapezium right angle=110, minimum width=3cm, minimum height=1cm, text centered, draw=black, fill=blue!30]
\tikzstyle{decision} = [diamond, minimum width=3cm, minimum height=1cm, text centered, draw=black, fill=green!30]
\tikzstyle{arrow} = [thick,->,>=stealth]
\tikzstyle{dottedline} = [thick, dotted]
\tikzstyle{process} = [rectangle, minimum width=2cm, minimum height=1cm, text centered, draw=black, fill=orange!30, rounded corners, text width=3cm]
\tikzstyle{smallbox} = [rectangle, minimum width=1cm, minimum height=1cm, text centered, draw=black, fill=orange!30, rounded corners, text width=1cm]
\tikzstyle{startstop} = [rectangle, minimum width=2cm, minimum height=1cm, text centered, draw=black, fill=blue!30, rounded corners, text width=6cm]
\tikzstyle{textbox} = [rectangle, minimum width=2cm, minimum height=1cm, text centered, rounded corners, text width=3cm]

\begin{figure}[t]
\small
    \centering
    \begin{tikzpicture}[node distance=2cm]

\node (in1) [startstop] {\textbf{Two-loop massive form factor}};

\node (topoB) [process, below of=in1, xshift=0cm] {Topo B};
\node (topoA) [process, left of=topoB, xshift=-2.5cm] {Topo A};
\node (topoX) [process, right of=topoB, xshift=2.5cm] {Topo X};

\node (topoAreg) [process, below of=topoA] {$\eps$ regulated};
\node (topoBreg) [process, below of=topoB] {$\eps + \nu$ regulated};
\node (topoXreg) [process, below of=topoX] {$\eps + \nu_1 + \nu_2$ regulated};

\node (region1A) [smallbox, below of=topoAreg, xshift=-2cm] {$hh$};
\node (region2A) [smallbox, below of=topoAreg, xshift=0cm] {$\cdots$};
\node (region3A) [smallbox, below of=topoAreg, xshift=2cm] {$cc$};
\node (test) [textbox, right of=region3A, xshift=1.8cm] {Expansion by regions + IBP (Kira)};
\node (test) [textbox, right of=region3A, xshift=6.3cm] {Expansion by regions + IBP (Kira)};

\node (out2) [startstop, below of=in1, text width = .9\linewidth, yshift=-6cm] {Solve master integrals and collect results};

\draw [arrow] (in1) -- (topoA);
\draw [arrow] (in1) -- (topoB);
\draw [arrow] (in1) -- (topoX);

\draw [arrow] (topoA) -- (topoAreg) ;
\draw [arrow] (topoB) -- (topoBreg)  ;
\draw [arrow] (topoX) -- (topoXreg) ;

\draw [arrow] (topoAreg) -- (region1A);
\draw [arrow] (topoAreg) -- (region2A);
\draw [arrow] (topoAreg) -- (region3A);

\draw [arrow] (region1A.south) -- ++(0cm,-1cm) node[midway, right] {IBP};
\draw [arrow] (region2A.south) -- ++(0cm,-1cm) node[midway, right] {IBP};
\draw [arrow] (region3A.south)-- ++(0cm,-1cm) node[midway, right] {IBP (Kira)};

\draw [dottedline] (topoBreg.south) -- ++(0cm,-3cm) ;
\draw [dottedline] (topoXreg.south) -- ++(0cm,-3cm) ;

\end{tikzpicture}
    \caption{Flow diagram displaying the pipeline of our NLP region analysis of the QED massive form factor. The diagrams in Fig.~\ref{fig:QED_diagrams} are classified into either topology $A$, $B$ or $X$ depending on their momentum flow. We regulate each topology using dimensional regularization, denoted by $\eps$, plus additional rapidity regulators, denoted $\nu$, in case of topology $B$, and $\nu_1, \nu_2$ in case of topology $X$, as it turns out to need two rapidity regulators. For each topology, we expand the relevant integrals by regions and then reduce the result into a simpler set of master integrals with IBP reduction using \texttt{Kira} \cite{Klappert:2020nbg}.
    }
    \label{fig:pipeline_flowchart}
\end{figure}
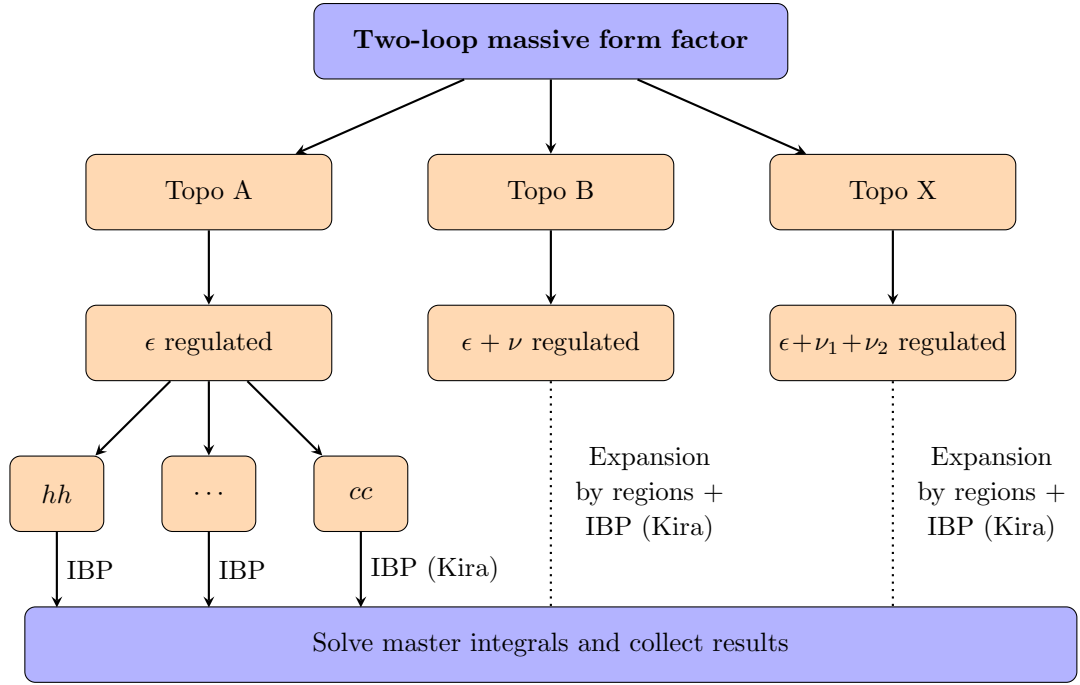

%%%%%%%%%%%%%%%%%%%%%%%%%%%%%%%%%%%%%%%%%%%%%%%%%%%%%%%%%%%%%%
\section{Region expansions}
\label{sec:details}
%%%%%%%%%%%%%%%%%%%%%%%%%%%%%%%%%%%%%%%%%%%%%%%%%%%%%%%%%%%%%%

Having presented the computational scheme in Sec.~\ref{sec:flow}, we now move to the technical details of the calculation of the integrals in topology $A$, $B$ and $X$ using the method of regions. An important remark from the outset concerns the distinction between the regions present at the level of the diagrams in Fig.~\ref{fig:QED_diagrams} on the one hand, and the integral level on the other hand;  these do not necessarily coincide as non-vanishing regions at the integral level can cancel when combined to constitute the diagrams. The results presented in this section should be understood at the integral level.

We shall now discuss topology $A$, $B$ and $X$ in turn. For each topology, we analyze first the associated regions (at the integral level), followed by a discussion of its IBP relations. At the end of each topology subsection, we provide a brief summary of the aspects that enter its computation.

%%%%%%%%%%%%%%%%%%%%%%%%%%%%%%%%%%%%%%%%%%%%%%%%%%%%%%%%%%%%%%
\subsection{Region expansion of topology \texorpdfstring{$A$}{ }}
\label{sec:topoA}
%%%%%%%%%%%%%%%%%%%%%%%%%%%%%%%%%%%%%%%%%%%%%%%%%%%%%%%%%%%%%%
As explained in Sec. \ref{sec:FF}, it is convenient to use light-cone coordinates, Eq.~\eqref{eq:SudakovDecomposition}, to identify the various momentum regions that lead to non-vanishing contributions. At one-loop we found that only momentum modes $h$, $c$ and $\bar{c}$ contributed. This picture changes as soon as we move to the two-loop level where we receive additional contributions coming from momentum modes such as $uc$ and $\overline{uc}$, as defined in Eq.~\eqref{eq:MomentumModesUc}. In total, there are 25 possible combinations of momentum modes at the two-loop level. However, many combinations vanish because they lead to scaleless integrals. For the Feynman integrals contributing to diagrams (a)-(d) we find 11 non-vanishing regions: $hh$, $cc$, $\bar{c}\bar{c}$, $c\bar{c}$, $\bar{c}c$, $ch$, $hc$, $\bar{c}h$, $h\bar{c}$, $uc\bar{c}$ and $\overline{uc}c$, with the momentum flow as indicated in Tab.~\ref{tab:topoARegions} on page~\pageref{tab:topoARegions}.
\begin{table}[t]
\centering
\begin{tabular}{||c|c|c|c|c|c||}
\hline\hline
    & $k^2$ & $k^+p_1^-$ & $k^-p_1^+$ & $k^+p_2^-$ & $k^-p_2^+$  \\\hline
$(h)$ & 1     & $\la^2$    & 1          & 1          & $\la^2$     \\
$(c)$ & $\la^2$ & $\la^2$  & $\la^2$    & 1          & $\la^4$    \\
($\bar{c}$) & $\la^2$ & $\la^4$  & 1       & $\la^2$   & $\la^2$\\
$(uc)$ & $\la^6$ & $\la^4$ & $\la^4$    & $\la^2$          & $\la^6$ \\
($\overline{uc}$) & $\la^6$ & $\la^6$ & $\la^2$ & $\la^4$  & $\la^4$ \\
\hline\hline
\end{tabular}
\caption{Scaling associated to the different momentum regions.  }
\label{tab:momentumRegions}
\end{table}
Even though the power expansion for the momentum modes is straightforward using e.g.~Tab.~\ref{tab:momentumRegions}, the resulting integrals can  in general become quite involved. Let us illustrate this by highlighting several subtleties that enter here. 

The first subtlety we want to discuss concerns the interplay between the usual IBP reduction and the region expansion. To this end, we consider as an example the $hh$ region of topology $A$, Eq.~\eqref{eq:topoA_full}, which up to LP reads 
\begin{align} 
I_A\Big|_{hh} 
&= \int[dk_1][dk_2]\frac{1}{k_1^{2n_1}}\frac{1}{k_2^{2n_2}}
\frac{1}{[(k_1-k_2)^2]^{n_3}}
\frac{1}{[k_1^2+k_1^-p_1^+]^{n_4}}
\frac{1}{[k_2^2+k_2^-p_1^+]^{n_5}} \nn\\
&\hspace{5cm}
\frac{1}{[k_1^2-k_1^+p_2^-]^{n_6}}
\frac{1}{[k_2^2-k_2^+p_2^-]^{n_7}}+ \cO\left(\la^2\right)\, ,\label{eq:topoA_hh}
\end{align}
while many additional terms occur beyond LP. To see this, we expand the fourth propagator of Eq.~\eqref{eq:topoA_full} in the $hh$-region up to NLP
\begin{align}
\frac{1}{(k_1+p_1)-m^2} 
&= \frac{1}{k_1^2+ k_1^-p_1^+} - \frac{k_1^+p_1^-}{[k_1^2+ k_1^-p_1^+]^2} + \cO\left(\la^4\right)\nn\\
&= \frac{1}{k_1^2+ k_1^-p_1^+} - \frac{m^2}{\hat{s}}\frac{k_1^2 - [k_1^2-k_1^+p_2^-]}{[k_1^2+ k_1^-p_1^+]^2} + \cO\left(\la^4\right)\, .
\label{eq:hh-NLP}
\end{align}
On the last line of Eq.~\eqref{eq:hh-NLP} we used the identity
$p_1^- = \left(m^2/\hat{s}\right) p_2^-$ to rewrite the power expansion in terms of the first, fourth and sixth LP (inverse) propagators appearing in Eq.~\eqref{eq:topoA_hh}.
Typically, one can perform an IBP reduction on the full integrals for a given diagram and then expand by regions. However, Eq.~\eqref{eq:hh-NLP} shows that after the regions expansion, the number of integrals increases considerably again beyond LP. Therefore, a new IBP reduction applied to the $hh$-region is called for. 
Instead, one might just as well expand the full integrals in the momentum region, and perform a single IBP-reduction on region integrals at the end. 

As a second subtlety, note that, in order to set up the IBP reduction for the expanded topology, the LP propagators of Eq.~\eqref{eq:topoA_hh} appear in the last line of Eq.~\eqref{eq:hh-NLP}. Similarly, as shown for the fourth propagator in Eq.~\eqref{eq:hh-NLP}, we can rewrite the power expansion of the fifth, sixth and seventh propagator in terms of the corresponding LP propagators, where we can use the identity $p_2^+ 
= \left(m^2/\hat{s}\right)p_1^+$
for the sixth and seventh propagator. This implies that Eq.~\eqref{eq:topoA_hh} defines a closed topology for the $hh$-region up to arbitrary order in the power expansion. This is particularly useful when applying IBP relations, because it leads to the least number of master integrals to solve. Along similar lines, the full power expansion of the $cc$ and $\bar{c}\bar{c}$ region can also be written in terms LP propagators only.

Another subtlety concerns regions where the loop momenta $k_1$ and $k_2$ scale according to different momentum modes, which requires extending the expanded topology by an additional propagator. For example, consider the expansion of the denominator of the third propagator in Eq.~\eqref{eq:topoA_full} in the $\bar{c}c$ region
\begin{equation}
(k_1-k_2)^2 
= \underbrace{k_1^2-2k_{1,\perp}\cdot k_{2,\perp}+k_2^2}_{\sim\la^2}
-\underbrace{k_1^-k_2^+}_{\sim 1}
-\underbrace{k_1^+k_2^-}_{\sim \la^4}\,.
\label{eq:cbarc-NLP}
\end{equation}
The perpendicular components in Eq.~\eqref{eq:cbarc-NLP} cannot be rewritten in terms of the LP propagators from Eq.~\eqref{eq:topoA_full}. Rather than adding $-2k_{1,\perp}\cdot k_{2,\perp}$ to the $\bar{c}c$ topology, we instead rewrite this as
\begin{equation}\label{eq:k1cdotk2}
-2k_{1,\perp}\cdot k_{2,\perp} = -2k_1\cdot k_2 + k_1^+k_2^- + k_1^-k_2^+\,,
\end{equation}
and add $[-2k_1\cdot k_2]^{-n_8}$ as an additional propagator to the $\bar{c}c$ topology. The same logic can be applied to other regions where $k_1$ and $k_2$ scale differently. Eq.~\eqref{eq:k1cdotk2} shows that standard propagators may turn into non-standard propagators of the form $k_1^-k_2^+$ which cannot be given as input to the current IBP programs directly. We treat these in the following way, which we will refer to as a loop-by-loop approach. First, we rewrite  $k_1^-k_2^+$ as $k_1\cdot (k_2^+ n^-)$ and perform an IBP reduction over $k_1$ while considering $k_2$ and $k_2^+ n^-$ as external momenta similar to $p_1$ and $p_2$. 
By doing so, the integrals over $k_1$ get reduced to a smaller set of integrals.
Next, we repeat the first step but now switching the roles of $k_2$ and $k_1$, i.e. we perform IBP over $k_2$ rewriting $k_1^-k_2^+$ as $(k_1^- n^+)\cdot k_2$, while considering $k_1$ and $k_1^- n^+$ as external momentum. Again, the number of integrals over $k_2$ gets reduced. The combination of both IBP reductions over $k_1$ and $k_2$ leaves us with a smaller set of (two-loop) integrals to solve.

Finally, one must be careful when dealing with regions where one of the loop momenta has hard scaling and the other has \mbox{(anti-)collinear} scaling. As discussed above, one can define a closed topology containing the LP propagators and the addition of an eighth propagator $[-2k_1\cdot k_2]^{-n_8}$. However, because a loop-by-loop IBP reduction may lead to new propagators that are not part of the (expanded) topology, adding these propagators to the topology does not work as this leads to an over-determined topology. A possible solution, which we used for topology $A$, is to perform the IBP reduction over the loop with hard loop momentum and then compute the masters. After that, the left-over one-loop integrals with \mbox{(anti-)collinear} loop momentum will be simple enough to calculate directly.

Let us summarize our strategy for topology $A$:
\begin{enumerate}
\item Expand  Eq.~\eqref{eq:topoA_full} in a given momentum region and rewrite subleading corrections in terms of the LP propagators.
\item Use IBP relations that can either handle both loops over $k_1$ and $k_2$ at the same time or adopt a loop-by-loop approach by introducing a non-standard additional propagator $[-2 k_1\cdot k_2]^{-n_8}$.
\item Solve the resulting master integrals and repeat steps 1-3 for the remaining momentum regions.
\end{enumerate}

%%%%%%%%%%%%%%%%%%%%%%%%%%%%%%%%%%%%%%%%%%%%%%%%%%%%%%%%%%%%%%
\subsection{Region expansion of topology  \texorpdfstring{$B$}{ }}
\label{sec:topoB}
%%%%%%%%%%%%%%%%%%%%%%%%%%%%%%%%%%%%%%%%%%%%%%%%%%%%%%%%%%%%%%
As we already stated in Sec.~\ref{sec:flow}, the Feynman integrals needed to calculate diagrams (e), (f) and (g) in Fig.~\ref{fig:QED_diagrams} can be classified as part of topology $B$, defined in Eq.~\eqref{eq:topoB_full}. More specifically, all integrals obtained from diagram (e) have the form $I_{B;\{n_i\}}^{b_3,b_4,b_5,b_6}$ with both $n_3\le 0$ and $n_5\le 0$, the integrals of diagrams (f) satisfy $n_3 \le 0$, while the integrals of diagrams (g) correspond to $n_5\le 0$. The integrals of diagrams (e) thus belong to a subclass of the integrals associated to diagrams (f) and (g). Consequently, the regions contributing to diagram (e) form a subset of those contributing to diagram (f) and (g). In addition, the integrals of diagram (f) and diagram (g) can be related to each other by the transformation $p_1 \leftrightarrow p_2$, $k_1\leftrightarrow -k_1$ and $k_2\leftrightarrow -k_2$. In the following we will discuss the regions obtained in diagrams (e), (f) and (g).

\subsubsection*{Diagram (e)}
Starting with the integrals of diagram (e), we find that the regions $hh$, $hc$, $\bar{c}h$, $cc$ and $\bar{c}\bar{c}$ contribute.\footnote{Note that the $h\bar{c}$-region and $hc$-region, as well as the $\bar{c}h$-region and $ch$-region are equivalent for this diagram. Furthermore, these two regions only appear at the Feynman integral level, but cancel at the form factor level as will become clear in Sec.~\ref{sec:results}. Similar cancellations occur for diagrams (f), (g) and (h).} Of these, the first three regions are free of rapidity divergences, so that we can set the analytic regulators $b_i=0$ in Eq.~\eqref{eq:topoB_full} either at the beginning or at the end of the calculation (both leading to the 
same results). Taking $b_i=0$ from the start, we can treat these three regions similar to the corresponding regions in topology $A$, as we discussed in Sec.~\ref{sec:topoA}. However, there are two other regions, the $cc$ and $\bar{c}\bar{c}$ regions, that do have rapidity divergences, so that here the regulators $b_i$ must be kept. 
However, we do not need to include four regulators $b_i$  in the calculation. Because $n_3\le 0$ and $n_5\le 0$ we can safely take $b_3 = 0$ and $b_5 = 0$ at the beginning of the calculation. Interestingly, we find that we cannot take $b_4=b_6$ to regulate the rapidity divergences of all integrals\footnote{Indeed if we take $b_4\neq b_6$, the integrals are proportional to $\Gamma(b_4-b_6)$. A similar situation was encountered in Ref.~\cite{Chiu:2007dg}.}, although taking either $b_4=0$ or $b_6=0$ is possible. We therefore choose $b_6=\nu$ and $b_3=b_4=b_5=0$ with corresponding scale $\mu_6 = \tilde{\mu}$ as our scheme to regulate the rapidity divergences in both the $cc$ and $\bar{c}\bar{c}$ regions of diagram (e). 
Note that this particular choice breaks the symmetry between the $cc$ and $\bar{c}\bar{c}$ regions. Nevertheless, after combining all of the above five regions, our result up to NLP for each integral,  has no rapidity divergence. Moreover we find agreement with the corresponding result obtained by expanding the full result in Ref.~\cite{Bonciani:2003te,Gluza:2009yy} in the small mass limit.

\subsubsection*{Diagram (f)}
The regions needed to calculate the integrals of diagram (f) are more complicated. First, the integrals of diagram (f) satisfy $n_3\le 0$. Similar to diagram (e), we encounter rapidity divergences in the $cc$ and $\bar{c}\bar{c}$ regions, and in order to handle those we take $b_3=b_4=0$ and $b_5=b_6=\nu$ with corresponding scale $\mu_5=\mu_6=\tilde{\mu}$. However, we find that the rapidity divergences do not cancel after summing the $cc$ and $\bar{c}\bar{c}$ regions. We therefore expect that there is at least one more region with rapidity divergences. 

Indeed, in addition to the five regions  for diagram (e) ($hh$, $hc$, $\bar{c}h$, $cc$ and $\bar{c}\bar{c}$), we find three additional contributing regions, although it is not straightforward to define these three regions in momentum space using the definition for $I_{B;\{n_i\}}^{b_3,b_4,b_5,b_6}$ in Eq.~\eqref{eq:topoB_full}. We exploit the freedom to shift $k_2\rightarrow -k_1-k_2$ to redefine topology $B$ as 
\begin{align}
{I'\,}_{B;\{n_i\}}^{b_3,b_4,b_5,b_6} 
&= \int[dk_1][dk_2]\frac{1}{[k_1^2-m^2]^{n_1}}\frac{1}{[(k_1+k_2)^2-m^2]^{n_2}}\frac{\mu_3^{2b_3}}{[(k_1-p_1)^2]^{n_3+b_3}}\label{eq:topoB_full_2}\\
&\hspace{0.5cm}
\times\frac{\mu_4^{2b_4}}{[(k_2+p_1)^2-m^2]^{n_4+b_4}}\frac{\mu_5^{2b_5}}{[(k_1+p_2)^2]^{n_5+b_5}}\frac{\mu_6^{2b_6}}{[(k_2-p_2)^2-m^2]^{n_6+b_6}}\frac{1}{k_2^{2n_7}}\nn\,.
\end{align}
We stress that $I_{B;\{n_i\}}^{b_3,b_4,b_5,b_6}$ and ${I'\,}_{B;\{n_i\}}^{b_3,b_4,b_5,b_6}$ are equivalent \emph{before} region expansion due to the Lorentz invariance, but this is not always the case for a given region, i.e. after expansion. For example, in the $cc$-region, the loop momenta $k_1$ and $k_2$ have the same momentum mode and as a result, the shift $k_2\rightarrow -k_1-k_2$ does not change the scale of the propagators nor the results of the integrals. However, in the $hc$-region, the shift $k_2\rightarrow -k_1-k_2$ changes the leading behavior of the second, fourth, sixth and seventh propagator of $I_{B;\{n_i\}}^{b_3,b_4,b_5,b_6}$ and as a result we find a different $hc$-region through this shift. In general, one must be aware that different momentum flows can lead to a different scaling of the leading term in the propagator and uncover additional regions as a result.
This illustrates the alternative viewpoint that regions correspond to the scaling of the leading term in the propagators rather than the loop momenta itself, a reasoning which connects also to the geometric approach in parameter space. However, in view of factorization, it is more convenient to still think about the scaling of the momentum modes of the loop momenta, rather than the scaling of the leading term in the propagators. 

\begin{figure}
    \centering
    \includegraphics[width=\linewidth]{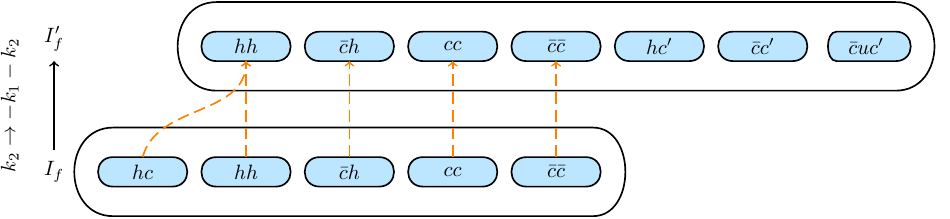}
    \caption{Momentum regions that contribute to the integrals of diagram (f) before and after applying the transformation $k_2 \rightarrow -k_1 -k_2$, corresponding to $I_f$ and $I_f'$ respectively. The dashed arrows represent how the regions in $I_f$ transform accordingly, e.g.~the original $hc$-region maps onto a new $hh$-region after the collinear mode $k_2$ mixes with the dominant hard scale associated to $k_1$. Note how two previously uncovered regions, $\bar{c}c'$ and $\bar{c}uc'$, and a different $hc$-region appear after the shift, where the $\bar{c}c'$-region removes the rapidity divergences present in the $cc$ and $\bar{c}\bar{c}$ regions. Regions that remain invariant are displayed on top of each other, while additional regions are shifted outwards such that all regions are found by collapsing the top row onto its base $I_f$.  }
    \label{fig:regions_shift_diag_f}
\end{figure}

Based on the new definition ${I'\,}_{B;\{n_i\}}^{b_3,b_4,b_5,b_6}$, we find three additional regions: $\bar{c}c'$, $hc'$ and $\bar{c}uc'$, as illustrated in Fig.~\ref{fig:regions_shift_diag_f}.
Besides a modified $hc$-region, we also find $\bar{c}c'$ and $\bar{c}uc'$ as complete new regions. Apart from these three regions, we do not need other regions based on ${I'\,}_{B;\{n_i\}}^{b_3,b_4,b_5,b_6}$, as these are the same as the corresponding regions given based on ${I}_{B;\{n_i\}}^{b_3,b_4,b_5,b_6}$.  The appearance of the $\bar{c}uc'$-region for example can be understood as follows (the $\bar{c}c'$-region following similarly). First, note that the last propagator, $1/k_2^2$, in \Eqn{eq:topoB_full_2} has ultra-collinear scaling in the $\bar{c}uc'$-region. Having found this region in this way, it is also clear that we could not have found it in the original momentum routing. First, it is not possible to select scalings of both loop momenta such that the last momentum factor is ultra-collinear. Second, it is only possible to make the last propagator have an ultra-collinear scaling unless one considers the $ucuc$-region, which leads to scaleless integrals. This is because the masses and external momenta in the propagators of \Eqn{eq:topoB_full} have harder scales than the loop momentum with an ultra-\mbox{(anti-)collinear} mode, thus kinematic configurations where one of the loop momenta is \mbox{ultra-(anti-)collinear} are always scaleless. In other words, in the parametrization of \Eqn{eq:topoB_full}, the propagator $(k_1+k_2)^2$ can produce a leading term with $uc$ scaling only if $k_1$ and $k_2$ are large and opposite, such as to almost cancel. Thus this ultra-collinear kinematic configuration can only be revealed by the shift leading to the parametrization in \Eqn{eq:topoB_full_2}. A similar circumstance has been discussed in Ref.~\cite{Bahjat-Abbas:2018hpv} at one-loop, where a soft region arises in the kinematic configuration in which the loop momentum is large and opposite to an external momentum, such that their sum is soft. In general, revealing such regions by means of momentum shifts in order to find the scaling of the leading term may become ever more intricate at higher loops, due to an increasing number of loop momenta that can conspire to yield new regions. We can still validate our results in another way though. Combining the new $\bar{c}c'$-region with the $cc$-region and $\bar{c}\bar{c}$-region from before the shift, we remove all the rapidity divergences belonging to the integrals of diagram (f). Furthermore, combining all of the above eight regions, we obtain the result up to NLP for each integral of diagram (f), reproducing the corresponding result found by expanding the exact result  in Refs.~\cite{Bonciani:2003te,Gluza:2009yy} in the small mass limit.

\subsubsection*{Diagram (g)}
All the Feynman integrals for diagram (g) fall in the category of Eq.~\eqref{eq:topoB_full} with $n_5\le 0$. Diagram (g) is related to diagram (f) by the transformation $p_1 \leftrightarrow p_2$, $k_1\leftrightarrow -k_1$ and $k_2\leftrightarrow -k_2$. Naturally the rapidity regulators should also be exchanged: $b_3 \leftrightarrow b_5$ and $b_4 \leftrightarrow b_6$. Thus we choose as rapidity regulators $b_3=b_4=\nu$ and $b_5=b_6=0$ with corresponding scales $\mu_3=\mu_4 = \tilde{\mu}$. All the regions from diagram (f), corresponding to the $hh$, $h\bar{c}$, $ch$, $cc$ and $\bar{c}\bar{c}$, $c\bar{c}'$, $h\bar{c}'$ and $c\overline{uc}'$ regions can then be copied.

\subsubsection*{Remarks}
Above we discussed how one can apply the rapidity regulators and find all the contributing regions per diagram. 
However, the calculation of the integrals in each region is not always straightforward and we therefore finish the discussion of topology $B$ with a few technical remarks on these calculations. 

First, regarding regions without rapidity divergences, such as the $hh$, $hc$ and $c\overline{uc}'$ regions, we can safely set $\nu=0$ and calculate the resulting integrals in each region following the same methods as for topology $A$. We emphasize however that in topology $B$ (and also in topology $X$ below) for the region including a hard loop momentum and a (anti-)collinear momentum, one should be quite careful when calculating the expanded integrals. Specifically, one should first expand the full integral $I_{B;\{n_i\}}^{b_3,b_4,b_5,b_6}$ into regions, e.g. the $hc$-region, and only then perform IBP reduction on the part with hard loop momentum. Then $I_{B;\{n_i\}}^{b_3,b_4,b_5,b_6}$ will be expressed as a linear combination of one-loop integrals, which we generically denote by $I^c_i$, which include only the collinear-mode loop momentum. Although in the $hc$-region $I_{B;\{n_i\}}^{b_3,b_4,b_5,b_6}$ is free of rapidity divergences and the $b_i$ have been set to zero, we find that rapidity divergences reappear in some of these $I^c_i$. The rapidity divergences are however expected to cancel among different $I^c_i$ leading to a finite $I_{B;\{n_i\}}^{b_3,b_4,b_5,b_6}$. In this particular case, one should choose an auxiliary regulator to regulate $I^c_i$. The poles in this auxiliary regulator will then cancel to yet yield a finite $I_{B;\{n_i\}}^{b_3,b_4,b_5,b_6}$. Alternatively one can rearrange the integrands of $I^c_i$ at the level of Feynman parametrization such that the integrations over the Feynman parameters are well-defined and finite without introducing extra regulators. 

We use the first method to calculate the integrals in the region including a hard loop momentum and a (anti-)collinear momentum, which is more convenient than the second one when dealing with a large number of such integrals. We also use the second method to calculate several integrals, always leading to the same results. Note that such complexity does not appear in topology $A$.

Then, for regions with rapidity divergences, we first expand the integrals with the rapidity regulator $\nu$ and perform IBP reduction using \texttt{Kira} to obtain a set of master integrals in each region. As a result, we only need to calculate the master integrals with up to 2-fold Mellin-Barnes (MB) representations  after expanding in the rapidity regulator $\nu$ and the dimensional regulator $\epsilon$ to the required order.

Before moving to the last topology, we summarise the subtleties we have encountered in topology $B$:
\begin{enumerate}
    \item Shifts in the loop momenta that leave the full integral invariant, can lead to additional regions nonetheless. These are needed to find all regions, and remove all rapidity divergences in a consistent manner.
    \item The introduction of rapidity regulators requires detailed inspection on a case by case basis depending on the given diagram. 
    \item One must expand in $\nu$ before $\eps$ as the rapidity regulator is a secondary regulator.
\end{enumerate}

%%%%%%%%%%%%%%%%%%%%%%%%%%%%%%%%%%%%%%%%%%%%%%%%%%%%%%%%%%%%%%
\subsection{Region expansion of topology \texorpdfstring{$X$}{ }}
\label{sec:topoX}
%%%%%%%%%%%%%%%%%%%%%%%%%%%%%%%%%%%%%%%%%%%%%%%%%%%%%%%%%%%%%%
The Feynman integrals needed for the last diagram in Fig.~\ref{fig:QED_diagrams}, diagram (h), belong to a new topology we denote as $X$, reflecting the shape of (h)  defined in Eq.~\eqref{eq:topoX_full} with $n_7\le 0$. Due to a new pattern of rapidity divergences, which we will see when analysing the $cc$ and $\bar{c}\bar{c}$ regions, $X$ is the most complicated of the three topologies. 

Focusing on the $\bar{c}\bar{c}$-region first, it suffices to set $b_3=b_4=\nu_1$ and $b_5=b_6=0$ in order to regulate the corresponding rapidity divergence, but this choice does not regulate the rapidity divergence in the $cc$-region. However, we note that the integrand in Eq.~\eqref{eq:topoX_full} is invariant under exchanging $p_1 \leftrightarrow p_2$, $k_1\leftrightarrow -k_1$ and $k_2\leftrightarrow -k_2$, which leads to a symmetry between the $cc$ and $\bar{c}\bar{c}$-regions. 
Motivated by this symmetry one may thus set $b_3=b_4=0$ and $b_5=b_6=\nu_2$ to regulate the rapidity divergence in the $cc$-region. So in order to regulate \emph{simultaneously} the rapidity divergences in the $cc$-region and the $\bar{c}\bar{c}$-region, we choose $b_3=b_4=\nu_1$ and $b_5=b_6=\nu_2$.  The associated scales we shall denote by $\tilde{\mu}_1$ and $\tilde{\mu}_2$ for $\nu_1$ and $\nu_2$ respectively. Note that $\nu_1=\nu_2$ leads to an unregulated divergence, similar to what we saw in topology $B$ for diagram (e).

According to the definition of topology $X$ as given in Eq.~\eqref{eq:topoX_full}, together with the choice of rapidity regulators $\nu_1$ and $\nu_2$ as argued above, we find 8 regions in total: $hh$, $hc$, $\bar{c}h$, $\overline{uc}c$, $\bar{c}uc$, $cc$, $\bar{c}\bar{c}$ and $\bar{c}c$. The rapidity divergences only appear in the last three regions. However, their sum does not yet lead to a finite result in the rapidity regulators $\nu_1$ and $\nu_2$, as can for example be checked for the integral with $n_i=1$ for $i\le 6$ and $n_7=0$, which is discussed in more detail in App.~\ref{sec:allRegionsTopoX}. This requires us to look for other regions with rapidity divergences, and this we do again by redefining the loop momenta. Let us adopt the shifts $k_1\rightarrow -k_1-p_2$ and $k_2\rightarrow -k_2+p_1$ to redefine topology $X$ as
\begin{align}
{I'\,}_{X;\{n_i\}}^{\nu_1,\nu_1,\nu_2,\nu_2}
&= \int[dk_1][dk_2]\frac{1}{(k_1+p_2)^{2n_1}}\frac{1}{(k_2-p_1)^{2n_2}}\frac{1}{[k_2^2-m^2]^{n_3+\nu_1}}\label{eq:topoX_full2}\\
&\hspace{-1.5cm}
\times\frac{\mu_1^{2\nu_1}}{[(k_1+k_2+p_2)^2-m^2]^{n_4+\nu_1}}\frac{\mu_1^{2\nu_1}}{[k_1^2-m^2]^{n_5+\nu_2}}\frac{\mu_2^{2\nu_2}}{[(k_1+k_2-p_1)^2-m^2]^{n_6+\nu_2}}\frac{\mu_2^{2\nu_2}}{[(k_1+k_2)^2]^{n_7}}\,.\nn
\end{align}
Note that the momentum in the last propagator should be $(k_1+k_2-p_1+p_2)$ rather than $(k_1+k_2)$ according to the above shifts $k_1\rightarrow -k_1-p_2$ and $k_2\rightarrow -k_2+p_1$. However, this does not affect the analysis of the regions in diagram (h) as all the integrals associated to diagram (h) satisfy $n_7\le 0$, meaning that $(k_1+k_2-p_1+p_2)^2$ appears in the numerator. After shifting, the integral $I_{X;\{n_i\}}^{\nu_1,\nu_1,\nu_2,\nu_2}$ with $n_7<0$ can always be rewritten as a linear combination of ${I'\,}_{X;\{n'_i\}}^{\nu_1,\nu_1,\nu_2,\nu_2}$. 

\begin{figure}
    \centering
    \includegraphics[width=\linewidth]{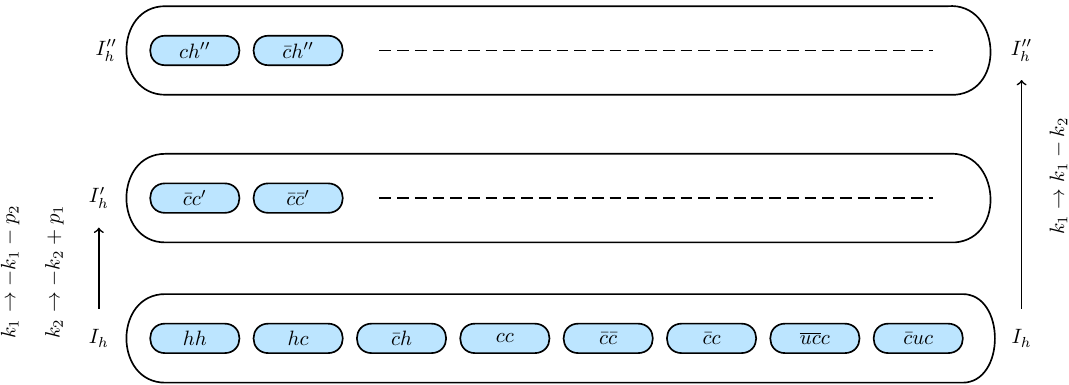}
    \caption{Similar to Fig.~\ref{fig:regions_shift_diag_f}, now displaying the momentum shifts performed in diagram (h) of topology $X$. Two shifts starting from $I_h$ are needed, to $I_h'$ and $I_h''$, in order to uncover all regions. In contrast to Fig.~\ref{fig:regions_shift_diag_f}, we only display new regions with respect to parametrization $I_h$. }
    \label{fig:regions_shift_diag_h}
\end{figure}

Adopting definition ${I'\,}_{X;\{n_i\}}^{\nu_1,\nu_1,\nu_2,\nu_2}$, we find two new regions: the $cc'$-region and the $\bar{c}\bar{c}'$-region, as shown in Fig.~\ref{fig:regions_shift_diag_h}. In these momentum regions, $(k_1+k_2)^2$ scales homogeneously, while $(k_1+k_2-p_1+p_2)^2$ does not, which provides another reason to choose the last propagator in the form of Eq.~\eqref{eq:topoX_full2}. 
We have checked that the rapidity divergences cancel after combining the $cc$ $\bar{c}\bar{c}$, $\bar{c}c$, $cc'$ and $\bar{c}\bar{c}'$ regions. However, to obtain the correct result after combining all regions, we find that we need yet another two regions. As it turns out these are the $ch''$-region and the $\bar{c}h''$-region, without rapidity divergences. These can be revealed by adopting the following 
parametrization
\begin{align}
\nn {I''\,}_{X;\{n_i\}}^{\nu_1,\nu_1,\nu_2,\nu_2}
&= \int[dk_1][dk_2]\frac{1}{(k_1-k_2)^{2n_1}}\frac{1}{k_2^{2n_2}}\frac{\mu_1^{2\nu_1}}{[(k_2-p_1)^2-m^2]^{n_3+\nu_1}}\frac{\mu_1^{2\nu_1}}{[(k_1-p_1)^2-m^2]^{n_4+\nu_1}}\\
&\hspace{1cm}
\times\frac{\mu_2^{2\nu_2}}{[(k_1-k_2+p_2)^2-m^2]^{n_5+\nu_2}}\frac{\mu_2^{2\nu_2}}{[(k_1+p_2)^2-m^2]^{n_6+\nu_2}}\frac{1}{k_1^{2n_7}}\,,
\label{eq:topoX_full3}
\end{align}
as obtained from $I_{X;\{n_i\}}^{\nu_1,\nu_1,\nu_2,\nu_2}$
after shifting $k_1\rightarrow k_1-k_2$, as also shown in Fig.~\ref{fig:regions_shift_diag_h}. Combining all of the above 12 regions ($hh$, $hc$, $\bar{c}h$, $u\bar{c}c$, $\bar{c}uc$, $cc$, $\bar{c}\bar{c}$, $\bar{c}c$, $cc'$, $\bar{c}\bar{c}'$, $ch''$ and $\bar{c}h''$) we indeed obtain the result up to NLP for each integral of diagram (h), consistent with expanding the full result in Ref.~\cite{Bonciani:2003te,Gluza:2009yy} in the small mass limit\footnote{\label{ColemanNortonEqFn}The identification of the missing contribution may depend on the momentum shift considered. In the case at hand, the shift $k_1\rightarrow k_1-k_2$ applied onto $I_{X;\{n_i\}}^{\nu_1,\nu_1,\nu_2,\nu_2}$ leads to the missing contribution being identified with the $ch''$- and the $\bar{c}h''$-regions. Note, however, that we always have the freedom to apply two further shifts $k_1 \to k_1 + p_1$ and $k_2 \to k_2 - p_2$ onto ${I''\,}_{X;\{n_i\}}^{\nu_1,\nu_1,\nu_2,\nu_2}$ which make the contribution due to the $ch''$- and $\bar{c}h''$ region unchanged, respectively. However, in this case, the $ch''$ and $\bar{c}h''$ can also be regarded as two $sh\,h''$ regions (where by $sh$ we indicate the semi-hard scaling introduced in Eq.~\eqref{eq:MomentumModesS}) without changing the scaling of each propagator and the final results of the integrals. From the point of view of a factorization analysis, the second shift is more meaningful: this is because interpreting the new regions as $ch''$- and $\bar{c}h''$-regions, one has a momentum configuration in which there is a lightlike edge with both endpoints in the hard subgraph, which does not conform with the Coleman-Norton interpretation, which is instead consistent with the $sh\,h''$ regions interpretation. We refer to section 2.3 of \cite{Gardi:2022khw} for further discussions. Here we do not explore this issue further, because, as indicated in Table \ref{tab:topoXRegions}, these additional regions (either identified as $ch''$- and $\bar{c}h''$-regions or $sh\, h''$-regions) do not contribute at the form factor level. From the point of view of a factorization analysis, this indicates that the relevant regions at two loops are still just the hard, collinear and anticollinear regions identified at one loop, which is consistent with the Coleman-Norton analysis developed in section II of \cite{Laenen:2020nrt}.}.

Compared to the topology $A$ and $B$, the calculation of the integrals in topology $X$ is more involved due to the appearance of two different rapidity regulators $\nu_1$ and $\nu_2$. As remarked already at the end of Sec.~\ref{subsec:topoB}, one should expand the integrals first in the rapidity regulator(s) followed by the dimensional regulator $\epsilon$, as the $\nu_i$ are secondary regulators. However, in the case of more than one rapidity regulator, we need to further fix the expansion order in the $\nu_i$. Here we choose to expand in $\nu_1$ before expanding in $\nu_2$. We emphasize that the expansion order in $\nu_i$ does not affect the final results once all regions are combined as the rapidity divergences cancel after all. Of course, additional rapidity regulators make the IBP reduction more complex. Even though the rapidity divergences are fully regularized by $\nu_1$ ($\nu_2$) in the $\bar{c}\bar{c}$-region ($cc$-region), meaning that we can choose $\nu_2=0$ ($\nu_1=0$), this is not the case in the $cc'$-region and $\bar{c}\bar{c}'$-region, where both $\nu_1$ and $\nu_2$ are necessary.\footnote{The integrals in these two regions were among the most challenging to calculate.}

Summarising the subtleties we encountered in topology $X$, we find that
\begin{enumerate}
    \item In contrast to topology $B$, topology $X$ requires two unique rapidity regulators $\nu_1$ and $\nu_2$. One must adopt a consistent order with respect expanding in $\nu_1$ and $\nu_2$. 
    \item We find altogether 12 regions, some of which only show up after rerouting the internal momenta.
\end{enumerate}

%%%%%%%%%%%%%%%%%%%%%%%%%%%%%%%%%%%%%%%%%%%%%%%%%%%%%%%%%%%%%%
\section{Results}
\label{sec:results}
%%%%%%%%%%%%%%%%%%%%%%%%%%%%%%%%%%%%%%%%%%%%%%%%%%%%%%%%%%%%%%
We now present the main results of this work and list the various momentum regions that contribute to the two-loop massive form factors $F_1$ and $F_2$. We switch viewpoint from Sec.~\ref{sec:details} and emphasize that the results here are at the level of the diagrams rather than integrals. Recall that some momentum regions may not contribute at the diagrammatic level even though they contribute at the integral level. An overview of the various regions that contribute to each diagram is provided below in Tables \ref{tab:topoARegions}-\ref{tab:topoXRegions} for topologies A, B and X respectively.

As discussed in Sec.~\ref{sec:details}, diagrams belonging to topology $B$ and X may require the introduction of rapidity regulators. Consequently, the corresponding  diagrams acquire poles in $1/\nu$ (in case of topology $B$), and poles in $1/\nu_1$ and $1/\nu_2$ (in case of topology $X$). Because the full form factor is independent of any rapidity divergences, the regulator dependence must cancel after combining all regions; we check this explicitly in the results we provide below. In this respect, two remarks are in order. 

First, as we already observed in the one-loop result given by Eqs.~\eqref{eq:FF1-1l-h} and \eqref{eq:FF1-1l-c}, it is natural to factor out the overall scaling per momentum region, i.e. 
$\left(-\mu^2/\hat{s}\right)^\eps$ and $\left(\mu^2/m^2\right)^\eps$ for each hard and \mbox{(anti-)collinear} loop respectively. In contrast to the one-loop case, we now receive an additional contribution coming from the \mbox{ultra-(anti-)collinear} region which appear with a factor $\left(\mu^2\hat s^2/m^6\right)^\eps$. 
Note that these scales also appear as powers of $\nu$ depending on the specific momentum region and as a result of  regulated propagators. For example, in case of collinear scaling in $k_2$, one expands 
\be
\nonumber
[k_2^2-2k_2 p_2]^\nu = [-2k_2 p_2^-]^\nu + \ord{\lambda^2},
\ee
which has hard scaling and thus leads to an overall factor $\left(\tilde\mu^2/\hat{s}\right)^\nu$.

Secondly, it is important to note that any power of $\nu$ that we factor out is irrelevant for carrying out the check whether the rapidity regulators cancel in the full result. This is due to the fact that except for the leading order term, all terms lead to finite terms in $\nu$ and thus vanish upon setting $\nu$ to zero, e.g.
\be 
\left( \frac{\tilde{\mu}^2}{\hat{s}}\right)^\nu \frac{1}{\nu} - \left( \frac{\tilde{\mu}^2}{-m^2}\right)^\nu \frac{1}{\nu} = \ln\left(-\frac{m^2}{\hat{s}}\right) + \mathcal{O}\left(\nu\right),
\ee
which shows how the regulator dependence indeed cancels. A side remark concerns the opposite behavior of the signs associated to the hard and collinear scales in case of rapidity regulators, i.e. $\left(\mu^2/\hat{s}\right)^\eps$ and $\left(-\mu^2/m^2\right)^\eps$, as compared to the scenario in which the rapidity regulator is absent, i.e. $\left(-\mu^2/\hat{s}\right)^\eps$ and $\left(\mu^2/m^2\right)^\eps$. This is purely an automatic consequence of a Wick rotation, as we explain in more detail in App.~\ref{sec:analyticregulator1}.
In topology $X$, we have two independent rapidity regulators, $\nu_1$ and $\nu_2$, and therefore double poles may arise. Similar to the single pole case, these cancel as can for example be seen by considering
\begin{align}
    &\left(\frac{\tilde{\mu}_2^2}{-m^2}\right)^{\nu_2}\frac{1}{\nu_2^2}
    -\left(\frac{\tilde{\mu}_2^2}{\hat{s}}\right)^{\nu_2}\left(\frac{1}{\nu_2^2}\right) \nn\\
    &\hspace{2cm}= -\frac{1}{\nu_2}\ln\left(-\frac{m^2}{\hat{s}}\right) + \frac{1}{2}\ln^2\left(-\frac{m^2}{\hat{s}}\right) - \ln\left(-\frac{m^2}{\hat{s}}\right)\ln\left(\frac{\tilde{\mu}_2^2}{\hat{s}}\right) + \mathcal{O}\left(\nu_2\right) \,
    \label{eq:single_pole_nu2}
\end{align}
where the remaining single pole in $\nu_2$ cancels against terms that have simultaneous poles in $\nu_1$ and $\nu_2$, 
\begin{align}
     \nonumber&\left(\frac{\tilde{\mu_1}^2}{\hat{s}}\right)^{\nu_1}\left(\frac{\tilde{\mu}_2^2}{\hat{s}}\right)^{\nu_2}\frac{1}{\nu_1\nu_2} -\left(\frac{\tilde{\mu_1}^2}{-m^2}\right)^{\nu_1}\left(\frac{\tilde{\mu}_2^2}{\hat{s}}\right)^{\nu_2}\frac{1}{\nu_1\nu_2} \\
     &\hspace{2cm}= \frac{1}{\nu_2}\ln\left(-\frac{m^2}{\hat{s}}\right) + \ln\left(-\frac{m^2}{\hat{s}}\right)\ln\left(\frac{\tilde{\mu}_2^2}{\hat{s}}\right) + \mathcal{O}\left(\nu_1, \nu_2\right) \, ,
     \label{eq:sim_poles_nu12}
\end{align}
while the finite terms in Eq.~\eqref{eq:single_pole_nu2} and Eq.~\eqref{eq:sim_poles_nu12} combine to a double logarithm of $(-m^2/\hat{s})$.

The rest of this section is structured as follows. First, we provide in Sec.~\ref{subsec:topoA} our results for $F_1$ and $F_2$ for all diagrams contributing to topology $A$, split by the various regions as specified in Tab.~\ref{tab:topoARegions}. Sec.~\ref{subsec:topoB} then contains results for topology $B$ with diagrams and regions provided in Tab.~\ref{tab:topoBRegions}. Results of topology $X$ are presented in Sec.~\ref{subsec:topoX}, corresponding to the diagrams and regions given in Tab.~\ref{tab:topoXRegions}. For further checks and in anticipation for a QCD generalization, we also list the QCD color factor for each diagram, which would follow if the virtual photons were gluons. Finally, in Sec.~\ref{subsec:cross-checks} we comment on the series of checks we have performed to validate our results against existing results  in the literature.

\begin{table}[]
\centering
\def\arraystretch{1.5}%
\begin{tabular}{||c|c|c|c|c|c|c|c|c||}
% \hline
\hline\hline
\multicolumn{2}{||c|}{topology $A$} & $hh$ & $cc$ & $\bar{c}\bar{c}$ & $ch$ & $\bar{c}h$ & $uc\bar{c}$ & $\overline{uc}c$ \\
\hline
(a) & $\vcenter{\vspace{.2cm}\hbox{\includegraphics[width=.23\textwidth]{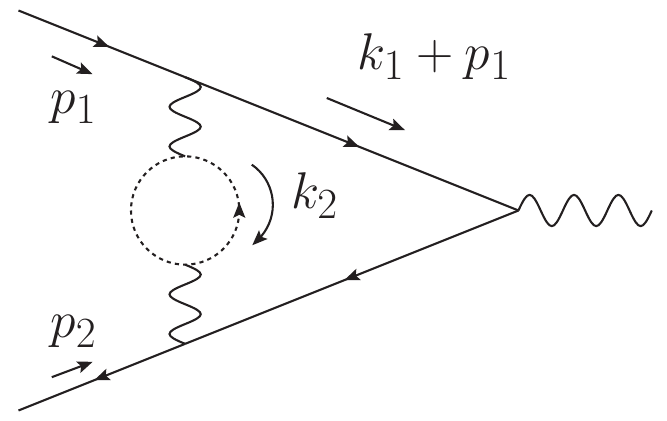}}\vspace{.2cm}}$ & \checkmark &\checkmark & \checkmark &   &  & &   \\
\hline
(b) & $\vcenter{\vspace{.2cm}\hbox{\includegraphics[width=.23\textwidth]{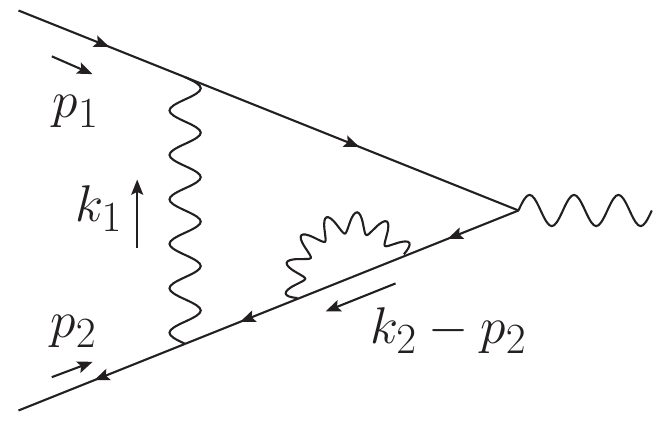}}\vspace{.2cm}}$& \checkmark & &\checkmark  & \checkmark &  &  \checkmark & \\
\hline
(c) &$\vcenter{\vspace{.2cm}\hbox{\includegraphics[width=.23\textwidth]{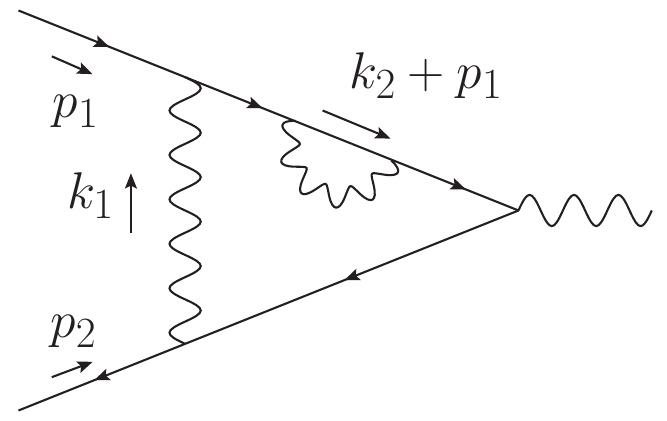}}\vspace{.2cm}}$  & \checkmark & \checkmark &  & &  \checkmark & & \checkmark\\
\hline
(d) &$\vcenter{\vspace{.2cm}\hbox{\includegraphics[width=.23\textwidth]{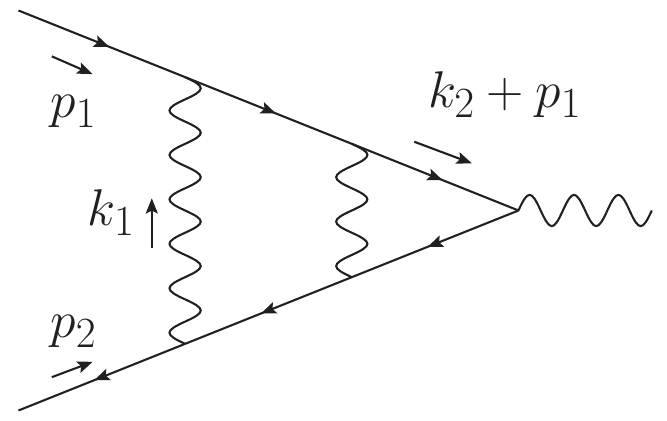}}\vspace{.2cm}}$  & \checkmark &\checkmark &\checkmark & \checkmark  & \checkmark & \checkmark & \checkmark\\
\hline
\hline
\end{tabular}
\caption{Overview of the regions that contribute up to NLP per diagram in topology $A$. 
}
\label{tab:topoARegions}
\end{table}

\begin{table}[]
\centering
\def\arraystretch{1.5}%
\begin{tabular}{||c|c|c|c|c|c|c|c|c|c|c|c|c||}
% \hline
\hline\hline
\multicolumn{2}{||c|}{topology $B$} & $hh$ & $cc$ & $\bar{c}\bar{c}$ & {\color{blue} $c\bar{c}'$} & {\color{blue} $\bar{c} c'$} & $ch$ & $\bar{c}h$ & {\color{blue} $h\bar{c}'$} & {\color{blue} $hc'$} & {\color{blue} $c\overline{uc}'$} & {\color{blue} $\bar{c}uc'$}  \\
\hline
(e) & $\vcenter{\vspace{.3cm}\hbox{\includegraphics[width=.23\textwidth]{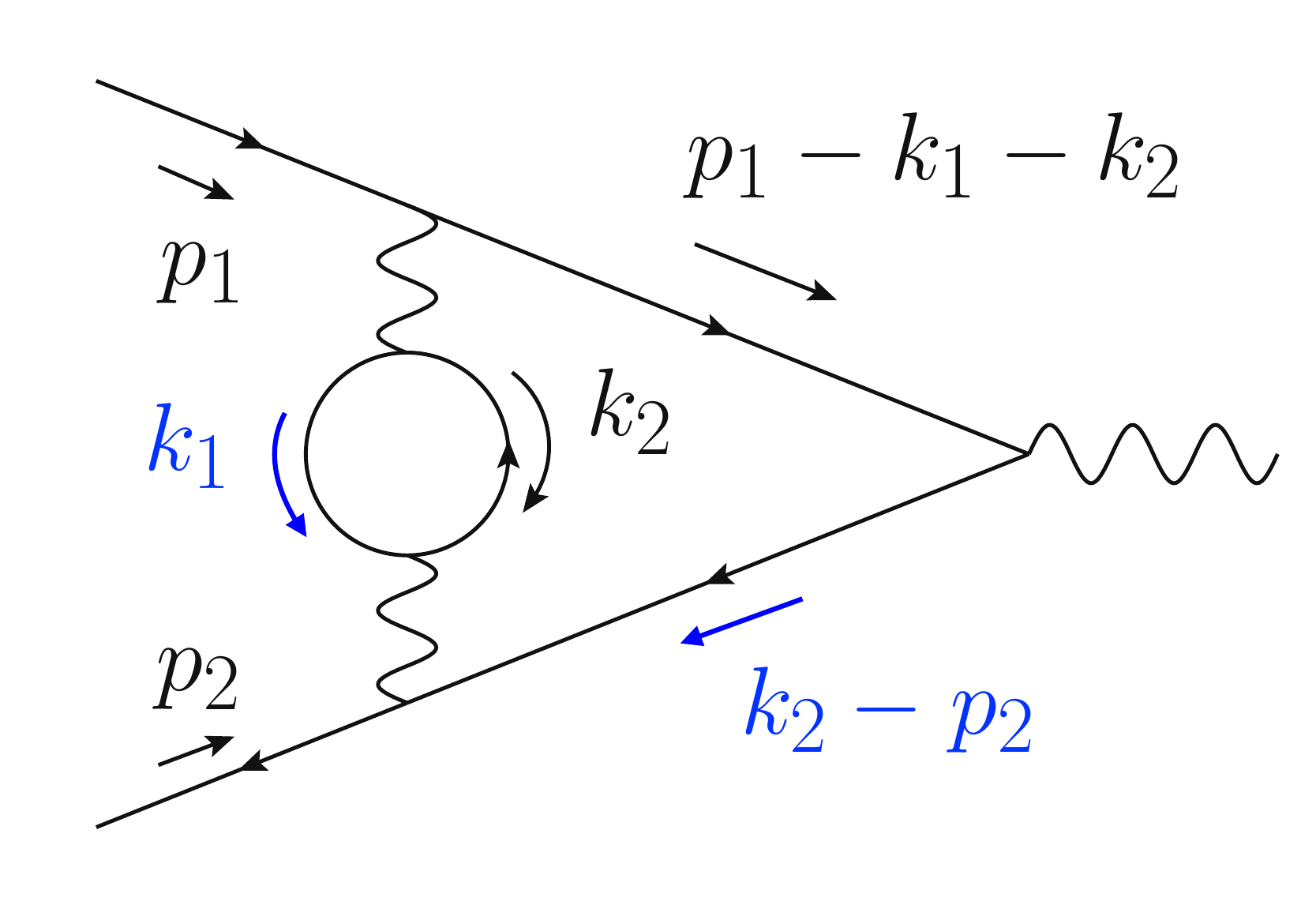}}\vspace{.3cm}}$ & \checkmark &$\nu$ & $\nu$ &  &  & &  & &  & & \\
\hline
(f) & $\vcenter{\vspace{.3cm}\hbox{\includegraphics[width=.23\textwidth]{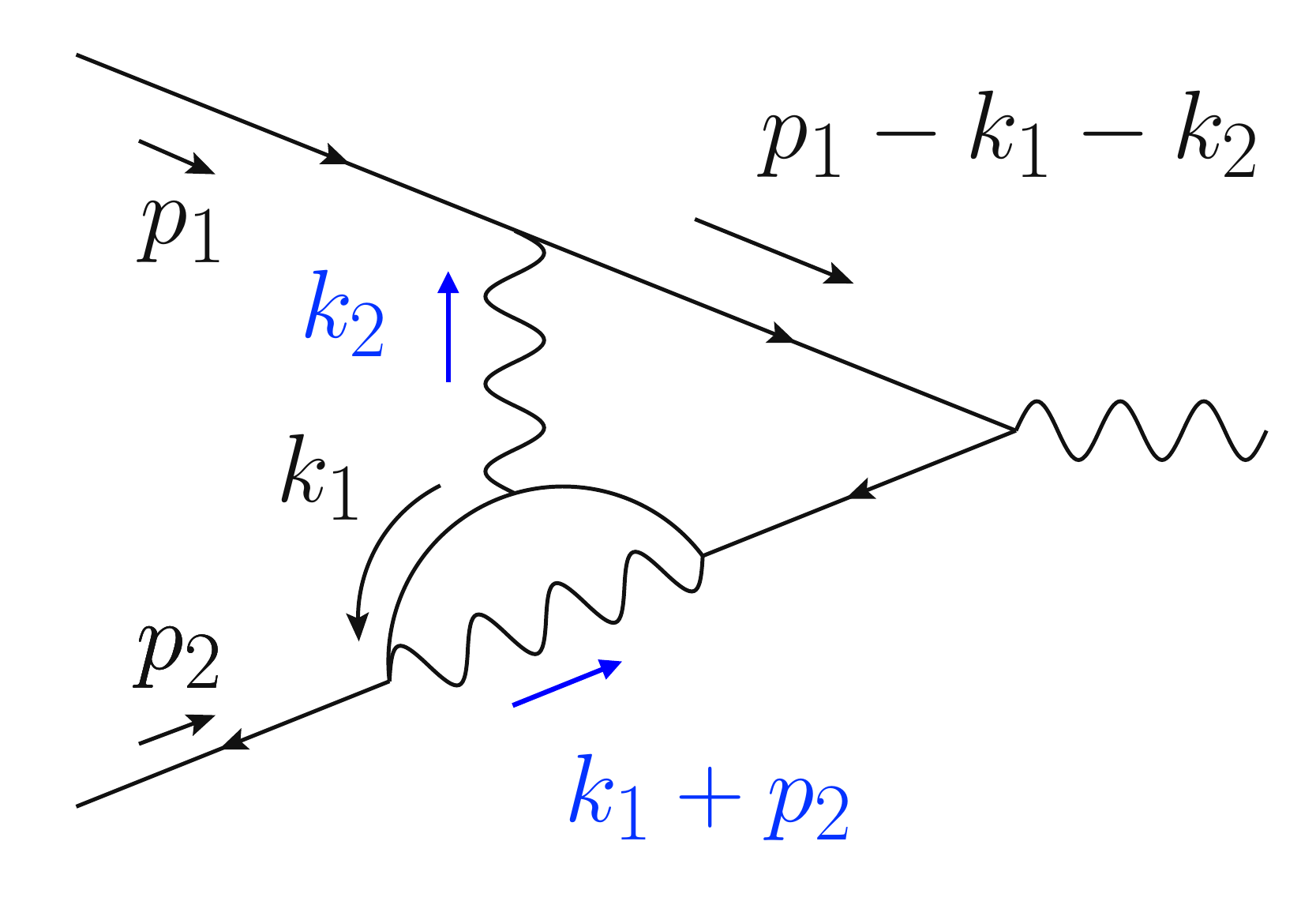}}\vspace{.3cm}}$ & \checkmark & $\nu$ & $\nu$ &  & $\nu$ & & \checkmark & & \checkmark &  & \checkmark \\
\hline
(g) & $\vcenter{\vspace{.3cm}\hbox{\includegraphics[width=.23\textwidth]{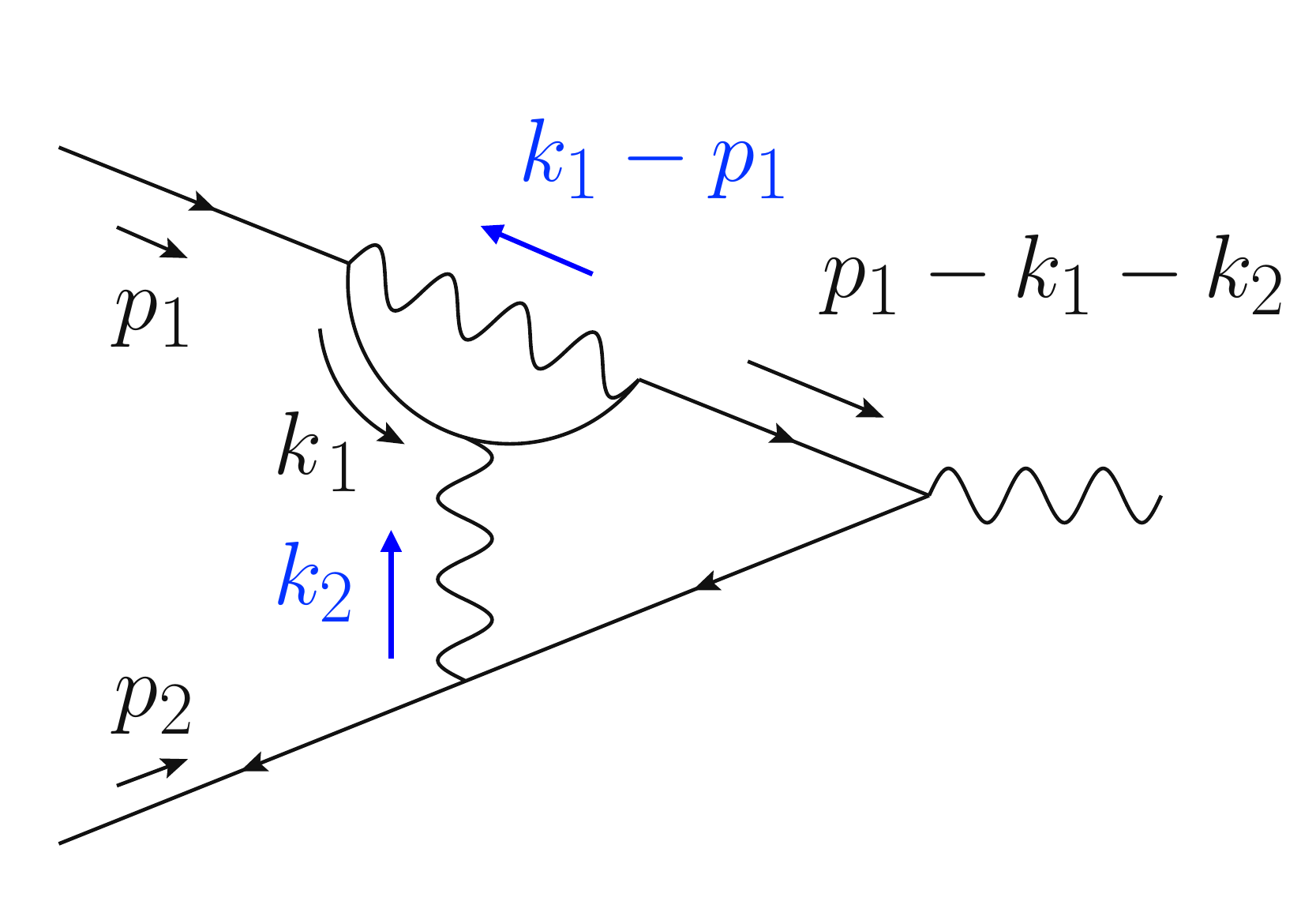}}\vspace{.3cm}}$ & \checkmark & $\nu$ & $\nu$ & $\nu$ &  & \checkmark &  & \checkmark &  & \checkmark & \\
\hline
\hline
\end{tabular}
\caption{Overview of the regions that contribute up to NLP per diagram in topology $B$. We denote with $\nu$ regions that require a rapidity regulator. In black (blue) we show the flow of loop momenta $k_1$, $k_2$ corresponding to the parametrization $I_B$ ($I_B'$). }
\label{tab:topoBRegions}
\end{table}

\begin{table}[]
\centering
\def\arraystretch{1.5}%
\begin{tabular}{||c|c|c|c|c|c|c|c|c|c|c|c||}
% \hline
\hline\hline
\multicolumn{2}{||c|}{topology $X$} & $hh$ & $cc$ & $\bar{c}\bar{c}$ & $\bar{c} c$ & $hc$ & $\bar{c}h$ & $\bar{c}uc$ & $\overline{uc}c$ & {\color{blue} $cc'$} & {\color{blue} $\bar{c}\bar{c}'$}\\
\hline
(h) & $\vcenter{\vspace{.3cm}\hbox{\includegraphics[width=.23\textwidth]{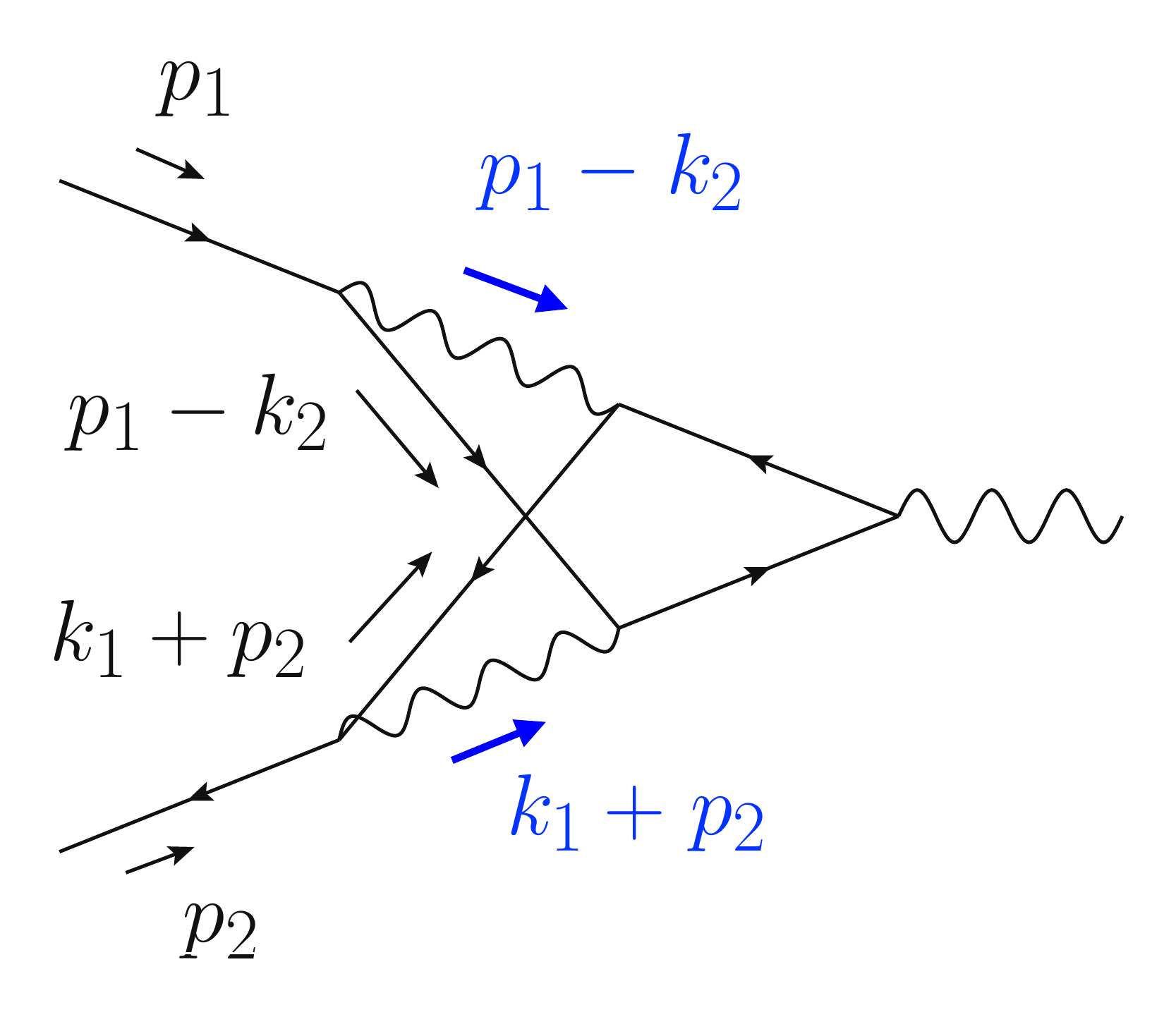}}\vspace{.3cm}}$ & \checkmark &  $\nu_2$ &  $\nu_1$ &  $\nu_1,\nu_2$ & \checkmark & \checkmark & \checkmark & \checkmark &$\nu_1, \nu_2$&$\nu_1, \nu_2$\\
\hline
\hline
\end{tabular}
\caption{Overview of the regions that contribute up to NLP in the topology $X$. Note the presence of only one diagram here. As opposed to the topology $B$, two rapidity regulators are needed to make all regions well-defined. Regions to which this applies are denoted by $\nu_1$ and/or $\nu_2$. In black (blue) we show the flow of loop momenta $k_1$, $k_2$ corresponding to the parametrization $I_{X}$ ($I_X'$). We omit the momentum flow of $I_X''$ as it does not contribute at the level of the form factor.
}
\label{tab:topoXRegions}
\end{table}

%%%%%%%%%%%%%%%%%%%%%%%%%%%%%%%%%%%%%%%%%%%%%%%%%%%%%%%%%%%%%%
\subsection{Topology \texorpdfstring{$A$}{ }}
\label{subsec:topoA}
%%%%%%%%%%%%%%%%%%%%%%%%%%%%%%%%%%%%%%%%%%%%%%%%%%%%%%%%%%%%%%
%---------------------------
\subsubsection*{Diagram (a)}
%---------------------------
\noindent QCD color factor: $C_FT_RN_f$, with $N_f$ the number of light flavors. 
For the QED massive form factors, we can also allow for multiple light flavors, with different charges, and therefore we add an overall factor to diagram $(a)$
\begin{equation}
C = \frac{N_f}{e_q^2}\sum_{l=1}^{N_f}e_{q,l}^2\,,
\end{equation}
with $e_{q,l}$ the fractional charges of the light flavors. We divided out the factor $e_q^2$ as it was explicitly extracted from the form factors in Eq.~\eqref{asexpansion}.

We get the following results for diagram (a):
\begin{align}
F^{(2l,a)}_1\Big|_{hh}
=&\, C\,\left(\frac{\mu^2}{-\hat{s}}\right)^{2\eps}\Bigg[
\frac{2}{3\epsilon^3}
+\frac{28}{9\epsilon^2}
+\frac{18\zeta_2+353}{27\epsilon}
+\frac{28\zeta_2}{9}-\frac{52\zeta_3}{9}+\frac{7541}{162}\nn\\
&\hspace{0cm}+\frac{m^2}{\hat{s}}\left(
\frac{4}{3\epsilon^2}
+\frac{110}{9\epsilon}
+\frac{4\zeta_2}{3}+\frac{1615}{27}
\right)\Bigg]\,,\label{eq:FF1da_hh} \\ 
%%%
F^{(2l,a)}_1\Big|_{cc}
=&\,C\,\left(\frac{\mu^2}{m^2}\right)^{2\eps}\Bigg[
-\frac{1}{3\epsilon^3}
-\frac{17}{9\epsilon^2}
+\frac{-45\zeta_2-196}{27\epsilon}
-\frac{85\zeta_2}{9}-\frac{22\zeta_3}{9}-\frac{2012}{81}\nn\\
&\hspace{0cm}+\frac{m^2}{\hat{s}}\left(
-\frac{2}{3\epsilon^2}
-\frac{79}{9\epsilon}
-\frac{10\zeta_2}{3}-\frac{2575}{54}
\right)\Bigg]\,.\label{eq:FF1da_cc} 
\end{align}
\unskip
By symmetry, we have
\begin{equation}
F^{(2l, a)}_1\Big|_{\bar{c}\bar{c}} = F^{(2l, a)}_1\Big|_{cc}\,.
\end{equation}
For $F_2$ we get
\begin{align}
F^{(2l,a)}_2\Big|_{hh}
&=C\,\left(\frac{\mu^2}{-\hat{s}}\right)^{2\eps}\frac{m^2}{\hat{s}}\Bigg[
-\frac{8}{3\epsilon^2}
-\frac{196}{9\epsilon}
-\frac{8\zeta_2}{3}-\frac{2498}{27}
\Bigg]\,,\label{eq:FF2da_hh} \\ 
%%%
F^{(2l,a)}_2\Big|_{cc}
&=C\,\left(\frac{\mu^2}{m^2}\right)^{2\eps}\frac{m^2}{\hat{s}}\Bigg[
\frac{4}{3\epsilon^2}
+\frac{98}{9\epsilon}
+\frac{20\zeta_2}{3}+\frac{1249}{27}
\Bigg]\,.\label{eq:FF2da_cc} 
\end{align}
\unskip
By symmetry, we have
\begin{equation}
F^{(2l, a)}_2\Big|_{\bar{c}\bar{c}} = F^{(2l, a)}_2\Big|_{cc}\,.
\end{equation}

%---------------------------
\subsubsection*{Diagram (b)}
%---------------------------
\noindent QCD color factor: $C_F^2$
\begin{align}
F^{(2l,b)}_1\Big|_{hh}
=&\left(\frac{\mu^2}{-\hat{s}}\right)^{2\eps}\Bigg[
\frac{1}{\epsilon^3}
+\frac{7}{2\epsilon^2}
+\frac{53-4\zeta_2}{4\epsilon}
-\frac{7\zeta_2}{2}-\frac{32\zeta_3}{3}+\frac{355}{8}\nn\\
&\hspace{0cm}
+\frac{m^2}{\hat{s}}\left(
\frac{9}{\epsilon^3}
+\frac{55}{2\epsilon^2}
-\frac{36\zeta_2-513}{4\epsilon}
-\frac{55\zeta_2}{2}-96\zeta_3+\frac{3171}{8}
\right)\Bigg]\,,\label{eq:FF1db_hh} \\ 
%%%
F^{(2l,b)}_1\Big|_{\bar{c}\bar{c}}
=&\left(\frac{\mu^2}{m^2}\right)^{2\eps}\Bigg[
\frac{1}{\epsilon^3}
+\frac{6}{\epsilon^2}
+\frac{17\zeta_2-6}{\epsilon}
+46\zeta_2+\frac{94\zeta_3}{3}+14\nn\\
&\hspace{0cm}
+\frac{m^2}{\hat{s}}\left(
-\frac{5}{2\epsilon^2}
+\frac{144\zeta_2-63}{4\epsilon}
+\frac{191\zeta_2}{2}+72\zeta_3-\frac{629}{8}
\right)\Bigg]\,,\label{eq:FF1db_cbcb} \\ 
%%%
F^{(2l,b)}_1\Big|_{ch}
=&\left(\frac{\mu^2}{m^2}\right)^{\eps}
\left(\frac{\mu^2}{-\hat{s}}\right)^{\eps}\Bigg[
-\frac{2}{3\epsilon^3}
-\frac{8}{3\epsilon^2}
-\frac{28}{3\epsilon}
+\frac{16\zeta_3}{9}-\frac{92}{3}\nn\\
&\hspace{0cm}
+\frac{m^2}{\hat{s}}\left(
-\frac{6}{\epsilon^3}
-\frac{31}{3\epsilon^2}
-\frac{575}{6\epsilon}
+16\zeta_3-\frac{1997}{12}
\right)\Bigg]\,,\label{eq:FF1db_ch} \\ 
%%%
F^{(2l,b)}_1\Big|_{uc\bar{c}}
=&\left(\frac{\mu^2\hat{s}^2}{m^6}\right)^{\eps}
\left(\frac{\mu^2}{m^2}\right)^{\eps}\Bigg[
-\frac{4}{3\epsilon^3}
-\frac{4}{3\epsilon^2}
-\frac{60\zeta_2+8}{3\epsilon}
-20\zeta_2-\frac{112\zeta_3}{9}-\frac{16}{3}\nn\\
&\hspace{0cm}
+\frac{m^2}{\hat{s}}\left(
-\frac{3}{\epsilon^3}
-\frac{8}{3\epsilon^2}
+\frac{-135\zeta_2-14}{3\epsilon}
-40\zeta_2-28\zeta_3-\frac{22}{3}
\right)\Bigg]\,.\label{eq:FF1db_uccb} 
\end{align}
\unskip
For $F_2$ we get
\begin{align}
F^{(2l,b)}_2\Big|_{hh}
&=\left(\frac{\mu^2}{-\hat{s}}\right)^{2\eps}\frac{m^2}{\hat{s}}\Bigg[
-\frac{6}{\epsilon^3}
-\frac{31}{\epsilon^2}
+\frac{12\zeta_2-235}{2\epsilon}
+31\zeta_2+64\zeta_3-\frac{1593}{4}
\Bigg]\,,\label{eq:FF2db_hh} \\ 
%%%
F^{(2l,b)}_2\Big|_{\bar{c}\bar{c}}
&=\left(\frac{\mu^2}{m^2}\right)^{2\eps}\frac{m^2}{\hat{s}}\Bigg[
-\frac{3}{\epsilon^2}
-\frac{48\zeta_2+15}{2\epsilon}
-71\zeta_2-48\zeta_3+\frac{27}{4}
\Bigg]\,,\label{eq:FF2db_cbcb} \\ 
%%%
F^{(2l,b)}_2\Big|_{ch}
&=\left(\frac{\mu^2}{m^2}\right)^{\eps}
\left(\frac{\mu^2}{-\hat{s}}\right)^{\eps}
\frac{m^2}{\hat{s}}\Bigg[
\frac{4}{\epsilon^3}
+\frac{62}{3\epsilon^2}
+\frac{247}{3\epsilon}
+\frac{1705}{6}-\frac{32\zeta_3}{3}
\Bigg]\,,\label{eq:FF2db_ch} \\ 
%%%
F^{(2l,b)}_2\Big|_{uc\bar{c}}
&=\left(\frac{\mu^2\hat{s}^2}{m^6}\right)^{\eps}
\left(\frac{\mu^2\hat{s}}{m^2}\right)^{\eps}
\frac{m^2}{\hat{s}}\Bigg[
\frac{2}{\epsilon^3}
+\frac{4}{3\epsilon^2}
+\frac{90\zeta_2+8}{3\epsilon}
+20\zeta_2+\frac{56\zeta_3}{3}+\frac{16}{3}
\Bigg]\,.\label{eq:FF2db_uccb} 
\end{align}

%---------------------------
\subsubsection*{Diagram (c)}
%---------------------------
Diagram (c) is related to diagram (b) via the symmetry $c\leftrightarrow\bar{c}$.

%---------------------------
\subsubsection*{Diagram (d)}
%---------------------------
\noindent QCD color factor: $C_F^2$

\begin{align}
F^{(2l,d)}_1\Big|_{hh}
=&\left(\frac{\mu^2}{-\hat{s}}\right)^{2\eps}\Bigg[
\frac{1}{\epsilon^4}
+\frac{2}{\epsilon^3}
+\frac{2\zeta_2+17}{2\epsilon^2}
+\frac{-24\zeta_2+184\zeta_3+303}{12\epsilon}
+\frac{103\zeta_2^2}{10}-\frac{35\zeta_2}{2}\nn\\
&\hspace{0cm}
+\frac{152\zeta_3}{3}+\frac{631}{8} +\frac{m^2}{\hat{s}}\left(
\frac{2}{\epsilon^3}
-\frac{16\zeta_2-32}{\epsilon^2}
+\frac{-6\zeta_2-16\zeta_3+73}{\epsilon}\right.
\nn\\
&\hspace{0cm}\left.
-\frac{64\zeta_2^2}{5}-174\zeta_2+\frac{68\zeta_3}{3}+\frac{893}{2}
\right)\Bigg]\,,\label{eq:FF1dd_hh} \\ 
%%%
F^{(2l,d)}_1\Big|_{cc}
=&\left(\frac{\mu^2}{m^2}\right)^{2\eps}\Bigg[
\frac{2-3\zeta_2}{\epsilon^2}
+\frac{-8\zeta_2-13\zeta_3+10}{\epsilon}
-\frac{163\zeta_2^2}{5}-22\zeta_2-16\zeta_3+38\nn\\
&\hspace{0cm}
+\frac{m^2}{\hat{s}}\bigg(
\frac{12-8\zeta_2}{\epsilon^2}
+\frac{12\zeta_2-72\zeta_3+60}{\epsilon}
-\frac{712\zeta_2^2}{5}-44\zeta_2+136\zeta_3+228
\bigg)\Bigg]\,,\label{eq:FF1dd_cc} \\ 
%%%
F^{(2l,d)}_1\Big|_{ch}
=&\left(\frac{\mu^2}{m^2}\right)^{\eps}
\left(\frac{\mu^2}{-\hat{s}}\right)^{\eps}\Bigg[
-\frac{2}{3\epsilon^4}
-\frac{5}{3\epsilon^3}
-\frac{22}{3\epsilon^2}
+\frac{2\left(8\zeta_3-111\right)}{9\epsilon}
+\frac{4\zeta_2^2}{5}+\frac{40\zeta_3}{9}-\frac{238}{3}\nn\\
&\hspace{0cm}
+\frac{m^2}{\hat{s}}\left(
-\frac{4}{3\epsilon^3}
+\frac{16\zeta_2-31}{\epsilon^2}
+\frac{-84\zeta_2+480\zeta_3-613}{6\epsilon}\right.
\nn\\
&\hspace{0cm}
+\left.\frac{608\zeta_2^2}{5}+20\zeta_2-\frac{598\zeta_3}{9}-\frac{4411}{12}
\right)\Bigg]\,,\label{eq:FF1dd_ch} \\ 
%%%
F^{(2l,d)}_1\Big|_{uc\bar{c}}
=&\left(\frac{\mu^2\hat{s}^2}{m^6}\right)^{\eps}
\left(\frac{\mu^2}{m^2}\right)^{\eps}\Bigg[
\frac{1}{6\epsilon^4}
+\frac{2}{3\epsilon^3}
+\frac{15\zeta_2+8}{6\epsilon^2}
+\frac{90\zeta_2+14\zeta_3+24}{9\epsilon}\nn\\
&\hspace{0cm}
+\frac{493\zeta_2^2}{20}+20\zeta_2+\frac{56\zeta_3}{9}+\frac{16}{3}
+\frac{m^2}{\hat{s}}\left(
\frac{1}{3\epsilon^3}
+\frac{3}{\epsilon^2}
+\frac{15\zeta_2+23}{3\epsilon}\right.\nn\\
&\hspace{0cm}\left.
+45\zeta_2+\frac{28\zeta_3}{9}+\frac{61}{3}
\right)\Bigg]\,.\label{eq:FF1dd_uccb} 
\end{align}
\unskip
By symmetry, we have
\begin{equation}
F^{(2l, d)}_1\Big|_{\bar{c}\bar{c}} = F^{(2l, d)}_1\Big|_{cc}\,,
\qquad\quad
F^{(2l, d)}_1\Big|_{\bar{c}h} = F^{(2l, d)}_1\Big|_{ch}\,,
\qquad\quad
F^{(2l, d)}_1\Big|_{\overline{uc}c} = F^{(2l, d)}_1\Big|_{uc\bar{c}}\,.    
\end{equation}
For $F_2$ we get
\begin{align}
F^{(2l,d)}_2\Big|_{hh}
&=\left(\frac{\mu^2}{-\hat{s}}\right)^{2\eps}\frac{m^2}{\hat{s}}\Bigg[
-\frac{4}{\epsilon^3}
-\frac{20}{\epsilon^2}
+\frac{4\zeta_2-70}{\epsilon}
+48\zeta_2-\frac{160\zeta_3}{3}-249
\Bigg]\,,\label{eq:FF2dd_hh} \\ 
%%%
F^{(2l,d)}_2\Big|_{cc}
&=\left(\frac{\mu^2}{m^2}\right)^{2\eps}\frac{m^2}{\hat{s}}\Bigg[
\frac{8\zeta_2-8}{\epsilon}
+48\zeta_2+16\zeta_3-40
\Bigg]\,,\label{eq:FF2dd_cc} \\ 
%%%
F^{(2l,d)}_2\Big|_{ch}
&=\left(\frac{\mu^2}{m^2}\right)^{\eps}
\left(\frac{\mu^2}{-\hat{s}}\right)^{\eps}
\frac{m^2}{\hat{s}}\Bigg[
\frac{8}{3\epsilon^3}
+\frac{14}{\epsilon^2}
+\frac{51}{\epsilon}
+8\zeta_2-\frac{64\zeta_3}{9}+\frac{349}{2}
\Bigg]\,,\label{eq:FF2dd_ch} \\ 
%%%
F^{(2l,d)}_2\Big|_{uc\bar{c}}
&=\left(\frac{\mu^2\hat{s}^2}{m^6}\right)^{\eps}
\left(\frac{\mu^2}{m^2}\right)^{\eps}\frac{m^2}{\hat{s}}\Bigg[
-\frac{2}{3\epsilon^3}
-\frac{4}{\epsilon^2}
-\frac{10\zeta_2+8}{\epsilon}
-60\zeta_2-\frac{56\zeta_3}{9}-16
\Bigg]\,.\label{eq:FF2dd_uccb} 
\end{align}
\unskip
By symmetry, we have
\begin{equation}
F^{(2l, d)}_2\Big|_{\bar{c}\bar{c}} = F^{(2l, d)}_2\Big|_{cc}\,,
\qquad\quad
F^{(2l, d)}_2\Big|_{\bar{c}h} = F^{(2l, d)}_2\Big|_{ch}\,,
\qquad\quad
F^{(2l, d)}_2\Big|_{\overline{uc}c} = F^{(2l, d)}_2\Big|_{uc\bar{c}}\,.    
\end{equation}
%

%%%%%%%%%%%%%%%%%%%%%%%%%%%%%%%%%%%%%%%%%%%%%%%%%%%%%%%%%%%%%%
\subsection{Topology \texorpdfstring{$B$}{ }}
\label{subsec:topoB}
%%%%%%%%%%%%%%%%%%%%%%%%%%%%%%%%%%%%%%%%%%%%%%%%%%%%%%%%%%%%%%
%---------------------------
\subsubsection*{Diagram (e)}
%---------------------------
\noindent QCD color factor: $C_FT_R$
\begin{align}
F^{(2l,e)}_1\Big|_{hh}
=&\left(\frac{\mu^2}{-\hat{s}}\right)^{2\eps}\Bigg[
\frac{2}{3\epsilon^3}
+\frac{28}{9\epsilon^2}
+\frac{18\zeta_2+353}{27\epsilon}
+\frac{28\zeta_2}{9}-\frac{52\zeta_3}{9}+\frac{7541}{162}\nn\\
&\hspace{0cm}
+\frac{m^2}{\hat{s}}\left(
\frac{28}{3\epsilon^2}
+\frac{254}{9\epsilon}
+\frac{28\zeta_2}{3}+\frac{3775}{27}
\right)\Bigg]\,,\label{eq:FF1de_hh} \\ 
%%%
F^{(2l,e)}_1\Big|_{cc}
=&\left(\frac{\mu^2}{m^2}\right)^{2\eps}\left(\frac{\tilde{\mu}^2}{\hat{s}}\right)^{\nu}\Bigg[
-\frac{4}{3\epsilon^3}
+\frac{1}{\epsilon^2}\left(\frac{4}{3\nu}-\frac{4}{3}\right)
-\frac{1}{\epsilon}\left(\frac{20}{9\nu}+\frac{8\zeta_2}{3}+\frac{56}{9}\right)\nn\\
&\hspace{0cm}
+\frac{36\zeta_2+112}{27\nu}+\frac{32\zeta_2}{3}+\frac{20\zeta_3}{9}-\frac{2144}{81}\nn\\
&\hspace{0cm}
+\frac{m^2}{\hat{s}}\left(
-\frac{2}{3\epsilon^2}
-\frac{1}{\epsilon}\left(\frac{16}{\nu}+\frac{151}{9}\right)
+42\zeta_2-\frac{3511}{54}
\right)\Bigg]\,,\label{eq:FF1de_cc} \\ 
%%%
F^{(2l,e)}_1\Big|_{\bar{c}\bar{c}}
=&\left(\frac{\mu^2}{m^2}\right)^{2\eps}\left(\frac{\tilde{\mu}^2}{-m^2}\right)^{\nu}\Bigg[
\frac{2}{3\epsilon^3}
-\frac{1}{\epsilon^2}\left(\frac{4}{3\nu}+\frac{28}{9}\right)
+\frac{1}{\epsilon}\left(\frac{20}{9\nu}-\frac{2\zeta_2}{3}-\frac{212}{27}\right)\nn\\
&\hspace{0cm}
-\frac{36\zeta_2+112}{27\nu}+\frac{80\zeta_2}{9}-\frac{16\zeta_3}{9}-\frac{1292}{81}\nn\\
&\hspace{0cm}
+\frac{m^2}{\hat{s}}\left(
-\frac{26}{3\epsilon^2}
+\frac{1}{\epsilon}\left(\frac{16}{\nu}-\frac{151}{9}\right)
+34\zeta_2-\frac{55}{54}
\right)\Bigg]\,.\label{eq:FF1de_cbcb} 
\end{align}
\unskip
Notice that the LP contribution of the $hh$ region is the same as for diagram (a), Eq.~\eqref{eq:FF1da_hh}.
For $F_2$ we get
\begin{align}
F^{(2l,e)}_2\Big|_{hh}
&=\left(\frac{\mu^2}{-\hat{s}}\right)^{2\eps}\frac{m^2}{\hat{s}}\Bigg[
-\frac{8}{3\epsilon^2}
-\frac{196}{9\epsilon}
-\frac{8\zeta_2}{3}-\frac{2498}{27}
\Bigg]\,,\label{eq:FF2de_hh} \\ 
%%%
F^{(2l,e)}_2\Big|_{cc}
&=\left(\frac{\mu^2}{m^2}\right)^{2\eps}\frac{m^2}{\hat{s}}\Bigg[
\frac{4}{3\epsilon^2}
+\frac{98}{9\epsilon}
+12\zeta_2+\frac{25}{27}
\Bigg]\,.\label{eq:FF2de_cc} 
\end{align}
\unskip
By symmetry, we have
\begin{equation}
F^{(2l, e)}_2\Big|_{\bar{c}\bar{c}} = F^{(2l, e)}_2\Big|_{cc}\,. 
\end{equation}

%---------------------------
\subsubsection*{Diagram (f)}
%---------------------------
\noindent QCD color factor: $C_F^2-\frac{1}{2}C_FC_A$
\begin{align}
F^{(2l,f)}_1\Big|_{hh}
=&\left(\frac{\mu^2}{-\hat{s}}\right)^{2\eps}\Bigg[
-\frac{1}{\epsilon^3}
+\frac{4\zeta_2-11}{2\epsilon^2}
+\frac{40\zeta_2+8\zeta_3-109}{4\epsilon}
+\frac{8\zeta_2^2}{5}+\frac{91\zeta_2}{2}+\frac{59\zeta_3}{3}-\frac{911}{8}\nn\\
&\hspace{0cm}
+\frac{m^2}{\hat{s}}\left(
-\frac{7}{\epsilon^3}
-\frac{59}{2\epsilon^2}
+\frac{108\zeta_2-695}{4\epsilon}
+\frac{231\zeta_2}{2}+\frac{284\zeta_3}{3}-\frac{5129}{8}
\right)\Bigg]\,,\label{eq:FF1df_hh} \\ 
%%%
F^{(2l,f)}_1\Big|_{cc}
=&\left(\frac{\mu^2}{m^2}\right)^{2\eps}\left(\frac{\tilde{\mu}^2}{\hat{s}}\right)^{2\nu}\Bigg[\frac{m^2}{\hat{s}}\left(
\frac{2}{\epsilon^3}
+\frac{1}{\epsilon^2}\left(\frac{2}{\nu}-4\right)
+\frac{6}{\epsilon}
+\frac{2\zeta_2+6}{\nu}\right.\nn\\
&\hspace{0cm}\left.
-\frac{10\zeta_3}{3}-20\zeta_2+22
\right)\Bigg]\,,\label{eq:FF1df_cc} \\ 
%%%
F^{(2l,f)}_1\Big|_{\bar{c}\bar{c}}
=&\left(\frac{\mu^2}{m^2}\right)^{2\eps}\left(\frac{\tilde{\mu}^2}{-m^2}\right)^{2\nu}\Bigg[
-\frac{1}{\epsilon^3}
+\frac{2\left(\zeta_2-1\right)}{\epsilon^2}
-\frac{2\left(6\zeta_2-5\zeta_3+5\right)}{\epsilon}\nn\\
&\hspace{0cm}
+\frac{32\zeta_2^2}{5}+24\zeta_2-\frac{19\zeta_3}{3}-72\zeta_2\log(2)-42 
+\frac{m^2}{\hat{s}}\left(
-\frac{5}{\epsilon^3}
-\frac{1}{\epsilon^2}\left(\frac{2}{\nu}+\frac{35}{2}\right)\right.\nn\\
&\hspace{0cm}\left.
-\frac{92\zeta_2+279}{4\epsilon}
-\frac{2\zeta_2+2}{\nu}+\frac{133\zeta_2}{2}+\frac{16\zeta_3}{3}-144\zeta_2\log(2)-\frac{2281}{8}
\right)\Bigg]\,,\label{eq:FF1df_cbcb} \\ 
%%%
F^{(2l,f)}_1\Big|_{\bar{c}h}
=&\left(\frac{\mu^2}{m^2}\right)^{\eps}
\left(\frac{\mu^2}{-\hat{s}}\right)^{\eps}\Bigg[
-\frac{4\zeta_2-4}{\epsilon^2}
-\frac{14\zeta_2+12\zeta_3-26}{\epsilon}
-\frac{44\zeta_2^2}{5}-52\zeta_2-42\zeta_3+122\nn\\
&\hspace{0cm}
+\frac{m^2}{\hat{s}}\left(
\frac{10}{\epsilon^3}
+\frac{50}{\epsilon^2}
+\frac{207-40\zeta_2}{\epsilon}
-200\zeta_2-\frac{440\zeta_3}{3}+\frac{1633}{2}
\right)\Bigg]\,,\label{eq:FF1df_cbh} \\ 
%%%
F^{(2l,f)}_1\Big|_{\bar{c}c'}
=&\left(\frac{\mu^2}{m^2}\right)^{2\eps}
\left(\frac{\tilde{\mu}^2}{-m^2}\right)^{\nu}
\left(\frac{\tilde{\mu}^2}{\hat{s}}\right)^{\nu}
\Bigg[\frac{m^2}{\hat{s}}\left(
-\frac{9}{\epsilon^3}
-\frac{14}{\epsilon^2}
-\frac{3\left(3\zeta_2+22\right)}{\epsilon}
-\frac{4}{\nu}\right.\nn\\
&\hspace{0cm}
-14\zeta_2+6\zeta_3-82
\bigg)\Bigg]\,,\label{eq:FF1df_cbc} \\ 
%%%
F^{(2l,f)}_1\Big|_{hc'}
=&\left(\frac{\mu^2}{-\hat{s}}\right)^{\eps}
\left(\frac{\mu^2}{m^2}\right)^{\eps}\Bigg[
\frac{2}{3\epsilon^3}
+\frac{8}{3\epsilon^2}
+\frac{28}{3\epsilon}
+\frac{92}{3}-\frac{16\zeta_3}{9}\nn\\
&\hspace{0cm}
+\frac{m^2}{\hat{s}}\left(
\frac{6}{\epsilon^3}
+\frac{37}{3\epsilon^2}
+\frac{575}{6\epsilon}
+\frac{1637}{12}-16\zeta_3
\right)\Bigg]\,,\label{eq:FF1df_hc2} \\ 
%%%
F^{(2l,f)}_1\Big|_{\bar{c}uc'}
=&\left(\frac{\mu^2}{m^2}\right)^{\eps}
\left(\frac{\mu^2\hat{s}^2}{m^6}\right)^{\eps}\Bigg[
\frac{4}{3\epsilon^3}
+\frac{4}{3\epsilon^2}
+\frac{60\zeta_2+8}{3\epsilon}
+20\zeta_2+\frac{112\zeta_3}{9}+\frac{16}{3}\nn\\
&\hspace{0cm}
+\frac{m^2}{\hat{s}}\left(
\frac{3}{\epsilon^3}
+\frac{8}{3\epsilon^2}
+\frac{135\zeta_2+14}{3\epsilon}
+40\zeta_2+28\zeta_3+\frac{22}{3}
\right)\Bigg]\,.\label{eq:FF1df_cbuc} 
\end{align}
\unskip
For $F_2$ we get
\begin{align}
F^{(2l,f)}_2\Big|_{hh}
=&\left(\frac{\mu^2}{-\hat{s}}\right)^{2\eps}\frac{m^2}{\hat{s}}\Bigg[
\frac{6}{\epsilon^3}
+\frac{39}{\epsilon^2}
+\frac{377-44\zeta_2}{2\epsilon}
-151\zeta_2-80\zeta_3+\frac{3075}{4}
\Bigg]\,,\label{eq:FF2df_hh} \\ 
%%%
F^{(2l,f)}_2\Big|_{\bar{c}\bar{c}}
=&\left(\frac{\mu^2}{m^2}\right)^{2\eps}\frac{m^2}{\hat{s}}\Bigg[
\frac{6}{\epsilon^3}
+\frac{15}{\epsilon^2}
+\frac{133-4\zeta_2}{2\epsilon}\nn\\
&\hspace{0cm}
-61\zeta_2-68\zeta_3+96\zeta_2\log(2)+\frac{1135}{4}
\Bigg]\,,\label{eq:FF2df_cbcb} \\ 
%%%
F^{(2l,f)}_2\Big|_{\bar{c}h}
=&\left(\frac{\mu^2}{m^2}\right)^{\eps}
\left(\frac{\mu^2}{-\hat{s}}\right)^{\eps}
\frac{m^2}{\hat{s}}\Bigg[
-\frac{12}{\epsilon^3}
-\frac{52}{\epsilon^2}
+\frac{48\zeta_2-222}{\epsilon}\nn\\
&\hspace{0cm}
+232\zeta_2+176\zeta_3-869
\Bigg]\,,\label{eq:FF2df_cbh} \\ 
%%%
F^{(2l,f)}_2\Big|_{\bar{c}c'}
=&\left(\frac{\mu^2}{m^2}\right)^{2\eps}\frac{m^2}{\hat{s}}\Bigg[
\frac{6}{\epsilon^3}
+\frac{24}{\epsilon^2}
+\frac{6\zeta_2+64}{\epsilon}
+24\zeta_2-4\zeta_3+160
\Bigg]\,,\label{eq:FF2df_cbc} \\ 
%%%
F^{(2l,f)}_2\Big|_{hc'}
=&\left(\frac{\mu^2}{-\hat{s}}\right)^{\eps}
\left(\frac{\mu^2}{m^2}\right)^{\eps}
\frac{m^2}{\hat{s}}\Bigg[
-\frac{4}{\epsilon^3}
-\frac{74}{3\epsilon^2}
-\frac{283}{3\epsilon}
+\frac{32\zeta_3}{3}-\frac{1873}{6}
\Bigg]\,,\label{eq:FF2df_hc2} \\ 
%%%
F^{(2l,f)}_2\Big|_{\bar{c}uc'}
=&\left(\frac{\mu^2}{m^2}\right)^{\eps}
\left(\frac{\mu^2\hat{s}^2}{m^6}\right)^{\eps}
\frac{m^2}{\hat{s}}\Bigg[
-\frac{2}{\epsilon^3}
-\frac{4}{3\epsilon^2}
-\frac{90\zeta_2+8}{3\epsilon}
-20\zeta_2-\frac{56\zeta_3}{3}-\frac{16}{3}
\Bigg]\,.\label{eq:FF2df_cbuc} 
\end{align}

%---------------------------
\subsubsection*{Diagram (g)}
%---------------------------
\noindent QCD color factor: $C_F^2-\frac{1}{2}C_FC_A$.
Diagram (g) is related to diagram (f) via the symmetry $c\leftrightarrow\bar{c}$.

%%%%%%%%%%%%%%%%%%%%%%%%%%%%%%%%%%%%%%%%%%%%%%%%%%%%%%%%%%%%%%
\subsection{Topology \texorpdfstring{$X$}{ }}
\label{subsec:topoX}
%%%%%%%%%%%%%%%%%%%%%%%%%%%%%%%%%%%%%%%%%%%%%%%%%%%%%%%%%%%%%%
%---------------------------
\subsubsection*{Diagram (h)}
%---------------------------
\noindent QCD color factor: $C_F^2-\frac{1}{2}C_FC_A$
\begin{align}
F^{(2l,h)}_1\Big|_{hh}
=&\left(\frac{\mu^2}{-\hat{s}}\right)^{2\eps}\Bigg[
\frac{1}{\epsilon^4}
+\frac{4}{\epsilon^3}
+\frac{16-7\zeta_2}{\epsilon^2}
-\frac{48\zeta_2+122\zeta_3-174}{3\epsilon}\nn\\
&\hspace{0cm}-\frac{53\zeta_2^2}{2}-58\zeta_2-\frac{380\zeta_3}{3}+204
+\frac{m^2}{\hat{s}}\left(
\frac{1}{\epsilon^4}
+\frac{5}{\epsilon^3}
+\frac{\zeta_2+18}{\epsilon^2}\right.\nn\\
&\hspace{0cm}\left.
-\frac{186\zeta_2+196\zeta_3-843}{6\epsilon}
-\frac{201\zeta_2^2}{10}-64\zeta_2-\frac{382\zeta_3}{3}+\frac{1491}{4}
\right)\Bigg]\,,\label{eq:FF1dh_hh} \\ 
%%%
F^{(2l,h)}_1\Big|_{hc}
=&\left(\frac{\mu^2}{-\hat{s}}\right)^{\eps}
\left(\frac{\mu^2}{m^2}\right)^{\eps}\Bigg[
-\frac{4}{3\epsilon^4}
-\frac{16}{3\epsilon^3}
+\frac{12\zeta_2-56}{3\epsilon^2}
+\frac{126\zeta_2+140\zeta_3-552}{9\epsilon}\nn\\
&\hspace{0cm}
+\frac{52\zeta_2^2}{5}+52\zeta_2+\frac{506\zeta_3}{9}-\frac{584}{3}
+\frac{m^2}{\hat{s}}\left(
-\frac{4}{\epsilon^4}
-\frac{20}{3\epsilon^3}
+\frac{4\zeta_2-29}{\epsilon^2}\right.\nn\\
&\hspace{0cm}\left.
+\frac{144\zeta_2+88\zeta_3-761}{6\epsilon}
-8\zeta_2^2+88\zeta_2+\frac{952\zeta_3}{9}-\frac{4955}{12}
\right)\Bigg]\,,\label{eq:FF1dh_hc} \\ 
%%%
F^{(2l,h)}_1\Big|_{\overline{uc}c}
=&\left(\frac{\mu^2\hat{s}^2}{m^6}\right)^{\eps}
\left(\frac{\mu^2}{m^2}\right)^{\eps}\Bigg[
-\frac{1}{6\epsilon^4}
-\frac{2}{3\epsilon^3}
-\frac{15\zeta_2+8}{6\epsilon^2}
-\frac{90\zeta_2+14\zeta_3+24}{9\epsilon}\nn\\
&\hspace{0cm}
-\frac{493\zeta_2^2}{20}-20\zeta_2-\frac{56\zeta_3}{9}-\frac{16}{3}
+\frac{m^2}{\hat{s}}\left(
-\frac{1}{3\epsilon^3}
-\frac{3}{\epsilon^2}
-\frac{15\zeta_2+23}{3\epsilon}\right.\nn\\
&\hspace{0cm}\left.
-45\zeta_2-\frac{28\zeta_3}{9}-\frac{61}{3}
\right)\Bigg]\,,\label{eq:FF1dh_ucbc} \\ 
%%%
F^{(2l,h)}_1\Big|_{cc}
=&\left(\frac{\mu^2}{m^2}\right)^{2\eps}\left(\frac{\tilde{\mu}_2^2}{\hat{s}}\right)^{2\nu_2}\Bigg[
\frac{1}{2\epsilon^4}
+\frac{2}{\epsilon^3}
+\frac{3\zeta_2+12}{2\epsilon^2}
+\frac{6\zeta_2+26\zeta_3+54}{3\epsilon}\nn\\
&\hspace{0cm}
+\frac{363\zeta_2^2}{20}+14\zeta_2-\frac{4\zeta_3}{3}+54 
+\frac{m^2}{\hat{s}}\left(
\frac{1}{\epsilon^3}
+\frac{8\zeta_2-5}{2\epsilon^2}\right.\nn\\
&\hspace{0cm}\left.
+\frac{1}{\epsilon}\left(\frac{4}{\nu_2}-3\zeta_2+36\zeta_3-\frac{27}{4}\right)
+\frac{8}{\nu_2}+\frac{356\zeta_2^2}{5}+\frac{7\zeta_2}{2}-\frac{110\zeta_3}{3}-\frac{177}{8}
\right)\Bigg]\,,\label{eq:FF1dh_cc} \\ 
%%%
F^{(2l,h)}_1\Big|_{\bar{c}c}
=&\left(\frac{\mu^2}{m^2}\right)^{2\eps}
\left(\frac{\tilde{\mu}_1^2}{-m^2}\right)^{\nu_1}
\left(\frac{\tilde{\mu}_1^2}{\hat{s}}\right)^{\nu_1}
\left(\frac{\tilde{\mu}_2^2}{-m^2}\right)^{\nu_2}
\left(\frac{\tilde{\mu}_2^2}{\hat{s}}\right)^{\nu_2}\Bigg[
\frac{1}{\epsilon^4}
+\frac{4}{\epsilon^3}
+\frac{\zeta_2+12}{\epsilon^2}\nn\\
&\hspace{0cm}
+\frac{12\zeta_2-2\zeta_3+96}{3\epsilon}
+\frac{7\zeta_2^2}{10}+12\zeta_2-\frac{8\zeta_3}{3}+80 \nn\\
&\hspace{0cm}
+ \frac{m^2}{\hat{s}}\bigg(
\frac{4}{\epsilon^4}
-\frac{1}{\epsilon^3}\left(\frac{4}{\nu_2}+\frac{4}{\nu_1}+10\right)
+\frac{1}{\epsilon^2}\left(\frac{8}{\nu_1}+\frac{4}{\nu_1\nu_2}+\frac{8}{\nu_2}+4\zeta_2+30\right)\nn\\
&\hspace{0cm}
+\frac{1}{\epsilon}\left(\frac{8-8\zeta_2}{\nu_1}+\frac{8-8\zeta_2}{\nu_2}-\frac{4}{\nu_1\nu_2}-18\zeta_2+\frac{16\zeta_3}{3}+110\right)\nn\\
&\hspace{0cm}
+\frac{12+36\zeta_2-52\zeta_3}{3\nu_1}
+\frac{4\zeta_2-4}{\nu_1\nu_2}
+\frac{12+36\zeta_2-52\zeta_3}{3\nu_2}\nn\\
&\hspace{0cm}
-\frac{46\zeta_2^2}{5}-\frac{124\zeta_3}{3}+22\zeta_2+278
\bigg)\Bigg]\,,\label{eq:FF1dh_cbc} \\ 
%%%
F^{(2l,h)}_1\Big|_{cc'}
=&\left(\frac{\mu^2}{m^2}\right)^{2\eps}
\left(\frac{\tilde{\mu}_1^2}{-m^2}\right)^{\nu_1}
\left(\frac{\tilde{\mu}_1^2}{\hat{s}}\right)^{\nu_1}
\left(\frac{\tilde{\mu}_2^2}{-m^2}\right)^{2\nu_2}\Bigg[
\frac{m^2}{\hat{s}}\left(
\frac{20}{\epsilon^4}
-\frac{1}{\epsilon^3}\left(\frac{10}{\nu_2}+10\right)\right.\nn\\
&\hspace{0cm}
+\frac{1}{\epsilon^2}\left(\frac{4}{\nu_2^2}+\frac{6}{\nu_2}-8\zeta_2-8\right)
+\frac{1}{\epsilon}\left(-\frac{4}{\nu_2^2}+\frac{2\zeta_2+6}{\nu_2}-\frac{4}{\nu_1}+6\zeta_2-\frac{112\zeta_3}{3}+16\right)
\nn\\
&\hspace{0cm}\left.+\frac{4\zeta_2-4}{\nu_2^2}+\frac{80\zeta_3-18\zeta_2-12}{3\nu_2}-\frac{8}{\nu_1}-46\zeta_2^2+\frac{32\zeta_3}{3}+4\zeta_2+46
\right)\Bigg]\,,\label{eq:FF1dh_cc2} \\ 
%%%
F^{(2l,h)}_1\Big|_{\bar{c}\bar{c}'}
=&\left(\frac{\mu^2}{m^2}\right)^{2\eps}
\left(\frac{\tilde{\mu}_1^2}{-m^2}\right)^{2\nu_1}
\left(\frac{\tilde{\mu}_2^2}{-m^2}\right)^{\nu_2}
\left(\frac{\tilde{\mu}_2^2}{\hat{s}}\right)^{\nu_2}\Bigg[\frac{m^2}{\hat{s}}\left(
-\frac{17}{\epsilon^4}
+\frac{1}{\epsilon^3}\left(\frac{14}{\nu_2}+\frac{4}{\nu_1}+27\right)\right.\nn\\
&\hspace{0cm}
+\frac{1}{\epsilon^2}\left(-\frac{4}{\nu_2^2}-\frac{14}{\nu_2}-\frac{8}{\nu_1}-\frac{4}{\nu_1\nu_2}-13\zeta_2+29\right)
\nn\\
&\hspace{0cm}
+\frac{1}{\epsilon}\left(\frac{4}{\nu_2^2}+\frac{6\zeta_2-18}{\nu_2}+\frac{8\zeta_2-8}{\nu_1}+\frac{4}{\nu_1\nu_2}-\frac{110\zeta_3}{3}+11\zeta_2+16\right)
\nn\\
&\hspace{0cm}
+\frac{4-4\zeta_2}{\nu_2^2}-\frac{28\zeta_3+18+24}{3\nu_2}+\frac{52\zeta_3-36\zeta_2-12}{3\nu_1}+\frac{4-4\zeta_2}{\nu_1\nu_2}\nn\\
&\hspace{0cm}\left.
-\frac{223\zeta_2^2}{10}+10\zeta_3+9\zeta_2+46
\right)\Bigg]\,.\label{eq:FF1dh_cbcb2} 
\end{align}
\unskip
By symmetry, we have
\begin{equation}
F^{(2l,h)}_1\Big|_{\bar{c}h} = F^{(2l,h)}_1\Big|_{hc}\,,
\quad
F^{(2l,h)}_1\Big|_{\bar{c}uc} = F^{(2l,h)}_1\Big|_{\overline{uc}c}\,,
\quad
F^{(2l,h)}_1\Big|_{\bar{c}\bar{c}} = F^{(2l,h)}_1\Big|_{cc}(\nu_2\leftrightarrow\nu_1)\,.
\end{equation}
For $F_2$ we get
\begin{align}
F^{(2l,h)}_2\Big|_{hh}
=&\left(\frac{\mu^2}{-\hat{s}}\right)^{2\eps}\frac{m^2}{\hat{s}}\Bigg[
-\frac{4}{\epsilon^3}
-\frac{26}{\epsilon^2}
+\frac{5\left(4\zeta_2-23\right)}{\epsilon}
+114\zeta_2+\frac{320\zeta_3}{3}-\frac{977}{2}
\Bigg]\,,\label{eq:FF2dh_hh} \\ 
%%%
F^{(2l,h)}_2\Big|_{hc}
=&\left(\frac{\mu^2}{-\hat{s}}\right)^{\eps}
\left(\frac{\mu^2}{m^2}\right)^{\eps}
\frac{m^2}{\hat{s}}\Bigg[
\frac{16}{3\epsilon^3}
+\frac{34}{\epsilon^2}
+\frac{127-16\zeta_2}{\epsilon}
-104\zeta_2-\frac{560\zeta_3}{9}+\frac{885}{2}
\Bigg]\,,\label{eq:FF2dh_hc} \\ 
%%%
F^{(2l,h)}_2\Big|_{u\bar{c}c}
=&\left(\frac{\mu^2\hat{s}^2}{m^6}\right)^{\eps}
\left(\frac{\mu^2}{m^2}\right)^{\eps}
\frac{m^2}{\hat{s}}\Bigg[
\frac{2}{3\epsilon^3}
+\frac{4}{\epsilon^2}
+\frac{2\left(5\zeta_2+4\right)}{\epsilon}
+60\zeta_2+\frac{56\zeta_3}{9}+16
\Bigg]\,,\label{eq:FF2dh_ucbc} \\ 
%%%
F^{(2l,h)}_2\Big|_{cc}
=&\left(\frac{\mu^2}{m^2}\right)^{2\eps}\frac{m^2}{\hat{s}}\Bigg[
-\frac{2}{\epsilon^3}
-\frac{9}{\epsilon^2}
+\frac{-4\zeta_2-59}{2\epsilon}
-29\zeta_2+\frac{4\zeta_3}{3}-\frac{369}{4}
\Bigg]\,,\label{eq:FF2dh_cc} \\ 
%%%
F^{(2l,h)}_2\Big|_{\bar{c}c}
=&\left(\frac{\mu^2}{m^2}\right)^{2\eps}\frac{m^2}{\hat{s}}\Bigg[
-\frac{4}{\epsilon^3}
-\frac{32}{\epsilon^2}
-\frac{4\left(\zeta_2+24\right)}{\epsilon}
-32\zeta_2+\frac{8\zeta_3}{3}-256
\Bigg]\,.\label{eq:FF2dh_cbc} 
\end{align}
\unskip
By symmetry, we have
\begin{equation}
F^{(2l, h)}_2\Big|_{\bar{c}h} = F^{(2l, h)}_2\Big|_{hc}\,,
\qquad
F^{(2l, h)}_2\Big|_{\bar{c}uc} = F^{(2l, h)}_2\Big|_{\overline{uc}c}\,,
\qquad
F^{(2l, h)}_2\Big|_{\bar{c}\bar{c}} = F^{(2l, h)}_2\Big|_{cc}\,.
\end{equation}

%%%%%%%%%%%%%%%%%%%%%%%%%%%%%%%%%%%%%%%%%%%%%%%%%%%%%%%%%%%%%%
\subsection{Cross-checks}
\label{subsec:cross-checks}
%%%%%%%%%%%%%%%%%%%%%%%%%%%%%%%%%%%%%%%%%%%%%%%%%%%%%%%%%%%%%%
In the previous sections, we have listed all contributions to the massive form factor at two loop at NLP per momentum region. To validate our results and to make sure no region has been left unaccounted for, we have performed several cross-checks with results presented in \cite{Bonciani:2003ai} and \cite{Bernreuther:2004ih}. The former presents the full result of the two-loop QED massive form factor at the level of the individual diagrams, while the latter provides the corresponding QCD result for all diagrams combined, see Eqs.~(22) and (23) in \cite{Bernreuther:2004ih} for $F_1$ and $F_2$ respectively. 

In order to compare our results with \cite{Bonciani:2003ai} and \cite{Bernreuther:2004ih}, one needs to keep the following in mind. To begin with, one must define
\begin{equation}
\cF_i^{(2l)}(\eps,s)
= \frac{e^{-2\eps\ga_E}}{\Ga(1+\eps)^2}\left(\frac{\mu^2}{m^2}\right)^{-2\eps}F_i^{(2l)}\left(\eps,\frac{m^2}{\hat{s}}\right) \, ,
\end{equation}
as defined in Eq.~$(21)$ in \cite{Bernreuther:2004ih}, and second, expand the variable $x$ as defined in Eq.~$(14)$ in \cite{Bernreuther:2004ih} in powers of $m^2/\hat{s}$ to match our conventions. Finally, we also remark that one must reinstate the QCD color factors in the QED diagrams in Fig.~\ref{fig:QED_diagrams} before comparing against  Ref.~\cite{Bernreuther:2004ih}. With these conventions in mind, we now compare the sum of the momentum regions as given in Sec.~\ref{sec:topoA}-\ref{sec:topoX} to the expansion in $m^2/\hat{s}$ up to NLP of the full result of \cite{Bonciani:2003ai} and \cite{Bernreuther:2004ih}. We have made use of the \texttt{Mathematica} package \texttt{HPL} \cite{Maitre:2005uu,Maitre:2007kp} to expand the polylogs. 

First, we have checked that diagram (a), which is absent in \cite{Bonciani:2003ai}, reproduces the $C_F T_R N_f$ term in \cite{Bernreuther:2004ih}, while diagram (e) can be checked either directly against the result of \cite{Bonciani:2003ai} or against the $C_F T_R$ term in \cite{Bernreuther:2004ih}. Similarly, one can verify the result of the remaining diagrams (b)-(d) and (f)-(h) by checking directly with \cite{Bonciani:2003ai}, which we reproduce up to some small verified typos. 

In addition to an inspection at the individual diagram level, we have also performed checks at the level of the form factor itself. The sum of our results of diagrams (b)-(d) and (f)-(h) reproduces the term proportional to $C_F^2$ in \cite{Bernreuther:2004ih}. Additional diagrams appearing only in QCD do not contribute at $C_F^2$.

As a final check we remark that the LP part of the $hh$ region corresponds to the massless limit and we have verified this with the massless form factors at two loops as given in \cite{Moch:2005id,Gehrmann:2010ue}.

%%%%%%%%%%%%%%%%%%%%%%%%%%%%%%%%%%%%%%%%%%%%%%%%%%%%%%%%%%%%%%
\section{Discussion of results}
\label{sec:discussion}
%%%%%%%%%%%%%%%%%%%%%%%%%%%%%%%%%%%%%%%%%%%%%%%%%%%%%%%%%%%%%%
\noindent
The previous section contains 
the main result of this paper, namely, 
the two-loop massive form factors $F_1$ 
and $F_2$, in the limit $\hat s \gg m^2$, 
written as the sum of contributions arising
from all momentum regions.  
It may be useful to elaborate on this result 
a bit more, focusing in particular on what 
can be learnt in light of the computational 
technique itself, and  of
factorization. 

The first issue one encounters within the 
expansion by regions is of course 
identifying all contributing regions.
To this end, geometric methods have been 
developed, which identify the regions by 
associating them to certain scaling vectors 
in the parameter representation of a given 
Feynman graph $G$ \cite{Pilipp:2008ef,Pak:2010pt,Jantzen:2012mw,Semenova:2018cwy,Ananthanarayan:2018tog,Heinrich:2021dbf,Gardi:2022khw}.
From the perspective of exploiting the 
expansion by regions to reveal the underlying
factorization structure of a given physical
observable, it is important to be able to
associate a given region to the (hard, 
collinear, soft, etc) scaling of the loop
momenta, in order to reinterpret a given 
region as originating from the exchange
of a (hard, collinear, soft, etc) particle. 
Identifying all regions by assigning all 
possible momentum scalings to the loop momentum 
is however non-trivial;  because 
loop momenta can be routed in many different ways. As discussed
in the literature (see e.g. \cite{Beneke:1997zp,Bahjat-Abbas:2018hpv}),
starting from a given integral representation 
it may be necessary to shift the loop momenta 
one or more times in order to reveal all regions.
In case of the two-loop calculation considered 
here, we found it was necessary to perform such shifts 
in topologies $B$ and $X$, as discussed 
in Sec.~\ref{sec:topoB} and 
\ref{sec:topoX} respectively. An obvious 
question is whether loop momentum shifts are
sufficient to reveal all regions. The answer 
is positive for the two-loop problem at hand. 
In general, finding all 
regions by means of loop momentum shifts 
becomes increasingly involved as the loop
order grows; it remains an open question 
whether this approach can be effective at 
to all loop orders. Let us also mention that 
shifting the loop momenta does not provide 
per se a criterion to establish whether all 
regions have been taken into account. In case 
the result of the exact integral is unknown, 
we find it useful to consider other constraints, 
such as the requirement that rapidity divergences 
cancel in the sum of all regions, which is a strong 
constraint on whether all regions have been correctly 
considered. Another criterion that we have found 
quite useful is to determine, given a certain loop 
momentum parametrization, whether all possible 
scalings of the \emph{leading} term in a given 
propagator can be obtained, as explained in 
Sec.~\ref{sec:topoB}. If not, this gives
a good indication that some momentum regions are
missing, and a momentum shift is needed to reveal 
them. 

Once all regions have been found, the next 
question concerns their significance for a 
factorization approach. In this respect, 
one of the relevant results of our analysis 
is the observation that new momentum regions 
appearing at the two-loop level cancel in the physical 
observable, i.e. the form factors $F_1$ and 
$F_2$ in this case. Indeed, it was observed
already some time ago \cite{Smirnov:1998vk,Smirnov:1999bza}
that at higher-loop order new regions may 
appear, compared to the one already present
at lower loops. From  a
factorization viewpoint this may be problematic, 
because it could imply that an all-order 
factorization cannot be obtained. Indeed, 
one would have to add new contributions to 
the factorization theorem at each subsequent
order in perturbation theory. For our case 
we find new \mbox{ultra-(anti-)collinear} 
regions appearing at two loops, both at the 
level of master integrals and single diagrams, 
but these cancel in the form factors, such 
that only the  regions already appearing 
at one loop contribute. More specifically, 
we find that the sum of the $uc\bar c$, 
$\overline{uc} c$ regions of diagrams (b) and (c) 
cancels against the sum of the regions 
$\bar c uc'$, $c \overline{uc}'$ in diagrams 
(f) and (g); similarly, the sum of the 
$\overline{uc} c$, $uc \bar c$ regions in diagrams 
(d) cancels against the sum of the $\bar c uc$, 
$\overline{uc} c$ regions in diagram (h). 

Focusing now on the calculation of the 
loop integrals in a given region, as usual 
one has to deal with standard UV and IR 
singularities, which we regulate in dimensional
regularization, as well as rapidity divergences.
In this work we consider massive form factors, 
which means that collinear singularities are 
regulated by the masses on the external legs, 
and one is left with soft singularities, which 
give a single pole per loop, proportional to
$1/\eps$ in dimensional regularization. As 
is well-known, the expansion into regions 
generates additional poles in each region. 
For instance, the hard region at one loop and 
the hard-hard region at two loop correspond
to the massless form factor, because masses are
neglected when the momentum is hard and proportional 
to the large scale $\hat s$; as such, in the 
hard region we find double poles per loop, 
generated when the loop momentum becomes soft 
and collinear to one of the external momenta. 
The double poles cancel against additional UV 
singularities arising in the collinear momentum 
regions; indeed, the cancellation of spurious 
singularities, i.e., the fact that the two loop 
massive form factors contain at most $1/\eps^2$ 
poles, provides another check that all regions 
have been correctly taken into account. 

As mentioned above, the expansion by regions
generates also rapidity singularities, which we 
observe in topology $B$ and $X$. We discussed 
in Sect.~\ref{sec:flow} that the original master 
integrals do not have rapidity divergences, and 
we observed their cancellation at the integral level, 
when summing all regions of a given integral (details
have been given in Sect.~\ref{sec:details}). In the
context of the present discussion, it may be more 
interesting to note that rapidity divergences 
cancel \emph{per region} for the form factor $F_1$ 
at LP, and at NLP for the form factor $F_2$ 
(which is probably a consequence of the fact 
that $F_2$ starts at NLP). With the calculation 
at hand we are however not able to determine 
whether this is a general feature, valid 
at arbitrary order. We leave such questions
for further research. 

Concerning the specific structure of the rapidity 
divergences per diagram, we saw that for 
topology $B$ it was enough to use a single rapidity 
regulator, while for topology $X$ we needed up to 
two rapidity regulators. In general the addition 
of a rapidity regulator breaks the symmetry 
$p_1\leftrightarrow p_2$: for instance, in case 
of topology $B$ the $cc$ region ceases to be equal 
to the $\bar{c}\bar{c}$ region. In case 
of more than one rapidity regulator, symmetry 
between regions can also be broken by expanding 
the integral in the regulators in a given order. 
This is what happens in case of topology 
$X$: even if we choose the rapidity regulators 
$\nu_1$ and $\nu_2$ such that the symmetry 
between the $cc$ and $\bar c \bar c$ regions is respected, 
expanding in $\nu_1$ and $\nu_2$ in a chosen 
order breaks this symmetry.

%%%%%%%%%%%%%%%%%%%%%%%%%%%%%%%%%%%%%%%%%%%%%%%%%%%%%%%%%%%%%%
\section{Conclusion}
\label{sec:conclusion}
%%%%%%%%%%%%%%%%%%%%%%%%%%%%%%%%%%%%%%%%%%%%%%%%%%%%%%%%%%%%%%
\noindent
In this paper we performed a region analysis of the two-loop massive quark form factor in QED. We categorized all contributions per region up to next-to-leading power in the quark mass, paving the way towards factorization tests beyond LP. The calculation itself required the introduction of three topologies of master integrals and up to 12 different regions per topology, with rapidity divergences appearing in 
two out of three topologies, thus revealing the richness and subtle aspects that entered our analysis.
We demonstrated how the rapidity divergences canceled in the sum of all regions and subsequently validated our result by reproducing the form factor at NLP as known in literature. Topology $A$ was the least complex and could be solved by rewriting subleading corrections in terms of LP propagators. Topology $B$ introduced for the first time rapidity regulators in our analysis and this required a detailed inspection on a case by case basis depending on the given diagram. In case of topology $X$, we had to introduce two unique rapidity regulators, which we labeled $\nu_1$ and $\nu_2$, and we found that its expansion order did not affect the form factor provided that order was kept fixed. Our method also revealed the need of multiple momentum routings in topologies $B$ and $X$ to cancel not only all rapidity divergences, and uncover additional regions that would otherwise have remained hidden. 

To conclude, we found that the calculation 
of the massive form factors in the limit 
$\hat s \gg m^2$ by means of the method of regions
provides useful data in light of developing a 
factorization framework for scattering amplitudes
beyond leading power. It gave us the additional opportunity to
test features of the region expansion of complete 
form factors, giving new perspectives with respect to cases 
where the method is applied to single integrals.

%~~~~~~~~~~~~~~~~~~~~~~~~~~~~~~~~~~~~~~~~~~~~~~~~~~~~~~~~~~~~~
\acknowledgments
%~~~~~~~~~~~~~~~~~~~~~~~~~~~~~~~~~~~~~~~~~~~~~~~~~~~~~~~~~~~~~
\noindent 
We thank Nicol\`o Maresca for cross-checking some of the master integrals, Roberto Bonciani and Pierpaolo Mastrolia for helpful communications on Ref.~\cite{Bonciani:2003ai}, and Yao Ma for pointing out the region interpretation discussed in footnote \ref{ColemanNortonEqFn}. L.V. thanks Nikhef for hospitality and partial support during the completion of this work, and the Erwin-Schrödinger International Institute for Mathematics and Physics at the University of Vienna for partial support during the Program ``Quantum Field Theory at the Frontiers of the Strong Interactions", July 31 - September 1, 2023. E.L. and L.V. also thank the Galileo Galilei Institute for Theoretical Physics of INFN, Florenze, for partial support during the Program ``Theory Challenges in the Precision Era of the Large Hadron Collider", August 28 - October 13, 2023.

The work of J.t.H is supported by the Dutch Science Council (NWO) via an ENW-KLEIN-2 project. The work of L.V. has been partly supported by Fellini - Fellowship for Innovation at INFN, funded by the European Union's Horizon 2020 research program under the Marie Sk\l{}odowska-Curie Cofund Action, grant agreement no. 754496. The research of G.W. was supported in part by the International Postdoctoral Exchange Fellowship Program from China Postdoctoral Council under Grant No. PC2021066.

\appendix
%%%%%%%%%%%%%%%%%%%%%%%%%%%%%%%%%
\section{Rapidity regulators}
\label{sec:app-rapidity-reg}
%%%%%%%%%%%%%%%%%%%%%%%%%%%%%%%%%
In this appendix, we consider the following one-loop integral
\begin{align}\label{eq:oneloopfull}
R&= \int [dk]
\frac{1}{[k^2-m^2]} 
\frac{1}{[(k+p_1)^2]} 
\frac{1}{[(k-p_2)^2]}\,,
\end{align}
with $p_i^2=m^2$ and study three different regulators that can be used to regulate the rapidity divergences that show up once Eq.~\eqref{eq:oneloopfull} is expanded in momentum regions. 
Before we discuss these regulators, let us briefly discuss how rapidity divergences appear. 
For reasons that will become clear in a moment, let us put the mass of the first propagator in Eq.~\eqref{eq:oneloopfull} to $M^2$ for now and consider the collinear expansion
\begin{align}\label{eq:oneloopc}
R\Big|_{c}&= \int[dk]
\frac{1}{[k^2-M^2]} 
\frac{1}{[(k+p_1)^2]} 
\frac{1}{[-k^+ p_2^-]}\nn\\
&= \frac{i \mu^{2\eps} e^{\eps\gamma_E}}{(4\pi)^2} \frac{\Ga(\eps)}{\hat{s}}
\int_0^1 dx\,x^{-1}(1-x)^{-\eps} \left(-m^2 x+M^2\right)^{-\eps}\,,
\end{align}
where we performed the loop integral. One can obtain the remaining $x$-integral after a standard Feynman parametrization.
In the limit $M^2\to m^2$ that we are interested in, the remaining $x$-integral 
\begin{equation}\label{eq:rapidityDivergence}
\int_0^1 dx\, x^{-1}\, (1-x)^{-2\eps}
\end{equation}
diverges for $x\to0$.\footnote{In fact, the $x$-integral diverges for all $M\neq 0$.}
To be precise, we see that the dimensional regulator $\eps$ does not regulate all divergences present in the integral in Eq.~\eqref{eq:oneloopc}. This so-called rapidity divergence can be traced back to the fact that the eikonal propagator only contains the $k^+$ component and therefore new divergences may arise from the $k^+$-integral as $\eps$ only regulates the transverse momentum component $k_\perp$. 

Eq.~\eqref{eq:oneloopc} also shows that in the limit $M^2\to 0$, the divergence of the remaining $x$-integral
\begin{equation}\label{eq:NoRapidityDivergence}
\int_0^1 dx\, x^{-1-\eps}\, (1-x)^{-\eps}
= \frac{\Ga(-\eps)\Ga(1-\eps)}{\Ga(1-2\eps)}\,,
\end{equation}
is fully regulated by $\eps$ alone. 
In other words, the presence of rapidity divergences in a Feynman integral is sensitive to propagator masses. This example reflects therefore why topology $A$, being free of rapidity divergences, is so different fromtopology $B$ and $X$, which do have rapidity divergences.

%~~~~~~~~~~~~~~~~~~~~~~~~~~~~
\subsection{Full result}
%~~~~~~~~~~~~~~~~~~~~~~~~~~~~
To validate that rapidity regulators work, we need to calculate the full integral Eq.~\eqref{eq:oneloopfull}. A straightforward calculation (using e.g. the Schwinger parameterisation followed by one-fold Melin-Barnes integral) yields
\begin{align}
\label{eq:oneloopfullresult}
R^{(\mathrm{full})}
&= \frac{ie^{\epsilon\gamma_E}}{(4\pi)^2}\left(\frac{\mu^2}{m^2}\right)^{\epsilon}\bigg[\left(1-\frac{4m^2}{s}\right)^{-\frac{1}{2}+\epsilon}\frac{\pi^2}{s}\frac{\Gamma(1-2\epsilon)\Gamma(1+2\epsilon)}{\Gamma^2(1-\epsilon)\Gamma(1+\epsilon)} \nonumber \\
&- \frac{\Gamma(-1-2\epsilon)\Gamma(1+\epsilon)}{m^2\Gamma(1-2\epsilon)}\,_2F_1(1,1,\frac{3}{2}+\epsilon,\frac{s}{4m^2})\nonumber \\
&+ \left(\frac{-s}{m^2}\right)^{-\epsilon}\frac{\Gamma^2(1-\epsilon)\Gamma(\epsilon)}{2m^2\Gamma(1-2\epsilon)}{}\,_2F_1(1,1-\epsilon,\frac{3}{2},\frac{s}{4m^2})\bigg],
\end{align}
where we recall that $s=(p_1+p_2)^2=2m^2+\hat{s}+m^4/\hat{s}$. 
In the small mass limit $m^2 \ll \hat{s}$, we expand Eq.~\eqref{eq:oneloopfullresult} up to NNLP as
\begin{align}
R^{(\mathrm{full})}\Big|_{\text{NNLP}}
&= \frac{i}{(4\pi)^2\hat{s}}\bigg\{\left[ 4\zeta_2 + \frac{1}{2}\ln^2\left(-\frac{m^2}{\hat{s}}\right)\right] 
-\frac{2m^2}{\hat{s}} \nn\\
&\hspace{3cm}
+ \left(\frac{m^2}{\hat{s}}\right)^2\left[ \frac{1}{2}  + 4\zeta_2 + \frac{1}{2}\ln^2\left(-\frac{m^2}{\hat{s}}\right)\right]\bigg\} + \mathcal{O}\left(\epsilon\right)\,.
\label{eq:oneloopfullr}
\end{align}
This expansion can now be compared to the region expansion of Eq.~\eqref{eq:oneloopfull}.

%~~~~~~~~~~~~~~~~~~~~~~~~~~~~--------------------------------
\subsection{Analytic regulator}\label{sec:analyticregulator1}
%~~~~~~~~~~~~~~~~~~~~~~~~~~~~--------------------------------
As this is the regulator used throughout the main body of this work, we first discuss the analytic regulator, which is implemented by raising the last propagator to a fractional power $\nu$ such that Eq.~\eqref{eq:oneloopfull} is rewritten as \cite{Beneke:2003pa} 
\begin{align}\label{eq:oneloopfulla}
R^{(\mathrm{a.r.})}&= \int[dk]
\frac{1}{[k^2-m^2]} 
\frac{1}{[(k+p_1)^2]} 
\frac{\left(\tilde{\mu}^2\right)^{\nu}}{[(k-p_2)^2]^{1+\nu}}\,.
\end{align}
For simplicity, we only consider the LP contribution in the remainder of this appendix. The hard contribution is given by
\begin{align}
R^{(\mathrm{a.r.})}\Big|_{h}
&= \int[dk]
\frac{1}{[k^2]} 
\frac{1}{[k^2+k^-p_1^+]} 
\frac{\left(\tilde{\mu}^2\right)^{\nu}}{[k^2-k^+p_2^-]^{1+\nu}} \nn \\
&= \frac{i}{(4\pi)^2\hat{s}}\left(\frac{\mu^2}{-\hat{s}}\right)^{\epsilon}
\left[\frac{1}{\eps^2}-\frac{\zeta_2}{2}+\ord{\eps,\nu}\right]\,.\label{eq:analyticRegulatorh}
\end{align}
We note there is no rapidity divergence in the hard region, so we can safely set $\nu=0$ either at the beginning or at the end of the calculation.
In the collinear region, we have
\begin{align}
R^{(\mathrm{a.r.})}\Big|_{c}
&= \int[dk]
\frac{1}{[k^2-m^2]} 
\frac{1}{[(k+p_1)^2]} 
\frac{\left(\tilde{\mu}^2\right)^{\nu}}{[-k^+ p_2^-]^{1+\nu}} \nn \\
&=\frac{ie^{\epsilon\gamma_E}}{(4\pi)^2\hat{s}}\left(\frac{\mu^2}{m^2}\right)^{\epsilon}
\left(\frac{\tilde{\mu}^2}{\hat{s}}\right)^{\nu}
\Ga(\eps)\int_0^1dx\,x^{-1-\nu}(1-x)^{-2\eps}\,.
\end{align}
Comparing to Eq.~\eqref{eq:rapidityDivergence} we explicitly see how the power $\nu$ regulates the rapidity divergence in a similar manner as how $\eps$ regulates the IR and UV divergences in Eq.~\eqref{eq:NoRapidityDivergence}. Carrying out the integral over $x$ yields
\begin{align}
R^{(\mathrm{a.r.})}\Big|_c
&= \frac{i}{(4\pi)^2\hat{s}}\left(\frac{\mu^2}{m^2}\right)^{\epsilon}\left(\frac{\tilde{\mu}^2}{\hat{s}}\right)^{\nu}
\left[-\frac{1}{\eps\,\nu}+2\zeta_2 +\ord{\eps,\nu}\right]\,.\label{eq:R.a.r.h}
\end{align}
Similarly, after expanding Eq.~\eqref{eq:oneloopfulla} in the anti-collinear region, we obtain
\begin{align}
R^{(\mathrm{a.r.})}\Big|_{\bar{c}} &= \int[dk]
\frac{1}{[k^2-m^2]} 
\frac{1}{[k^- p_1^+]} 
\frac{\left(\tilde{\mu}^2\right)^{\nu}}{[(k-p_2)^2]^{1+\nu}} \nn \\
&= \frac{i}{(4\pi)^2\hat{s}}\left(\frac{\mu^2}{m^2}\right)^{\epsilon}\left(\frac{\tilde{\mu}^2}{-m^2}\right)^{\nu}
\left[-\frac{1}{\eps^2}+\frac{1}{\eps\,\nu}+\frac{5\zeta_2}{2}+\ord{\eps,\nu}\right]\, .\label{eq:R.a.r.cb}
\end{align}
Note that the symmetry of the collinear and anti-collinear gets broken due to the fact that we added the analytic regulator to the last propagator only.\footnote{This is similar to diagram (e) as is discussed in Sec.~\ref{sec:topoB}: adding a power $\nu$ in a symmetric way, i.e. to both the second and third propagator, does not regulate the rapidity divergences.} Here, we remark that the natural overall scales for a hard or collinear loop are $(\mu^2/m^2)^\eps$ and $(\mu^2/(-\hat{s}))^\eps$ respectively. However, with the analytic regulator, we get slightly different overall factors $(\tilde\mu^2/\hat{s})^\nu$ and $(\tilde\mu^2/(-m^2))^\nu$, see Eqs.~\eqref{eq:R.a.r.h} and \eqref{eq:R.a.r.cb} respectively. This is an effect of the usual Wick rotation to Euclidean space, which produces an additional factor $(-1)^\nu$.

Other regions do not contribute. For example, the semi-hard region leads to a scaleless integral and thus a vanishing contribution
\begin{align}
R^{(\mathrm{a.r.})}\Big|_{sh}= \int[dk]
\frac{1}{[k^2-m^2]} 
\frac{1}{[k^- p_1^+]} 
\frac{\left(\tilde{\mu}^2\right)^{\nu}}{[-k^+ p_2^-]^{1+\nu}} =0.\label{eq:analyticRegulatorSoft}
\end{align}
Now, after combining all contributing regions we find that
\begin{align}
R^{(\mathrm{a.r.})}\Big|_{\text{LP}} &= R^{(\mathrm{a.r.})}\Big|_{h} + R^{(\mathrm{a.r.})}\Big|_{c} + R^{(\mathrm{a.r.})}\Big|_{\bar{c}} \nonumber \\ 
  & = \frac{i}{(4\pi)^2\hat{s}}\left[ 4\zeta_2 + \frac{1}{2}\ln^2\left(-\frac{m^2}{\hat{s}}\right)\right] + \mathcal{O}\left(\epsilon\right),
\end{align}
which is the same as the LP result in Eq.~\eqref{eq:oneloopfullr} and we notice that all rapidity divergences have cancelled in the final result, as they should.

%~~~~~~~~~~~~~~~~~~~~~~~~~~~~--------------------------------
\subsection{Modified analytic regulator}\label{sec:analyticregulator2}
%~~~~~~~~~~~~~~~~~~~~~~~~~~~~--------------------------------
Instead of raising the power of a propagator by $\nu$, the analytic regulator can also be used to modify the phase space measure as \cite{Becher:2011dz}
\begin{align}
\int d^dk \, \delta\left(k^2\right)\theta\left(k^0\right)\; \rightarrow \; \int d^dk \, \delta\left(k^2\right)\theta\left(k^0\right) \left(\frac{\tilde{\mu}}{k^-}\right)^{\nu}\,.
\end{align}
Here the amplitude itself does not need to be modified. This has the advantage that fundamental properties such as gauge invariance and the eikonal form of the soft and collinear emissions are maintained. The modified analytic regulator is therefore convenient to construct factorization theorems to all orders. In this work, although we deal with loop integrals, it is possible to apply a similar scheme to regulate the rapidity divergence. To this end, we can modify the measure $[dk]$ as follows \cite{Gritschacher:2013tza}
\begin{align}\label{eq:regulatord2}
\int [dk]\; \rightarrow \; \int [dk]\left(\frac{\tilde{\mu}^2}{-k^+ p_2^-+i0^+}\right)^{\nu}  \,,
\end{align}
and define
\begin{align}\label{eq:oneloopfullb}
R^{(\mathrm{m.a.r.})}&= \int[dk]
\frac{1}{[k^2-m^2]} 
\frac{1}{[(k+p_1)^2]} 
\frac{1}{[(k-p_2)^2]} \left(\frac{\tilde{\mu}^2}{-k^+ p_2^-+i0^+}\right)^{\nu} \,.
\end{align}
Note that the chosen regulator on the right hand side of Eq.~\eqref{eq:regulatord2} leads to $R^{(\mathrm{m.a.r.})}|_h=R^{(\mathrm{a.r.})}|_h$, $R^{(\mathrm{m.a.r.})}|_c=R^{(\mathrm{a.r.})}|_c$ and $R^{(\mathrm{m.a.r.})}|_{sh}=R^{(\mathrm{a.r.})}|_{sh}$. 
After a direct calculation of $R^{(\mathrm{m.a.r.})}_{\bar{c}}$, we have 
\begin{align}
R^{(\mathrm{m.a.r.})}\Big|_{\bar{\text{c}}} 
&= \int[dk]
\frac{1}{[k^2-m^2]} 
\frac{1}{[k^-p_1^+]} 
\frac{1}{[(k-p_2)^2]}
\frac{\left(\tilde{\mu}\right)^2}{[-k^+p_2^-]^{\nu}}\nonumber \\
&=\frac{ie^{\epsilon\gamma_E}}{(4\pi)^2\hat{s}}\left(\frac{\mu^2}{m^2}\right)^{\epsilon}\left(\frac{\tilde{\mu}^2}{-m^2}\right)^{\nu}\left[\frac{\Gamma(\epsilon)}{\nu} - \Gamma(\epsilon) \left(\psi(1-2 \epsilon )-\psi(\epsilon) \right)  +\ord{\nu}\right]\,.\label{eq:modifiedAnalyticCb}
\end{align}
Comparing $R^{(\mathrm{m.a.r.})}|_{\bar{c}}$ with $R^{(\mathrm{a.r.})}|'_{\bar{c}}$, we find that
\begin{align}
 R^{(\mathrm{m.a.r.})}\Big|_{\bar{c}} - R^{(\mathrm{a.r.})}\Big|_{\bar{c}}= \mathcal{O}(\nu).
\end{align}
which means that this regulator is also sufficient to reproduce the result of Eq.~\eqref{eq:oneloopfullr}.

%~~~~~~~~~~~~~~~~~~~~~~~~~~~~---------------
\subsection{\texorpdfstring{$\delta$}{ }-regulator}\label{sec:deltaregulator}
%~~~~~~~~~~~~~~~~~~~~~~~~~~~~---------------
The final regulator we want to discuss is the so-called $\delta$-regulator \cite{Chiu:2009yx}.
It is implemented by adding a small mass to the propagator denominators,
\begin{equation}
R^{(\delta)}  = \int [dk]
\frac{1}{[k^2-m^2-\delta_1]} 
\frac{1}{[(k+p_1)^2]-\delta_2]} 
\frac{1}{[(k-p_2)^2-\delta_3]]}\,.
\end{equation}
The $\delta_i$ are regulator parameters that are set to zero unless they are needed to regulate any divergences.
Following the discussion below Eq.~\eqref{eq:oneloopc}, we see that we can immediately put $\delta_1 = 0$ from the beginning as Eq.~\eqref{eq:oneloopc} is divergent for all $M\neq0$. This means that the rapidity divergences have to be regulated by $\delta_2$ and/or $\delta_3$.

We recall from Sec.~\ref{sec:analyticregulator1} that the hard region is free of rapidity divergences, so $\delta_2$ and $\delta_3$ can be set to zero in this region as well, which reproduces Eq.~\eqref{eq:analyticRegulatorh}
\begin{equation}
R^{(\delta)}\Big|_h
= \frac{i}{(4\pi)^2\hat{s}}\left(\frac{\mu^2}{-\hat{s}}\right)^{\epsilon}
\left[\frac{1}{\eps^2}-\frac{\zeta_2}{2}+\ord{\eps}\right]\,.
\end{equation}
Next, we consider the collinear expansion
\begin{align}
R^{(\delta)}\Big|_c 
&= \int[dk]
\frac{1}{[k^2-m^2]} 
\frac{1}{[(k+p_1)^2-\delta_2]} 
\frac{1}{[-k^+p_2^- -\de_3]}\label{eq:rapidityTestCregion}\nn\\
&= \frac{i\mu^{2\eps}e^{\eps\gamma_E}}{(4\pi)^2}\frac{\Ga(\eps)}{\hat{s}}
\int_0^1 dx\ \left(x-\delta_3/\hat{s}\right)^{-1}\left(m^2(1-x)^2+\delta_2 x\right)^{-\eps}\,.
\end{align}
Comparing Eq.~\eqref{eq:rapidityTestCregion} to Eq.~\eqref{eq:rapidityDivergence} we see that $\delta_3$ regulates the rapidity divergence in the remaining $x$-integral. Furthermore, we notice that $\delta_2$ is not needed to regulate the divergence and can therefore be set to zero. Carrying out the remaining $x$-integral and performing the $\eps$-expansion yields
\begin{align}
R^{(\delta)}\Big|_c
&= \frac{i}{(4\pi)^{2}\hat{s}}
\bigg(\frac{\mu^2}{m^2}\bigg)^{\eps}
\bigg[
\frac{1}{\eps}\ln\Big(1-\frac{\hat{s}}{\de_3}\Big)
+2\text{Li}_2\Big(\frac{1}{1-\de_3/\hat{s}}\Big)
+\ord{\eps}\bigg]\,.
\end{align}
Similarly for the anti-collinear region we now need to keep $\delta_2$ to regulate the rapidity divergence and can set $\delta_3=0$.  This yields
\begin{equation}
R^{(\delta)}\Big|_{\bar{c}}
=\int[dk]
\frac{1}{[k^2-m^2]} 
\frac{1}{[k^-p_1^+-\delta_2]} 
\frac{1}{[(k-p_2)^2]}
= R^{(\delta)}\Big|_c\ (\delta_2\leftrightarrow\delta_3)\,.
\end{equation}
Next, we take the semi-hard expansion, where both $\delta_2$ and $\delta_3$ need to be kept to regulate the rapidity divergences. Because of the mass-like terms $\delta_2$ and $\delta_3$, this does not lead to a scaleless integral like for the analytic regulator, Eq.~\eqref{eq:analyticRegulatorSoft}. We get
\begin{align}
R^{(\delta)}\Big|_{sh} 
&= \int[dk]
\frac{1}{[k^2-m^2]} 
\frac{1}{[k^-p_1^+-\de_2]} 
\frac{1}{[-k^+p_2^--\de_3]} \label{eq:rapidityTestSregion}\nn\\
&= \frac{i}{(4\pi)^{2}\hat{s}} 
\left(\frac{\mu^2}{m^2}\right)^{\eps}
\left[
\frac{1}{\eps^2}
+\frac{1}{\eps}\ln\left(-\frac{m^2\hat{s}}{\de_2\de_3}\right)
+\frac{\zeta_2}{2} - \text{Li}_2\left(1+\frac{\de_2
\de_3}{m^2\hat{s}}\right)+\ord{\eps}\right]\,,
\end{align}
where we performed the $\eps$-expansion.

Unfortunately, taking the $h$, $c$, $\bar{c}$ and $sh$ regions together does not reproduce the LP part of the full result as given in Eq.~\eqref{eq:oneloopfullr} and in particular, the rapidity divergences are not canceled. The reason is that the semi-hard region has overlap with the collinear and anti-collinear regions. To be precise, taking the soft limit of the collinear region, Eq.~\eqref{eq:rapidityTestCregion} yields 
\begin{align}
R^{(\delta)}\Big|_{c,\emptyset}
=\int[dk]
\frac{1}{[k^2-m^2]} 
\frac{1}{[k^-p_1^+-\de_2]} 
\frac{1}{[-k^+p_2^--\de_3]} 
= R^{(\delta)}\Big|_{sh}\,,
\end{align}
where we recognize the semi-hard region $R^{(\delta)}|_{sh}$ of Eq.~\eqref{eq:rapidityTestSregion}. To get the full result, we have to perform a so-called zero-bin subtraction where one subtracts the overlapping semi-hard region from the collinear region. Similarly, for the anti-collinear region one has to subtract
\begin{equation}
R^{(\delta)}\Big|_{\bar{c},\emptyset}
= R^{(\delta)}\Big|_{sh}\,.
\end{equation}
Indeed, adding all regions together and including the zero-bin subtractions yields the correct full result of Eq.~\eqref{eq:oneloopfullr}
\begin{align}
R^{(\delta)}\Big|_{\text{LP}}
=&R^{(\delta)}\Big|_h+R^{(\delta)}\Big|_c+R^{(\delta)}\Big|_{\bar{c}}-R^{(\delta)}\Big|_{sh}\nn\\
=& \frac{i}{(4\pi)^2\hat{s}}\left[ 4\zeta_2 + \frac{1}{2}\ln^2\left(-\frac{m^2}{\hat{s}}\right)\right] 
+ \mathcal{O}\left(\epsilon,\delta_i\right)\,,
\end{align}
where in the last line we were able to take the $\delta_2\to0$ and $\delta_3\to0$ limits as the rapidity divergences cancel.

%~~~~~~~~~~~~~~~~~~~~~~~~~~~~-----------------
\subsection{Choosing a rapidity regulator}
%~~~~~~~~~~~~~~~~~~~~~~~~~~~~-----------------
We found that all three regulators can be used to regulate the rapidity divergence that shows up in the region expansion of the Feynman integral of Eq.~\eqref{eq:oneloopfull}. 
However, from a calculation point of view, the $\delta$-regulator given in Sec.~\ref{sec:deltaregulator} is the most complicated one as it leads to additional scales in the integral. Furthermore, it introduces an additional semi-hard region compared to the analytic and modified analytic regulators as discussed in Secs.~\ref{sec:analyticregulator1} and~\ref{sec:analyticregulator2}. Regarding the $
\delta$-regulator, we showed that a zero-bin subtraction was necessary to avoid double counting momentum regions, which complicated the region analysis even further. 

The analytic regulator and the modified analytic regulator are similar to each other. Neither of them increase the number of scales present in the Feynman integral and in case of the example discussed in this appendix, both lead to the same regions. With the specific choice we made for the analytic and modified analytic regulator, the only difference between the two at the integrand level comes from the anti-collinear region, Eq.~\eqref{eq:modifiedAnalyticCb}. As a result, due to the additional propagator, the calculation of $R^{(\mathrm{m.a.r.})}|_{\bar{c}}$ is more complex than $R^{(\mathrm{a.r.})}|_{\bar{c}}$. This additional complexity that arises from the modified analytic regulator would make the two-loop calculation of the form factor much more difficult as compared to when one would adopt the analytic regulator instead. In this work, we therefore used the analytic regulator whenever rapidity divergences showed up.
%%%%%%%%%%%%%%%%%%%%%%%%%%%%%%%%%%%%%%%%%%%%%%%%%%%%%%%%%%%%%%
\section{Regions in topology \texorpdfstring{$X$}{ }}
\label{sec:regions_X}
%%%%%%%%%%%%%%%%%%%%%%%%%%%%%%%%%%%%%%%%%%%%%%%%%%%%%%%%%%%%%%
As discussed in  detail in Sec.~\ref{sec:details}, finding all momentum regions for a given Feynman integral can be a subtle process. 
To supplement the discussion given in Sec.~\ref{sec:details}, we provide in this appendix some details about the momentum region analysis for one of the Feynman integrals needed in the calculation of diagram (h), which belongs to topology $X$. 
Topology $X$ in particular is difficult because of the complexity of the integrals - we needed two different rapidity regulators and had to route the momenta in three different ways in order to find all regions.
The example we consider is $I_{X;1,1,1,1,1,1,0}^{\nu_1,\nu_1,\nu_2,\nu_2}$, which according to the momentum routing given in Eq.~\eqref{eq:topoX_full} reads
\begin{align}\label{eq:topoX_full111111}
\mathcal{I}^X&=\frac{1}{C}\,I_{X;1,1,1,1,1,1,0}^{\nu_1,\nu_1,\nu_2,\nu_2} \nonumber \\
&=\frac{1}{C}\,\int[dk_1][dk_2]
\frac{1}{k_1^2}
\frac{1}{k_2^2}
\frac{\tilde{\mu}_1^{2\nu_1}}{[(k_2-p_1)^2-m^2]^{1+\nu_1}}
\frac{\tilde{\mu}_1^{2\nu_1}}{[(k_1+k_2-p_1)^2-m^2]^{1+\nu_1}}\nonumber \\
&\hspace{2cm}\times
\frac{\tilde{\mu}_2^{2\nu_2}}{[(k_1+p_2)^2-m^2]^{1+\nu_2}}
\frac{\tilde{\mu}_2^{2\nu_2}}{[(k_1+k_2+p_2)^2-m^2]^{1+\nu_2}} \,,
\end{align}
where for convenience we factored out
\begin{align}
C= \frac{1}{(4\pi)^4\hat{s}^2}\,.
\end{align}
As discussed in Sec.~\ref{sec:topoX}, both regulators $\nu_1$ and $\nu_2$ are needed to regulate the rapidity divergences once $\mathcal{I}^X$ is expanded in different momentum regions.
The full unexpanded integral on the contrary is free of rapidity divergences and therefore $\nu_1$ and $\nu_2$ can be set to zero, and the result can be found in Ref.~\cite{Bonciani:2003te,Gluza:2009yy}. In order to compare with the momentum region approach, we expand the full result in the small mass limit up to NLP:
\begin{align}\label{eq:Ifullr}
\mathcal{I}^X_{\text{full}}\Big|_{\text{NLP}}
=& \left(\frac{\mu^2}{m^2}\right)^{2\epsilon}\left[-\frac{1}{\epsilon}\left(
\frac{1}{3}L^3 + \zeta_2L + \zeta_3 \right) 
-\frac{1}{2}L^4 + \zeta_2L^2 - \zeta_3L -\frac{37\zeta_2^2}{10} \right.\nn \\
&\hspace{4cm}
\left.- \frac{m^2}{\hat{s}}\left(4L^2 - 8L + 4\zeta_2\right)
+ \mathcal{O}\left(\eps\right)\right]\,,
\end{align}
where we defined 
\begin{equation}
L = \ln\left(-\frac{m^2}{\hat{s}}\right)\,.
\end{equation}
In the remainder of this appendix we first present in App.~\ref{sec:allRegionsTopoX} all the momentum regions which contribute to the full result of Eq.~\eqref{eq:Ifullr} up to NLP. In App.~\ref{sec:asy}, we then compare these regions to the output of the software package \texttt{Asy.m}, which uses a geometric approach to reveal all the relevant regions for a given Feynman integral in parameter space. 
We finish this appendix with a discussion on a method that can be used to find all regions in momentum space.

%------------------------------------------
\subsection{Regions in momentum space}
\label{sec:allRegionsTopoX}
%------------------------------------------
As we have shown in Sec.~\ref{sec:topoX}, there are 12 regions needed for the integrals of topology $X$.\footnote{Note that we can safely limit ourselves to the Feynman integrals with $n_7\leq0$ as only these are needed in the computation of $F_1$ and $F_2$ for diagram (h).} 
To cover all these regions in momentum space, the region expansion given by the momentum routing of Eq.~\eqref{eq:topoX_full} is not sufficient, and one also needs to consider the routings as given in Eqs.\eqref{eq:topoX_full2} and~\eqref{eq:topoX_full3}. In what follows, we will show that indeed all three different momenta routings are needed in the region expansion of $\mathcal{I}^X$, Eq.~\eqref{eq:topoX_full111111}, to get all the 12 regions that make up the full result of Eq.~\eqref{eq:Ifullr}.
 
We first present all the eight different momentum regions given by the first parameterisation of topology $X$ defined by Eq.~\eqref{eq:topoX_full}. Up to NLP we get
\begin{align}
\mathcal{I}^X\Big|_{hh}
=& \left(\frac{\mu^2}{-\hat{s}}\right)^{2\epsilon}\Bigg[
-\frac{1}{\epsilon^4}
+\frac{6\zeta_2}{\epsilon^2}
+\frac{83\zeta_3}{3\epsilon}
+\frac{177\zeta_2^2}{10}
+\frac{m^2}{\hat{s}}\left(
-\frac{3}{\epsilon^2}
-\frac{6}{\epsilon}
+9\zeta_2+12\right)
\Bigg]\label{eq:IXhh}\\
%%%
\mathcal{I}^X\Big|_{cc}
=& \left(\frac{\mu^2}{m^2}\right)^{2\epsilon}\left(\frac{\tilde{\mu}_2^2}{\hat{s}}\right)^{2\nu_2}\Bigg[
\frac{3}{8\epsilon^4}
-\frac{1}{2\epsilon^3\nu_2}
+\frac{21\zeta_2}{8\epsilon^2}
+\frac{1}{\epsilon}\left(6\zeta_3-\frac{\zeta_2}{2\nu_2}\right)
+\frac{\zeta_3}{3\nu_2}+\frac{1177\zeta_2^2}{80}\nn\\
&
+\frac{m^2}{\hat{s}}\left(
\frac{1}{\epsilon}\left(\frac{2}{\nu_2}+7\right)
-14\zeta_2+\frac{10}{\nu_2}+21\right)
\Bigg]\\
%%%
\mathcal{I}^X\Big|_{\bar{c}c}
=& \left(\frac{\mu^2}{m^2}\right)^{2\epsilon}
\left(\frac{\tilde{\mu}_1^2}{-m^2}\right)^{\nu_1}
\left(\frac{\tilde{\mu}_1^2}{\hat{s}}\right)^{\nu_1}
\left(\frac{\tilde{\mu}_2^2}{-m^2}\right)^{\nu_2}
\left(\frac{\tilde{\mu}_2^2}{\hat{s}}\right)^{\nu_2}
\Bigg[
\frac{1}{4\epsilon^4}
+\frac{1}{\epsilon^3}\left(\frac{1}{2\nu_2}+\frac{1}{2\nu_1}\right)
\nn\\
&
+\frac{1}{\epsilon^2}\left(\frac{5\zeta_2}{4}-\frac{1}{\nu_1\nu_2}\right)+\frac{1}{\epsilon}\left(\frac{3\zeta_2}{2\nu_1}+\frac{3\zeta_2}{2\nu_2}+\frac{17\zeta_3}{6}\right)
-\frac{\zeta_2}{\nu_1\nu_2}+\frac{14\zeta_3}{3\nu_1}+\frac{14\zeta_3}{3\nu_2}+\frac{279\zeta_2^2}{40}\nn\\
&
+\frac{m^2}{\hat{s}}\left(
\frac{1}{\epsilon}\left(\frac{2}{\nu_2}+\frac{2}{\nu_1}+4\right)
-4\zeta_2+\frac{2}{\nu_1}+\frac{2}{\nu_2}-4\right)
\Bigg]\\
%%%
\mathcal{I}^X\Big|_{hc}
=& \left(\frac{\mu^2}{-\hat{s}}\right)^{\epsilon}\left(\frac{\mu^2}{m^2}\right)^{\epsilon}\Bigg[
\frac{8}{3\epsilon^4}
-\frac{8\zeta_2}{\epsilon^2}
-\frac{316\zeta_3}{9\epsilon}
-\frac{158\zeta_2^2}{5}\nn\\
&+\frac{m^2}{\hat{s}}\left(
-\frac{4}{3\epsilon^2}
+\frac{4}{3\epsilon}
+4\zeta_2-\frac{40}{3}\right)
\Bigg]\label{eq:IXhca}\\
%%%
\mathcal{I}^X\Big|_{\bar{c}uc}
=& \left(\frac{\mu^2}{m^2}\right)^{\epsilon}\left(\frac{\mu^2\hat{s}^2}{m^6}\right)^{\epsilon}\Bigg[
-\frac{1}{24\epsilon^4}
-\frac{5\zeta_2}{8\epsilon^2}
-\frac{7\zeta_3}{18\epsilon}
-\frac{493\zeta_2^2}{80}\nn\\
&+\frac{m^2}{\hat{s}}\left(
\frac{1}{3\epsilon^2}
+\frac{5}{3\epsilon}
+5\zeta_2+\frac{19}{3}\right)
\Bigg]
\end{align}
Note that the LP of $\cI^X|_{hh}$ can be found in \cite{Gonsalves:1983nq,Smirnov:1998vk}.
By symmetry, we have
\begin{align}
\mathcal{I}^X\Big|_{\bar{c}\bar{c}}
=\cI^X\Big|_{cc}(\nu_1\leftrightarrow\nu_2) 
\,,\qquad
\mathcal{I}^X\Big|_{\bar{c}h}
=\mathcal{I}^X\Big|_{hc}
\,,\qquad\text{and}\qquad
\mathcal{I}^X\Big|_{\overline{uc}c}
=\mathcal{I}^X\Big|_{\bar{c}uc}\,.
\end{align}
The second parameterisation of topology $X$, Eq.~\eqref{eq:topoX_full2}, gives two new regions: the $cc'$-region and the $\bar{c}\bar{c}'$-region. By defining\footnote{Recall that the difference between ${\cI'}^X$ and $\cI^X$ only arises in the momentum region expansion. The unexpanded integrals are the same.}
\begin{equation}
\mathcal{I'}^X=\frac{1}{C}\,{I'}_{X;1,1,1,1,1,1,0}^{\nu_1,\nu_1,\nu_2,\nu_2}\,,
\end{equation}
the NLP results of these two new regions read
\begin{align}
\mathcal{I'}^{X}\Big|_{cc'}
=& \left(\frac{\mu^2}{m^2}\right)^{2\epsilon}
\left(\frac{\tilde{\mu}_1^2}{-m^2}\right)^{\nu_1}\left(\frac{\tilde{\mu}_1^2}{\hat{s}}\right)^{\nu_1}
\left(\frac{\tilde{\mu}_2^2}{-m^2}\right)^{2\nu_2}\Bigg[
-\frac{29}{4\epsilon^4}
+\frac{1}{\epsilon^3}\left(\frac{3}{\nu_2}+\frac{1}{2\nu_1}\right)
\nn\\
&
+\frac{1}{\epsilon^2}\left(\frac{7\zeta_2}{4}-\frac{1}{\nu_2^2}\right)+\frac{1}{\epsilon}\left(\frac{\zeta_2}{2\nu_1}+\frac{83\zeta_3}{6}\right)
-\frac{\zeta_2}{\nu_2^2}-\frac{\zeta_3}{3\nu_1}-\frac{7\zeta_3}{\nu_2}+\frac{473\zeta_2^2}{40}\nn\\
&
+\frac{m^2}{\hat{s}}\left(
-\frac{3}{\epsilon^2}
+\frac{1}{\epsilon}\left(-\frac{2}{\nu_1}-14\right)
-\zeta_2-\frac{10}{\nu_1}-\frac{2}{\nu_2}-42\right)
\Bigg]\\
%%%
\mathcal{I'}^{X}\Big|_{\bar{c}\bar{c}'}
=& \left(\frac{\mu^2}{m^2}\right)^{2\epsilon}
\left(\frac{\tilde{\mu}_2^2}{-m^2}\right)^{2\nu_1}
\left(\frac{\tilde{\mu}_1^2}{-m^2}\right)^{\nu_2}\left(\frac{\tilde{\mu}_1^2}{\hat{s}}\right)^{\nu_2}
\Bigg[
\frac{2}{\epsilon^4}
+\frac{1}{\epsilon^3}\left(-\frac{3}{\nu_2}-\frac{1}{2\nu_1}\right)\nn\\
&
+\frac{1}{\epsilon^2}\left(3\zeta_2+\frac{1}{\nu_1\nu_2}+\frac{1}{\nu_2^2}\right)
+\frac{1}{\epsilon}\left(-\frac{3\zeta_2}{2\nu_1}-\frac{\zeta_2}{\nu_2}+\frac{41\zeta_3}{3}\right)
\nn\\
&
+\frac{\zeta_2}{\nu_1\nu_2}+\frac{\zeta_2}{\nu_2^2}-\frac{14\zeta_3}{3\nu_1}+\frac{2\zeta_3}{\nu_2}+\frac{59\zeta_2^2}{10}\nn\\
&
+\frac{m^2}{\hat{s}}\left(
\frac{4}{\epsilon^2}
+\frac{1}{\epsilon}\left(-\frac{4}{\nu_2}-\frac{2}{\nu_1}-4\right)
+2\zeta_2-\frac{2}{\nu_1}-\frac{10}{\nu_2}+2\right)
\Bigg]\,.
\end{align}
Finally, to get the full NLP result, two more regions are needed and can be found using the third parametersisation.
Recalling Eq.~\eqref{eq:topoX_full3}, we define 
\begin{equation}
\mathcal{I''}^{X}=\frac{1}{C}\,{I''}_{X;1,1,1,1,1,1,0}^{\nu_1,\nu_1,\nu_2,\nu_2}\,.
\end{equation}
These last two missing momentum regions are the $ch''$ and $\bar{c}h''$ regions and their expressions are given by
\begin{align}
\mathcal{I''}^{X}\Big|_{ch''}
&= \left(\frac{\mu^2}{m^2}\right)^{\epsilon}\left(\frac{\mu^2}{-\hat{s}}\right)^{\epsilon}\Bigg[
\frac{m^2}{\hat{s}}\left(
\frac{2}{\epsilon^2}
+2\right)
\Bigg]\,,
\end{align}
and by symmetry
\begin{align}\label{eq:IXcbarhpp}
\mathcal{I''}^{X}\Big|_{\bar{c}h}
&=\mathcal{I''}^{X}\Big|_{ch''}\,.
\end{align}
Note that the $ch''$ and $\bar{c}h''$ regions do not contribute at LP.

For notational simplicity, we denote $\cI^X\big|_\bullet$ as $\cI^X_\bullet$ from now on. Combining all of the above 12 regions, we obtain the result of the integral Eq.~\eqref{eq:topoX_full111111} up to NLP. That is,
\begin{align}\label{eq:Iregionr}
\mathcal{I}^{X}_{hh} 
&+ \mathcal{I}^{X}_{cc} 
+ \mathcal{I}^{X}_{\bar{c}\bar{c}} 
+ \mathcal{I}^{X}_{\bar{c}c}
+ \mathcal{I}^{X}_{hc}
+ \mathcal{I}^{X}_{\bar{c}h}
+ \mathcal{I}^{X}_{\bar{c}uc} 
+ \mathcal{I}^{X}_{\overline{uc}c}
+ \mathcal{I'}^{X}_{cc'} 
+ \mathcal{I'}^{X}_{\bar{c}\bar{c}'} 
+ \mathcal{I''}^{X}_{ch''} 
+ \mathcal{I''}^{X}_{\bar{c}h''}  \nonumber \\
&=\left(\frac{\mu^2}{m^2}\right)^{2\epsilon}\left[-\frac{1}{\epsilon}\left(
\frac{1}{3}L^3 + \zeta_2L + \zeta_3 \right) 
-\frac{1}{2}L^4 + \zeta_2L^2 - \zeta_3L -\frac{37\zeta_2^2}{10} \right.\nn \\
&\hspace{4cm}
\left.- \frac{m^2}{\hat{s}}\left(4L^2 - 8L + \frac{4\zeta_2}{3}\right) + \mathcal{O}(\nu_1, \nu_2, \epsilon) \right]\,,
\end{align}
which is the same as Eq.~\eqref{eq:Ifullr}. We notice that all rapidity divergences have canceled.

%------------------------------------------
\subsection{Regions in parameter space}
\label{sec:asy}
%------------------------------------------
To find all the above 12 regions for $\cI^X$, we can also use the \texttt{Mathematica} package $\texttt{Asy.m}$~\cite{Pak:2010pt,Jantzen:2012mw}, which implements a geometric approach to reveal the relevant regions for a given Feynman integral and a given limit of momenta and masses.
The program relies on the alpha-representation of $\mathcal{I}^X$ in Eq.~\eqref{eq:topoX_full111111}, which can be written in the following form
\begin{align}\label{eq:topoX_full111111a}
\cI^X=C_1\int_0^{\infty}[dy] \, (y_3y_4)^{\nu_1}(y_5y_6)^{\nu_2}A_3^{3\epsilon+2\nu_1+2\nu_2} \left(m^2A_1-s_{12}A_2-i\eta \right)^{-2-2\epsilon-2\nu_1-2\nu_2}\,,
\end{align}
where $y_i$ are the so-called alpha parameters and 
\begin{equation}
[dy]=\prod_{i=1}^6dy_i\delta(1-y_1-y_2-y_3-y_4-y_5-y_6)\,.
\end{equation}
For notational convenience, we defined
\begin{align}
C_1=\frac{\hat{s}^2\mu^{4\epsilon}\tilde{\mu}_1^{4\nu_1}\tilde{\mu}_2^{4\nu_2}e^{2\epsilon \gamma_E}e^{-2i(\nu_1+\nu_2)\pi}\Gamma(2+2\epsilon+2\nu_1+2\nu_2)}{\Gamma^2(1+\nu_1)\Gamma^2(1+\nu_2)} \,,
\end{align}
and
\begin{align}
A_1=& y_4y_5(y_4+y_5)+y_5y_6(y_5+y_6)+y_3^2(y_5+y_6)+y_1(y_3+y_4)^2+(y_2+y_3)(y_5+y_6)^2\nonumber \\
&+ y_3y_4(y_4+y_5)+y_3y_4(y_3+y_5)+y_1y_6^2+y_2y_4^2\,,\nonumber \\
A_2=&y_1y_6(y_3+y_4)+y_4y_6(y_3+y_5)+y_2y_4(y_5+y_6)\,, \nonumber \\
A_3=&y_5(y_4+y_6)+(y_2+y_3)(y_4+y_5+y_6)+y_1(y_2+y_3+y_4+y_6)\,.
\end{align}
The package \texttt{Asy.m} formulates the expansion by regions of a Feynman integral by studying the scaling of each alpha-parameter $y_i$, as opposed to the method used in this work where we defined the regions by studying the scaling behaviour of loop momentum components. The two methods are closely related, as the scaling of each parameter $y_i$ corresponds directly to the scale of the $i$-th denominator factor of the original Feynman integral.
To find the possible scalings of the parameters $y_i$ that lead to non-vanishing integrals, \texttt{Asy.m} uses a geometrical method based on convex hulls \cite{Pak:2010pt}. 
Using the package \texttt{Asy.m} for $\cI^X$, we get 12 regions listed as
\begin{align}\label{eq:regionsasy}
R=&\big(\{0, 0, 0, 0, 0,  0\}, \{0, 0, 0, 0, 2, 2\}, \{0, 0, 2, 2, 0, 0\}, \{0, 0, 0, 2, 0, 2\},\{0, -4, -2, 2, 0, 0\}, \nonumber \\
  &\{0, 4, 4, 4, 2, 6\}, \{0, -2, -2, 0, -2, -2\}, \{0, 
  2, 0, 0, 0, 2\}, \{0, 0, 0, -2, 0, 0\},  \nonumber \\
  & \{0, 0, 0, 0, 0, -2\}, \{0, -2, -2, 0, 0, 
 0\},\{0, 2, 2, 2, 0, 2\}\big)\,.
\end{align}
The $j$-th region of $\cI^X$ is now denoted by the vector $R_j$, which specifies the scales of the alpha parameters. To be precise, one scales $y_i\rightarrow y_i\lambda^{R_j^i}$ with $\lambda\ll 1$ and expands Eq.~\eqref{eq:topoX_full111111a} around $\lambda=0$. This yields the alpha representation of $\mathcal{I}^X$ in the $j$-th region which we denote as $\mathcal{I}^X_j$. 

We have checked that the regions as listed by \texttt{Asy.m} lead to the same regions as we found in momentum space and listed in App.~\ref{sec:allRegionsTopoX}.
For simplicity, we only calculated the LP term of $\mathcal{I}^X_j$, except for $\cI^X_9$ and $\cI^X_{10}$ as these start at NLP. We find that $\mathcal{I}^X_1$-$\mathcal{I}^X_{12}$ are the same as 
$\mathcal{I}^X_{hh}$, 
$\mathcal{I}^X_{cc}$, 
$\mathcal{I}^X_{\bar{c}\bar{c}}$, 
$\mathcal{I}^X_{\bar{c}c}$, 
$\mathcal{I}^X_{\bar{c}uc}$, 
$\mathcal{I}^X_{u\bar{c}c}$, 
$\mathcal{I'}^{X}_{cc'}$, 
$\mathcal{I'}^{X}_{\bar{c}\bar{c}'}$, 
$\mathcal{I''}^{X}_{ch''}$,  
$\mathcal{I''}^{X}_{\bar{c}h''}$, 
$\mathcal{I}^X_{hc}$ 
and $\mathcal{I}^X_{\bar{c}h}$, respectively.

%------------------------------------------
\subsection{Finding regions in momentum space}\label{sec:findregions3}
%------------------------------------------

\begin{figure}
\centering
\subfloat[]{
    \includegraphics[width=.40\textwidth]{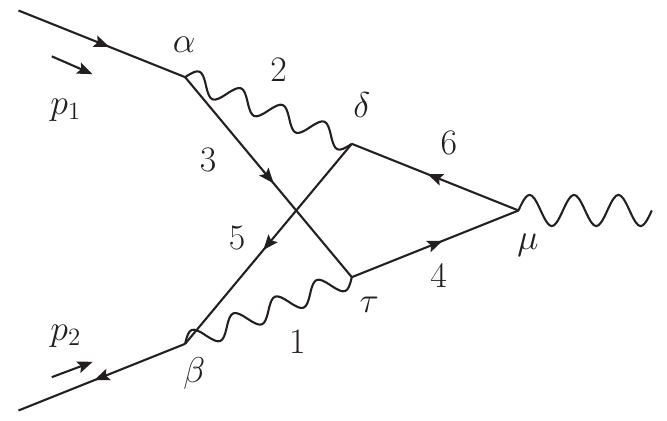}
\label{fig:QED_1_app_C3_a}}
\\
\subfloat[]{
\vspace*{1cm}
    \includegraphics[width=.25\textwidth]{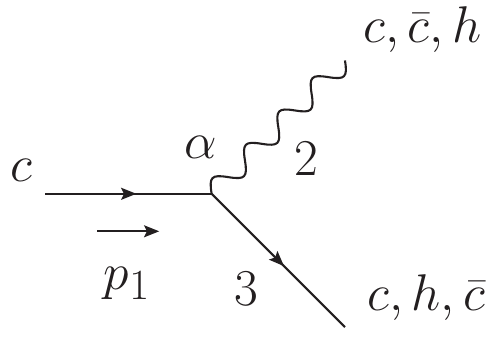}
\label{fig:QED_1_app_C3_b}}
 \qquad \qquad
\subfloat[]{
\vspace*{1cm}
    \includegraphics[width=.25\textwidth]{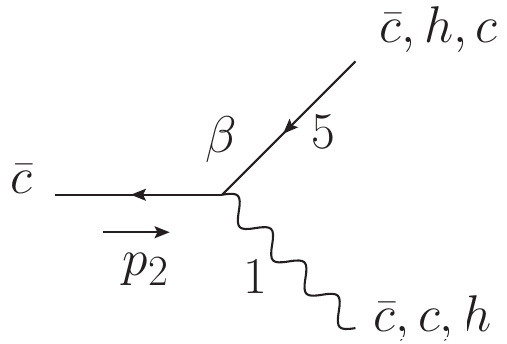}
\label{fig:QED_1_app_C3_c}}
\caption{We show diagram (h) including labels for the vertices and propagators in Fig.~\ref{fig:QED_1_app_C3_a}. The vertices $\alpha$ and $\beta$ are shown in Figs.~\ref{fig:QED_1_app_C3_b} and~\ref{fig:QED_1_app_C3_c} with possible momentum modes for each propagator when the loop momenta are regarded as \mbox{(anti-)collinear}.}
\label{fig:QED_1_app_C3}
\end{figure}

We finish this appendix by presenting a method that can be used to find all the regions in momentum space of a Feynman integral. 
In principle, we can use \texttt{Asy.m} to find the scale of each propagator in a given region and then obtain the corresponding modes of loop momenta. 
However, in our case, it is important to apply an independent cross-check to find the regions in momentum space. 
To illustrate our method, we focus again on the integral $\cI^X$, Eq.~\eqref{eq:topoX_full111111}, and use the collinear-type region -- which means the loop momenta $k_1$ and $k_2$ are both collinear or anti-collinear -- as an example.

Regardless of whether we perform expansion by region or not, the momentum flowing into a vertex of a Feynman diagram is conserved. 
We can use this fact to constrain the possible scales of the propagators connected to the same vertex. 
To make this precise, we labeled the vertices and propagators of $\cI^X$ in Fig.~\ref{fig:QED_1_app_C3_a}. Vertex $\alpha$, shown again in Fig.~\ref{fig:QED_1_app_C3_b}, includes three lines with one of them being the external fermion line with momentum $p_1$ which is regarded as the collinear momentum. At present, we only focus on the collinear-type regions such that the momentum of one of the remaining two lines should be collinear or anti-collinear. 
By momentum conservation, the momentum of the third line is now fixed. 
That is, if the photon line -- labeled by 2 -- has momentum with collinear scaling, then by momentum conservation the fermion line -- labeled by 3 -- has also collinear momentum. However, if the photon line has anti-collinear scaling, then the momentum of the fermion line should be hard.\footnote{Notice that hard here and $h$ in the following refer to the scaling $\sqrt{\hat{s}}(\la^0,\la^0,\la^1)$ as it is the sum of collinear, which scales as $\sqrt{\hat{s}}(\la^0,\la^2,\la^1)$, and anti-collinear, which scales as $\sqrt{\hat{s}}(\la^2,\la^0,\la^1)$. This is slightly different compared to the hard mode defined in Eq.~\eqref{eq:momentumModes1L}.} The only case left for the collinear-type region is when the fermion line has anti-collinear scaling, which leads to a hard momentum scaling for the photon line. In Fig.~\ref{fig:QED_1_app_C3_c}, all possibilities are listed for vertex $\beta$, which attaches to the fermion line with momentum $p_2$, regarded as the anti-collinear momentum.

We notice that once we determine the modes of the momenta that flow into the vertices $\alpha$ and $\beta$, the momentum scaling of every line in the diagram can be extracted.\footnote{In fact, one can also pick the vertices $\alpha$ and $\mu$ or $\beta$ and $\mu$.}
As discussed above, the momenta of the lines labelled 2 and 3 can have the modes $cc$, $\bar{c}h$ and $h\bar{c}$ and the momenta of the lines labelled 1 and 5 can have modes $\bar{c}\bar{c}$, $ch$ and $hc$.
Naively, we have 9 different configurations after considering these two vertices. We divide these 9 configurations into 3 categories, which are given by
\begin{align}
R_1^c&=\left(\bar{c}cc\bar{c}, \bar{c}\bar{c}h\bar{c}, ccch, c\bar{c}hh\right), \nonumber \\
R_2^c&=\left(\bar{c}h\bar{c}\bar{c}, hccc, hh\bar{c}c\right), \nonumber \\
R_3^c&=\left(ch\bar{c}h, h\bar{c}hc\right).
\end{align}
Each configuration, e.g. $\bar{c}cc\bar{c}$ in $R_1^c$, represents the momentum modes of the lines labelled $1$, $2$, $3$ and $5$ respectively. Note that the regions determined by the configurations in $R_1^c$ can be given using the definition of Eq.~\eqref{eq:topoX_full}, while those in $R_2^c$ can be given using Eq.~\eqref{eq:topoX_full2}. We did not show the definitions of the propagators that can be used to give the configurations in $R_3^c$. The reason is that these two configurations in $R_3^c$ only give scaleless integrals and hence do not contribute. It is also straightforward to check that the integrals in both configurations $c\bar{c}hh$ and $hh\bar{c}c$ are scaleless. 
Finally, we find 5 contributing configurations, $\bar{c}cc\bar{c}$, $\bar{c}\bar{c}h\bar{c}$, $ccch$, $\bar{c}h\bar{c}\bar{c}$ and $hccc$ which indeed correspond to the $\bar{c}c$, $\bar{c}\bar{c}$, $cc$, $\bar{c}\bar{c}'$ and $cc'$ regions given in Sec.~\ref{subsec:topoX}, respectively. 
In principle, one can also choose vertices $\alpha$ and $\mu$ or $\beta$ and $\mu$ to analyze the possible collinear-type regions in topology $X$. However, we did not find any additional collinear-type regions that contributed up to NLP. The other regions in topology $X$, where the loop momenta have hard or \mbox{ultra-(anti-)collinear} scaling, can be found following a similar procedure.

% === BIBLIO =====
%\bibliographystyle{JHEP}
%\bibliography{main.bib}

\begin{thebibliography}{10}

\bibitem{Sterman:1987aj}
G.~Sterman, \emph{Summation of large corrections to short distance hadronic
  cross-sections}, {\emph{Nucl. Phys.} {\bfseries B281} (1987) 310}.

\bibitem{Catani:1989ne}
S.~Catani and L.~Trentadue, \emph{{Resummation of the QCD perturbative series
  for hard processes}}, {\emph{Nucl. Phys.} {\bfseries B327} (1989) 323}.

\bibitem{Catani:1990rp}
S.~Catani and L.~Trentadue, \emph{Comment on qcd exponentiation at large x},
  {\emph{Nucl. Phys.} {\bfseries B353} (1991) 183}.

\bibitem{Korchemsky:1993xv}
G.P.~Korchemsky and G.~Marchesini, \emph{Structure function for large x and
  renormalization of wilson loop}, {\emph{Nucl. Phys.} {\bfseries B406} (1993)
  225} [\href{https://arxiv.org/abs/hep-ph/9210281}{{\ttfamily
  hep-ph/9210281}}].

\bibitem{Korchemsky:1993uz}
G.P.~Korchemsky and G.~Marchesini, \emph{{Resummation of large infrared
  corrections using Wilson loops}},
  \href{https://doi.org/10.1016/0370-2693(93)90015-A}{\emph{Phys. Lett.}
  {\bfseries B313} (1993) 433}.

\bibitem{Bauer:2000yr}
C.W.~Bauer, S.~Fleming, D.~Pirjol and I.W.~Stewart, \emph{An effective field
  theory for collinear and soft gluons: Heavy to light decays}, {\emph{Phys.
  Rev.} {\bfseries D63} (2001) 114020}
  [\href{https://arxiv.org/abs/hep-ph/0011336}{{\ttfamily hep-ph/0011336}}].

\bibitem{Bauer:2001yt}
C.W.~Bauer, D.~Pirjol and I.W.~Stewart, \emph{{Soft collinear factorization in
  effective field theory}},
  \href{https://doi.org/10.1103/PhysRevD.65.054022}{\emph{Phys. Rev. D}
  {\bfseries 65} (2002) 054022}
  [\href{https://arxiv.org/abs/hep-ph/0109045}{{\ttfamily hep-ph/0109045}}].

\bibitem{Beneke:2002ph}
M.~Beneke, A.P.~Chapovsky, M.~Diehl and T.~Feldmann, \emph{{Soft collinear
  effective theory and heavy to light currents beyond leading power}},
  \href{https://doi.org/10.1016/S0550-3213(02)00687-9}{\emph{Nucl. Phys. B}
  {\bfseries 643} (2002) 431}
  [\href{https://arxiv.org/abs/hep-ph/0206152}{{\ttfamily hep-ph/0206152}}].

\bibitem{DelDuca:1990gz}
V.~Del~Duca, \emph{High-energy bremsstrahlung theorems for soft photons},
  \href{https://doi.org/10.1016/0550-3213(90)90392-Q}{\emph{Nucl. Phys.}
  {\bfseries B345} (1990) 369}.

\bibitem{Laenen:2010uz}
E.~Laenen, L.~Magnea, G.~Stavenga and C.D.~White, \emph{{Next-to-eikonal
  corrections to soft gluon radiation: a diagrammatic approach}},
  \href{https://doi.org/10.1007/JHEP01(2011)141}{\emph{JHEP} {\bfseries 1101}
  (2011) 141} [\href{https://arxiv.org/abs/1010.1860}{{\ttfamily 1010.1860}}].

\bibitem{Bonocore:2015esa}
D.~Bonocore, E.~Laenen, L.~Magnea, S.~Melville, L.~Vernazza and C.D.~White,
  \emph{{A factorization approach to next-to-leading-power threshold
  logarithms}}, \href{https://doi.org/10.1007/JHEP06(2015)008}{\emph{JHEP}
  {\bfseries 06} (2015) 008}
  [\href{https://arxiv.org/abs/1503.05156}{{\ttfamily 1503.05156}}].

\bibitem{Bonocore:2016awd}
D.~Bonocore, E.~Laenen, L.~Magnea, L.~Vernazza and C.D.~White,
  \emph{{Non-abelian factorisation for next-to-leading-power threshold
  logarithms}}, \href{https://doi.org/10.1007/JHEP12(2016)121}{\emph{JHEP}
  {\bfseries 12} (2016) 121}
  [\href{https://arxiv.org/abs/1610.06842}{{\ttfamily 1610.06842}}].

\bibitem{Qiu:1990xxa}
J.-w.~Qiu and G.F.~Sterman, \emph{{Power corrections in hadronic scattering. 1.
  Leading 1/Q**2 corrections to the Drell-Yan cross-section}},
  \href{https://doi.org/10.1016/0550-3213(91)90503-P}{\emph{Nucl. Phys. B}
  {\bfseries 353} (1991) 105}.

\bibitem{Qiu:1990xy}
J.-w.~Qiu and G.F.~Sterman, \emph{{Power corrections to hadronic scattering. 2.
  Factorization}},
  \href{https://doi.org/10.1016/0550-3213(91)90504-Q}{\emph{Nucl. Phys. B}
  {\bfseries 353} (1991) 137}.

\bibitem{Laenen:2020nrt}
E.~Laenen, J.~Sinninghe~Damst\'e, L.~Vernazza, W.~Waalewijn and L.~Zoppi,
  \emph{{Towards all-order factorization of QED amplitudes at next-to-leading
  power}}, \href{https://doi.org/10.1103/PhysRevD.103.034022}{\emph{Phys. Rev.
  D} {\bfseries 103} (2021) 034022}
  [\href{https://arxiv.org/abs/2008.01736}{{\ttfamily 2008.01736}}].

\bibitem{Gervais:2017yxv}
H.~Gervais, \emph{{Soft photon theorem for high energy amplitudes in Yukawa and
  scalar theories}},
  \href{https://doi.org/10.1103/PhysRevD.95.125009}{\emph{Phys. Rev.}
  {\bfseries D95} (2017) 125009}
  [\href{https://arxiv.org/abs/1704.00806}{{\ttfamily 1704.00806}}].

\bibitem{Beneke:2019oqx}
M.~Beneke, A.~Broggio, S.~Jaskiewicz and L.~Vernazza, \emph{{Threshold
  factorization of the Drell-Yan process at next-to-leading power}},
  \href{https://doi.org/10.1007/JHEP07(2020)078}{\emph{JHEP} {\bfseries 07}
  (2020) 078} [\href{https://arxiv.org/abs/1912.01585}{{\ttfamily
  1912.01585}}].

\bibitem{Larkoski:2014bxa}
A.J.~Larkoski, D.~Neill and I.W.~Stewart, \emph{{Soft theorems from effective
  field theory}}, \href{https://doi.org/10.1007/JHEP06(2015)077}{\emph{JHEP}
  {\bfseries 1506} (2015) 077}
  [\href{https://arxiv.org/abs/1412.3108}{{\ttfamily 1412.3108}}].

\bibitem{Beneke:2017ztn}
M.~Beneke, M.~Garny, R.~Szafron and J.~Wang, \emph{{Anomalous dimension of
  subleading-power N-jet operators}},
  \href{https://doi.org/10.1007/JHEP03(2018)001}{\emph{JHEP} {\bfseries 03}
  (2018) 001} [\href{https://arxiv.org/abs/1712.04416}{{\ttfamily
  1712.04416}}].

\bibitem{Beneke:2018rbh}
M.~Beneke, M.~Garny, R.~Szafron and J.~Wang, \emph{{Anomalous dimension of
  subleading-power $N$-jet operators. Part II}},
  \href{https://doi.org/10.1007/JHEP11(2018)112}{\emph{JHEP} {\bfseries 11}
  (2018) 112} [\href{https://arxiv.org/abs/1808.04742}{{\ttfamily
  1808.04742}}].

\bibitem{Beneke:2019kgv}
M.~Beneke, M.~Garny, R.~Szafron and J.~Wang, \emph{{Violation of the
  Kluberg-Stern-Zuber theorem in SCET}},
  \href{https://doi.org/10.1007/JHEP09(2019)101}{\emph{JHEP} {\bfseries 09}
  (2019) 101} [\href{https://arxiv.org/abs/1907.05463}{{\ttfamily
  1907.05463}}].

\bibitem{Bernreuther:2004ih}
W.~Bernreuther, R.~Bonciani, T.~Gehrmann, R.~Heinesch, T.~Leineweber,
  P.~Mastrolia et~al., \emph{{Two-loop QCD corrections to the heavy quark
  form-factors: The Vector contributions}},
  \href{https://doi.org/10.1016/j.nuclphysb.2004.10.059}{\emph{Nucl. Phys. B}
  {\bfseries 706} (2005) 245}
  [\href{https://arxiv.org/abs/hep-ph/0406046}{{\ttfamily hep-ph/0406046}}].

\bibitem{Gluza:2009yy}
J.~Gluza, A.~Mitov, S.~Moch and T.~Riemann, \emph{{The QCD form factor of heavy
  quarks at NNLO}},
  \href{https://doi.org/10.1088/1126-6708/2009/07/001}{\emph{JHEP} {\bfseries
  07} (2009) 001} [\href{https://arxiv.org/abs/0905.1137}{{\ttfamily
  0905.1137}}].

\bibitem{Blumlein:2020jrf}
J.~Bl\"umlein, A.~De~Freitas, C.~Raab and K.~Sch\"onwald, \emph{{The
  $O(\alpha^2)$ initial state QED corrections to $e^+e^- \rightarrow
  \gamma^*/Z_0^*$}},
  \href{https://doi.org/10.1016/j.nuclphysb.2020.115055}{\emph{Nucl. Phys. B}
  {\bfseries 956} (2020) 115055}
  [\href{https://arxiv.org/abs/2003.14289}{{\ttfamily 2003.14289}}].

\bibitem{Henn:2016kjz}
J.M.~Henn, A.V.~Smirnov and V.A.~Smirnov, \emph{{Analytic results for planar
  three-loop integrals for massive form factors}},
  \href{https://doi.org/10.1007/JHEP12(2016)144}{\emph{JHEP} {\bfseries 12}
  (2016) 144} [\href{https://arxiv.org/abs/1611.06523}{{\ttfamily
  1611.06523}}].

\bibitem{Henn:2016tyf}
J.~Henn, A.V.~Smirnov, V.A.~Smirnov and M.~Steinhauser, \emph{{Massive
  three-loop form factor in the planar limit}},
  \href{https://doi.org/10.1007/JHEP01(2017)074}{\emph{JHEP} {\bfseries 01}
  (2017) 074} [\href{https://arxiv.org/abs/1611.07535}{{\ttfamily
  1611.07535}}].

\bibitem{Ablinger:2017hst}
J.~Ablinger, A.~Behring, J.~Bl\"umlein, G.~Falcioni, A.~De~Freitas, P.~Marquard
  et~al., \emph{{Heavy quark form factors at two loops}},
  \href{https://doi.org/10.1103/PhysRevD.97.094022}{\emph{Phys. Rev. D}
  {\bfseries 97} (2018) 094022}
  [\href{https://arxiv.org/abs/1712.09889}{{\ttfamily 1712.09889}}].

\bibitem{Lee:2018nxa}
R.N.~Lee, A.V.~Smirnov, V.A.~Smirnov and M.~Steinhauser, \emph{{Three-loop
  massive form factors: complete light-fermion corrections for the vector
  current}}, \href{https://doi.org/10.1007/JHEP03(2018)136}{\emph{JHEP}
  {\bfseries 03} (2018) 136}
  [\href{https://arxiv.org/abs/1801.08151}{{\ttfamily 1801.08151}}].

\bibitem{Lee:2018rgs}
R.N.~Lee, A.V.~Smirnov, V.A.~Smirnov and M.~Steinhauser, \emph{{Three-loop
  massive form factors: complete light-fermion and large-N$_{c}$ corrections
  for vector, axial-vector, scalar and pseudo-scalar currents}},
  \href{https://doi.org/10.1007/JHEP05(2018)187}{\emph{JHEP} {\bfseries 05}
  (2018) 187} [\href{https://arxiv.org/abs/1804.07310}{{\ttfamily
  1804.07310}}].

\bibitem{Ablinger:2018yae}
J.~Ablinger, J.~Bl\"umlein, P.~Marquard, N.~Rana and C.~Schneider, \emph{{Heavy
  quark form factors at three loops in the planar limit}},
  \href{https://doi.org/10.1016/j.physletb.2018.05.077}{\emph{Phys. Lett. B}
  {\bfseries 782} (2018) 528}
  [\href{https://arxiv.org/abs/1804.07313}{{\ttfamily 1804.07313}}].

\bibitem{Blumlein:2018tmz}
J.~Bl\"umlein, P.~Marquard and N.~Rana, \emph{{Asymptotic behavior of the heavy
  quark form factors at higher order}},
  \href{https://doi.org/10.1103/PhysRevD.99.016013}{\emph{Phys. Rev. D}
  {\bfseries 99} (2019) 016013}
  [\href{https://arxiv.org/abs/1810.08943}{{\ttfamily 1810.08943}}].

\bibitem{Blumlein:2019oas}
J.~Bl\"umlein, P.~Marquard, N.~Rana and C.~Schneider, \emph{{The Heavy Fermion
  Contributions to the Massive Three Loop Form Factors}},
  \href{https://doi.org/10.1016/j.nuclphysb.2019.114751}{\emph{Nucl. Phys. B}
  {\bfseries 949} (2019) 114751}
  [\href{https://arxiv.org/abs/1908.00357}{{\ttfamily 1908.00357}}].

\bibitem{Fael:2022rgm}
M.~Fael, F.~Lange, K.~Sch\"onwald and M.~Steinhauser, \emph{{Massive Vector
  Form Factors to Three Loops}},
  \href{https://doi.org/10.1103/PhysRevLett.128.172003}{\emph{Phys. Rev. Lett.}
  {\bfseries 128} (2022) 172003}
  [\href{https://arxiv.org/abs/2202.05276}{{\ttfamily 2202.05276}}].

\bibitem{Fael:2022miw}
M.~Fael, F.~Lange, K.~Sch\"onwald and M.~Steinhauser, \emph{{Singlet and
  nonsinglet three-loop massive form factors}},
  \href{https://doi.org/10.1103/PhysRevD.106.034029}{\emph{Phys. Rev. D}
  {\bfseries 106} (2022) 034029}
  [\href{https://arxiv.org/abs/2207.00027}{{\ttfamily 2207.00027}}].

\bibitem{Fael:2023zqr}
M.~Fael, F.~Lange, K.~Sch\"onwald and M.~Steinhauser, \emph{{Massive three-loop
  form factors: Anomaly contribution}},
  \href{https://doi.org/10.1103/PhysRevD.107.094017}{\emph{Phys. Rev. D}
  {\bfseries 107} (2023) 094017}
  [\href{https://arxiv.org/abs/2302.00693}{{\ttfamily 2302.00693}}].

\bibitem{Blumlein:2023uuq}
J.~Bl\"umlein, A.~De~Freitas, P.~Marquard, N.~Rana and C.~Schneider,
  \emph{{Analytic results on the massive three-loop form factors: Quarkonic
  contributions}},
  \href{https://doi.org/10.1103/PhysRevD.108.094003}{\emph{Phys. Rev. D}
  {\bfseries 108} (2023) 094003}
  [\href{https://arxiv.org/abs/2307.02983}{{\ttfamily 2307.02983}}].

\bibitem{Beneke:1997zp}
M.~Beneke and V.A.~Smirnov, \emph{{Asymptotic expansion of Feynman integrals
  near threshold}},
  \href{https://doi.org/10.1016/S0550-3213(98)00138-2}{\emph{Nucl. Phys.}
  {\bfseries B522} (1998) 321}
  [\href{https://arxiv.org/abs/hep-ph/9711391}{{\ttfamily hep-ph/9711391}}].

\bibitem{Smirnov:2002pj}
V.A.~Smirnov, \emph{{Applied asymptotic expansions in momenta and masses}},
  {\emph{Springer Tracts Mod. Phys.} {\bfseries 177} (2002) 1}.

\bibitem{Bonocore:2014wua}
D.~Bonocore, E.~Laenen, L.~Magnea, L.~Vernazza and C.D.~White, \emph{{The
  method of regions and next-to-soft corrections in Drell-Yan production}},
  \href{https://doi.org/10.1016/j.physletb.2015.02.008}{\emph{Phys. Lett.}
  {\bfseries B742} (2015) 375}
  [\href{https://arxiv.org/abs/1410.6406}{{\ttfamily 1410.6406}}].

\bibitem{Bahjat-Abbas:2018hpv}
N.~Bahjat-Abbas, J.~Sinninghe~Damst\'{e}, L.~Vernazza and C.D.~White, \emph{{On
  next-to-leading power threshold corrections in Drell-Yan production at
  N$^3$LO}}, \href{https://doi.org/10.1007/JHEP10(2018)144}{\emph{JHEP}
  {\bfseries 1810} (2018) 144}
  [\href{https://arxiv.org/abs/1807.09246}{{\ttfamily 1807.09246}}].

\bibitem{Smirnov:1998vk}
V.A.~Smirnov and E.R.~Rakhmetov, \emph{{The Strategy of regions for asymptotic
  expansion of two loop vertex Feynman diagrams}},
  \href{https://doi.org/10.1007/BF02557396}{\emph{Theor. Math. Phys.}
  {\bfseries 120} (1999) 870}
  [\href{https://arxiv.org/abs/hep-ph/9812529}{{\ttfamily hep-ph/9812529}}].

\bibitem{Smirnov:1999bza}
V.A.~Smirnov, \emph{{Problems of the strategy of regions}},
  \href{https://doi.org/10.1016/S0370-2693(99)01061-8}{\emph{Phys. Lett. B}
  {\bfseries 465} (1999) 226}
  [\href{https://arxiv.org/abs/hep-ph/9907471}{{\ttfamily hep-ph/9907471}}].

\bibitem{Engel:2018fsb}
T.~Engel, C.~Gnendiger, A.~Signer and Y.~Ulrich, \emph{{Small-mass effects in
  heavy-to-light form factors}},
  \href{https://doi.org/10.1007/JHEP02(2019)118}{\emph{JHEP} {\bfseries 02}
  (2019) 118} [\href{https://arxiv.org/abs/1811.06461}{{\ttfamily
  1811.06461}}].

\bibitem{Becher:2007cu}
T.~Becher and K.~Melnikov, \emph{{Two-loop QED corrections to Bhabha
  scattering}},
  \href{https://doi.org/10.1088/1126-6708/2007/06/084}{\emph{JHEP} {\bfseries
  06} (2007) 084} [\href{https://arxiv.org/abs/0704.3582}{{\ttfamily
  0704.3582}}].

\bibitem{Jantzen:2011nz}
B.~Jantzen, \emph{{Foundation and generalization of the expansion by regions}},
  \href{https://doi.org/10.1007/JHEP12(2011)076}{\emph{JHEP} {\bfseries 1112}
  (2011) 076} [\href{https://arxiv.org/abs/1111.2589}{{\ttfamily 1111.2589}}].

\bibitem{Pak:2010pt}
A.~Pak and A.~Smirnov, \emph{{Geometric approach to asymptotic expansion of
  Feynman integrals}},
  \href{https://doi.org/10.1140/epjc/s10052-011-1626-1}{\emph{Eur.Phys.J.}
  {\bfseries C71} (2011) 1626}
  [\href{https://arxiv.org/abs/1011.4863}{{\ttfamily 1011.4863}}].

\bibitem{Jantzen:2012mw}
B.~Jantzen, A.V.~Smirnov and V.A.~Smirnov, \emph{{Expansion by regions:
  revealing potential and Glauber regions automatically}},
  \href{https://doi.org/10.1140/epjc/s10052-012-2139-2}{\emph{Eur. Phys. J.}
  {\bfseries C72} (2012) 2139}
  [\href{https://arxiv.org/abs/1206.0546}{{\ttfamily 1206.0546}}].

\bibitem{Gardi:2022khw}
E.~Gardi, F.~Herzog, S.~Jones, Y.~Ma and J.~Schlenk, \emph{{The on-shell
  expansion: from Landau equations to the Newton polytope}},
  \href{https://doi.org/10.1007/JHEP07(2023)197}{\emph{JHEP} {\bfseries 07}
  (2023) 197} [\href{https://arxiv.org/abs/2211.14845}{{\ttfamily
  2211.14845}}].

\bibitem{Bonciani:2003ai}
R.~Bonciani, P.~Mastrolia and E.~Remiddi, \emph{{QED vertex form-factors at two
  loops}}, \href{https://doi.org/10.1016/j.nuclphysb.2003.10.031}{\emph{Nucl.
  Phys. B} {\bfseries 676} (2004) 399}
  [\href{https://arxiv.org/abs/hep-ph/0307295}{{\ttfamily hep-ph/0307295}}].

\bibitem{Studerus:2009ye}
C.~Studerus, \emph{{Reduze-Feynman Integral Reduction in C++}},
  \href{https://doi.org/10.1016/j.cpc.2010.03.012}{\emph{Comput. Phys. Commun.}
  {\bfseries 181} (2010) 1293}
  [\href{https://arxiv.org/abs/0912.2546}{{\ttfamily 0912.2546}}].

\bibitem{PhysRev.51.125}
W.H.~Furry, \emph{A symmetry theorem in the positron theory},
  \href{https://doi.org/10.1103/PhysRev.51.125}{\emph{Phys. Rev.} {\bfseries
  51} (1937) 125}.

\bibitem{collins_2023}
J.~Collins, \emph{Foundations of Perturbative QCD}, Cambridge Monographs on
  Particle Physics, Nuclear Physics and Cosmology, Cambridge University Press
  (2023), \href{https://doi.org/10.1017/9781009401845}{10.1017/9781009401845}.

\bibitem{Echevarria:2011epo}
M.G.~Echevarria, A.~Idilbi and I.~Scimemi, \emph{{Factorization Theorem For
  Drell-Yan At Low $q_T$ And Transverse Momentum Distributions
  On-The-Light-Cone}},
  \href{https://doi.org/10.1007/JHEP07(2012)002}{\emph{JHEP} {\bfseries 07}
  (2012) 002} [\href{https://arxiv.org/abs/1111.4996}{{\ttfamily 1111.4996}}].

\bibitem{Echevarria:2012js}
M.G.~Echevarr\'\i{}a, A.~Idilbi and I.~Scimemi, \emph{{Soft and Collinear
  Factorization and Transverse Momentum Dependent Parton Distribution
  Functions}},
  \href{https://doi.org/10.1016/j.physletb.2013.09.003}{\emph{Phys. Lett. B}
  {\bfseries 726} (2013) 795}
  [\href{https://arxiv.org/abs/1211.1947}{{\ttfamily 1211.1947}}].

\bibitem{Echevarria:2015usa}
M.G.~Echevarria, I.~Scimemi and A.~Vladimirov, \emph{{Transverse momentum
  dependent fragmentation function at
  next-to\textendash{}next-to\textendash{}leading order}},
  \href{https://doi.org/10.1103/PhysRevD.93.011502}{\emph{Phys. Rev. D}
  {\bfseries 93} (2016) 011502}
  [\href{https://arxiv.org/abs/1509.06392}{{\ttfamily 1509.06392}}].

\bibitem{Echevarria:2015byo}
M.G.~Echevarria, I.~Scimemi and A.~Vladimirov, \emph{{Universal transverse
  momentum dependent soft function at NNLO}},
  \href{https://doi.org/10.1103/PhysRevD.93.054004}{\emph{Phys. Rev. D}
  {\bfseries 93} (2016) 054004}
  [\href{https://arxiv.org/abs/1511.05590}{{\ttfamily 1511.05590}}].

\bibitem{Echevarria:2016scs}
M.G.~Echevarria, I.~Scimemi and A.~Vladimirov, \emph{{Unpolarized Transverse
  Momentum Dependent Parton Distribution and Fragmentation Functions at
  next-to-next-to-leading order}},
  \href{https://doi.org/10.1007/JHEP09(2016)004}{\emph{JHEP} {\bfseries 09}
  (2016) 004} [\href{https://arxiv.org/abs/1604.07869}{{\ttfamily
  1604.07869}}].

\bibitem{Chiu:2011qc}
J.-y.~Chiu, A.~Jain, D.~Neill and I.Z.~Rothstein, \emph{{The Rapidity
  Renormalization Group}},
  \href{https://doi.org/10.1103/PhysRevLett.108.151601}{\emph{Phys. Rev. Lett.}
  {\bfseries 108} (2012) 151601}
  [\href{https://arxiv.org/abs/1104.0881}{{\ttfamily 1104.0881}}].

\bibitem{Chiu:2012ir}
J.-Y.~Chiu, A.~Jain, D.~Neill and I.Z.~Rothstein, \emph{{A Formalism for the
  Systematic Treatment of Rapidity Logarithms in Quantum Field Theory}},
  \href{https://doi.org/10.1007/JHEP05(2012)084}{\emph{JHEP} {\bfseries 05}
  (2012) 084} [\href{https://arxiv.org/abs/1202.0814}{{\ttfamily 1202.0814}}].

\bibitem{Li:2016axz}
Y.~Li, D.~Neill and H.X.~Zhu, \emph{{An exponential regulator for rapidity
  divergences}},
  \href{https://doi.org/10.1016/j.nuclphysb.2020.115193}{\emph{Nucl. Phys. B}
  {\bfseries 960} (2020) 115193}
  [\href{https://arxiv.org/abs/1604.00392}{{\ttfamily 1604.00392}}].

\bibitem{Beneke:2003pa}
M.~Beneke and T.~Feldmann, \emph{{Factorization of heavy to light form-factors
  in soft collinear effective theory}},
  \href{https://doi.org/10.1016/j.nuclphysb.2004.02.033}{\emph{Nucl. Phys. B}
  {\bfseries 685} (2004) 249}
  [\href{https://arxiv.org/abs/hep-ph/0311335}{{\ttfamily hep-ph/0311335}}].

\bibitem{chiu:2007yn}
J.-y.~Chiu, F.~Golf, R.~Kelley and A.V.~Manohar, \emph{{Electroweak Sudakov
  corrections using effective field theory}},
  \href{https://doi.org/10.1103/PhysRevLett.100.021802}{\emph{Phys. Rev. Lett.}
  {\bfseries 100} (2008) 021802}
  [\href{https://arxiv.org/abs/0709.2377}{{\ttfamily 0709.2377}}].

\bibitem{Becher:2010tm}
T.~Becher and M.~Neubert, \emph{{Drell-Yan Production at Small $q_T$,
  Transverse Parton Distributions and the Collinear Anomaly}},
  \href{https://doi.org/10.1140/epjc/s10052-011-1665-7}{\emph{Eur. Phys. J. C}
  {\bfseries 71} (2011) 1665}
  [\href{https://arxiv.org/abs/1007.4005}{{\ttfamily 1007.4005}}].

\bibitem{Becher:2011dz}
T.~Becher and G.~Bell, \emph{{Analytic Regularization in Soft-Collinear
  Effective Theory}},
  \href{https://doi.org/10.1016/j.physletb.2012.05.016}{\emph{Phys. Lett. B}
  {\bfseries 713} (2012) 41} [\href{https://arxiv.org/abs/1112.3907}{{\ttfamily
  1112.3907}}].

\bibitem{Ebert:2018gsn}
M.A.~Ebert, I.~Moult, I.W.~Stewart, F.J.~Tackmann, G.~Vita and H.X.~Zhu,
  \emph{{Subleading power rapidity divergences and power corrections for
  q$_{T}$}}, \href{https://doi.org/10.1007/JHEP04(2019)123}{\emph{JHEP}
  {\bfseries 04} (2019) 123}
  [\href{https://arxiv.org/abs/1812.08189}{{\ttfamily 1812.08189}}].

\bibitem{Klappert:2020nbg}
J.~Klappert, F.~Lange, P.~Maierh\"ofer and J.~Usovitsch, \emph{{Integral
  reduction with Kira 2.0 and finite field methods}},
  \href{https://doi.org/10.1016/j.cpc.2021.108024}{\emph{Comput. Phys. Commun.}
  {\bfseries 266} (2021) 108024}
  [\href{https://arxiv.org/abs/2008.06494}{{\ttfamily 2008.06494}}].

\bibitem{Lee:2013mka}
R.N.~Lee, \emph{{LiteRed 1.4: a powerful tool for reduction of multiloop
  integrals}}, \href{https://doi.org/10.1088/1742-6596/523/1/012059}{\emph{J.
  Phys. Conf. Ser.} {\bfseries 523} (2014) 012059}
  [\href{https://arxiv.org/abs/1310.1145}{{\ttfamily 1310.1145}}].

\bibitem{Chiu:2007dg}
J.-y.~Chiu, F.~Golf, R.~Kelley and A.V.~Manohar, \emph{{Electroweak Corrections
  in High Energy Processes using Effective Field Theory}},
  \href{https://doi.org/10.1103/PhysRevD.77.053004}{\emph{Phys. Rev. D}
  {\bfseries 77} (2008) 053004}
  [\href{https://arxiv.org/abs/0712.0396}{{\ttfamily 0712.0396}}].

\bibitem{Bonciani:2003te}
R.~Bonciani, P.~Mastrolia and E.~Remiddi, \emph{{Vertex diagrams for the QED
  form-factors at the two loop level}},
  \href{https://doi.org/10.1016/j.nuclphysb.2004.08.009}{\emph{Nucl. Phys. B}
  {\bfseries 661} (2003) 289}
  [\href{https://arxiv.org/abs/hep-ph/0301170}{{\ttfamily hep-ph/0301170}}].

\bibitem{Maitre:2005uu}
D.~Maitre, \emph{{HPL, a mathematica implementation of the harmonic
  polylogarithms}},
  \href{https://doi.org/10.1016/j.cpc.2005.10.008}{\emph{Comput. Phys. Commun.}
  {\bfseries 174} (2006) 222}
  [\href{https://arxiv.org/abs/hep-ph/0507152}{{\ttfamily hep-ph/0507152}}].

\bibitem{Maitre:2007kp}
D.~Maitre, \emph{{Extension of HPL to complex arguments}},
  \href{https://doi.org/10.1016/j.cpc.2011.11.015}{\emph{Comput. Phys. Commun.}
  {\bfseries 183} (2012) 846}
  [\href{https://arxiv.org/abs/hep-ph/0703052}{{\ttfamily hep-ph/0703052}}].

\bibitem{Moch:2005id}
S.~Moch, J.~Vermaseren and A.~Vogt, \emph{{The Quark form-factor at higher
  orders}}, \href{https://doi.org/10.1088/1126-6708/2005/08/049}{\emph{JHEP}
  {\bfseries 0508} (2005) 049}
  [\href{https://arxiv.org/abs/hep-ph/0507039}{{\ttfamily hep-ph/0507039}}].

\bibitem{Gehrmann:2010ue}
T.~Gehrmann, E.W.N.~Glover, T.~Huber, N.~Ikizlerli and C.~Studerus,
  \emph{{Calculation of the quark and gluon form factors to three loops in
  QCD}}, \href{https://doi.org/10.1007/JHEP06(2010)094}{\emph{JHEP} {\bfseries
  06} (2010) 094} [\href{https://arxiv.org/abs/1004.3653}{{\ttfamily
  1004.3653}}].

\bibitem{Pilipp:2008ef}
V.~Pilipp, \emph{{Semi-numerical power expansion of Feynman integrals}},
  \href{https://doi.org/10.1088/1126-6708/2008/09/135}{\emph{JHEP} {\bfseries
  09} (2008) 135} [\href{https://arxiv.org/abs/0808.2555}{{\ttfamily
  0808.2555}}].

\bibitem{Semenova:2018cwy}
T.Y.~Semenova, A.V.~Smirnov and V.A.~Smirnov, \emph{{On the status of expansion
  by regions}},
  \href{https://doi.org/10.1140/epjc/s10052-019-6653-3}{\emph{Eur. Phys. J. C}
  {\bfseries 79} (2019) 136}
  [\href{https://arxiv.org/abs/1809.04325}{{\ttfamily 1809.04325}}].

\bibitem{Ananthanarayan:2018tog}
B.~Ananthanarayan, A.~Pal, S.~Ramanan and R.~Sarkar, \emph{{Unveiling Regions
  in multi-scale Feynman Integrals using Singularities and Power Geometry}},
  \href{https://doi.org/10.1140/epjc/s10052-019-6533-x}{\emph{Eur. Phys. J. C}
  {\bfseries 79} (2019) 57} [\href{https://arxiv.org/abs/1810.06270}{{\ttfamily
  1810.06270}}].

\bibitem{Heinrich:2021dbf}
G.~Heinrich, S.~Jahn, S.P.~Jones, M.~Kerner, F.~Langer, V.~Magerya et~al.,
  \emph{{Expansion by regions with pySecDec}},
  \href{https://doi.org/10.1016/j.cpc.2021.108267}{\emph{Comput. Phys. Commun.}
  {\bfseries 273} (2022) 108267}
  [\href{https://arxiv.org/abs/2108.10807}{{\ttfamily 2108.10807}}].

\bibitem{Gritschacher:2013tza}
S.~Gritschacher, A.~Hoang, I.~Jemos and P.~Pietrulewicz, \emph{{Two loop soft
  function for secondary massive quarks}},
  \href{https://doi.org/10.1103/PhysRevD.89.014035}{\emph{Phys. Rev. D}
  {\bfseries 89} (2014) 014035}
  [\href{https://arxiv.org/abs/1309.6251}{{\ttfamily 1309.6251}}].

\bibitem{Chiu:2009yx}
J.-y.~Chiu, A.~Fuhrer, A.H.~Hoang, R.~Kelley and A.V.~Manohar,
  \emph{{Soft-Collinear Factorization and Zero-Bin Subtractions}},
  \href{https://doi.org/10.1103/PhysRevD.79.053007}{\emph{Phys. Rev. D}
  {\bfseries 79} (2009) 053007}
  [\href{https://arxiv.org/abs/0901.1332}{{\ttfamily 0901.1332}}].

\bibitem{Gonsalves:1983nq}
R.J.~Gonsalves, \emph{Dimensionally regularized two-loop on-shell quark form
  factor}, \href{https://doi.org/10.1103/PhysRevD.28.1542}{\emph{Phys. Rev. D}
  {\bfseries 28} (1983) 1542}.

\end{thebibliography}

\providecommand{\href}[2]{#2}\begingroup\raggedright\endgroup

\end{document}